\documentclass[12pt]{article}
\usepackage{amsfonts}
\usepackage{verbatim,amssymb,amscd,amsmath,graphics,amsthm}
\usepackage{graphicx}
\usepackage{fix-cm}

\numberwithin{equation}{section}

\begin{document}
\title{\bf The Friedrichs Model and its use in resonance phenomena.}

\author{ M. Gadella$^1$,  G. Pronko$^2$.}

\maketitle

$^1$  Departamento de F\'{\i}sica Te\'orica. Facultad de Ciencias,
47071 Valladolid, Spain.

$^2$ Institute for High Energy Physics, Protvino 142284, Moscow
Region, Russia.

\begin{abstract}
We present here a relation of different types of Friedrichs models
and their use in the description and comprehension of resonance
phenomena. We first discuss the basic Friedrichs model and obtain
its resonance in the case that this is  simple or doubly
degenerated. Next, we discuss the model with $N$ levels and show how
the probability amplitude has an oscillatory behavior. Two
generalizations of the Friedrichs model are suitable to introduce
resonance behavior in quantum field theory. We also discuss a
discrete version of the Friedrichs model and also a resonant
interaction between two systems both with continuous spectrum. In an
Appendix, we review the mathematics of rigged Hilbert spaces.
\end{abstract}

\section{Introduction.}

The Friedrichs model is a model aimed to describe the basic features
of resonance phenomena. The basic idea is considering resonances
associated to a Hamiltonian pair $\{H_0,H\}$, where $H_0$ is the
Hamiltonian for the ``non perturbed'' dynamics. $H_0$ has a simple
non-degenerate absolutely continuous spectrum that coincides with
the positive semiaxis. In addition, $H_0$ has at least an eigenvalue
imbedded in the continuous spectrum. The total or ``perturbed''
Hamiltonian has the form $H=H_0+\lambda V$, where $V$ is a potential
and $\lambda$ a coupling parameter that it is usually taken to be
real (to preserve the self adjointness of $H$) and positive. The
potential depends on a form factor function $f(\omega)$, which
determines the existence and properties of the resonance. The action
of the potential is to transform the bound state into a resonance,
characterized by a point in the complex plane, as shall be described
below. This point depends analytically on the coupling parameter
$\lambda$. This is the basic description of the model as originally
introduced by Friedrichs in 1948 \cite{F}.

The first step to show that the Fridrichs model is an excellent
device in order to understand the machinery of decay in Quantum
Mechanics accessible to physicists was given by Horwitz and Marchand
\cite{HM}. After that, there were given several generalizations of
the original model for various purposes including a description of
unstable theory of fields.

In the present review, we intend to discuss most of the known
versions of The Friedrichs model together their applications to
model various situations in which quantum decay appears.

The Fridrichs model was conceived as mathematically rigorous and
exactly solvable so that it could well serve as a toy model for a
precise description of quantum decay. Also its possible
generalizations are enormous in number and vast in applications. The
present review is a first step to collect these generalizations. In
order not to make this paper excessively long, we have selected some
of these generalizations and not included a few ones. Our selection
has been biased by our own work in the field. Examples of
generalizations of the Fridrichs model that we have not included in
our review are:

i.) The Fano-Anderson model \cite{FA,AND,LO}. This is a Fridrichs
model in which the unperturbed Hamiltonian has a bounded absolutely
continuous spectrum. This model is useful in solid state physics to
analyze instabilities due to the presence of resonances (see
references in \cite{LO}).

ii.) The Cascade model \cite{CSB}. This is a model for unstable
field theory.

iii.) A typical example of a generalization of the Fridrichs model
that cannot be solved unless we make some rather severe
approximations is given in a description of the boson-fermion
interaction in nuclei developed by our group \cite{CG1,CGP}.

We also do not intend to discuss some special physical  features
concerning to decay, such as the Zeno effect that deserves a whole
monograph both by its extension and importance \cite{MI,PAS}.
Neither the possible relation of the Friedrichs model with a model
for quantum systems with diagonal singularity \cite{SUC1,SUC2,CGIL}.

We are mainly concern in the study of resonances in the Friedrichs
model ant its various generalizations and we have obtained these
resonances by means of the resolvent and not through the $S$ matrix.
Consequently, we did not attempt to obtain in our examples the $S$
matrix or the M{\o}ller operators, which being relevant in a study
of scattering are not necessary for our purposes\footnote{A study of
the Friedrichs model from the point of view of scattering theory is
given in \cite{E}. Among the big number of papers and textbooks on
the quantum theory of scattering let me quote some absolutely
essential books \cite{NEW,AJS,BAU}. A quite interesting review of
one dimensional quantum scattering is given in \cite{BOYA}.}.

The list of topics to be discussed in the present monograph goes as
follows: In Section 2, we introduce a the basic Friedrichs model
having a simple and as well as a double pole resonance. We define
basic features in a language which can be accessible to both
mathematicians and physicists. We devote 2.1, to the description of
the resonance that emerges along the construction and properties of
its corresponding Gamow vector, i.e., the state vector that decays
exponentially. We also include a situation in which the resonance
comes from a double pole of the resolvent. In this case, the
exponential decay of the resonance is multiplied by a linear
function of time.

The original Friedrichs model considers that the unperturbed
Hamiltonian $H_0$ has one bound state only. What if we assume that
$H_0$ has more than one bound states? Then, the solution to the
problem becomes more complex. In Section 3, we consider the case in
which $H_0$ has two bound states which become resonances due to the
interaction with the potential. This is the two level Friedrichs
model.

The study of the $N$ level Fridrichs model  is of particular
interest. Here, the decay behavior, given by the survival
probability, although modulated by an exponential, becomes
oscillatory. This fact has been observed experimentally \cite{Z},
which incidentally shows the interest of these particular model in
physics.

In Section 5, we introduce two generalizations of the Friedrichs
model to describe resonances in relativistic quantum field theory.
In the first case, two scalar relativistic quantum fields $\psi$ and
$\varphi$ with respective masses $M$ and $m$ have an interaction of
the type $\psi\varphi^2$. This problem can be solved on a sector in
which a particle $\psi$ decays into two particles $\varphi$. The
solution is provided by a generalization of the Friedrichs model,
which shows resonances if $M>2m$ and can be exactly solved.

In the second case, we study the interaction between two boson
fields. One is a local field and the other a bilocal field with a
continuous bounded mass spectrum. The interaction between these two
fields is quadratic. This situation can be solved by formulating it
in terms of a generalized Friedrichs model that can be exactly
solved due to the fact that the interaction in quadratic.

We conclude this section with a discussion of a Friedrichs model
suitable for virtual transitions, which is possibly the simplest
second quantization of the Friedrichs model. It studies the
formation of a photon cloud around an atom.

In Section 6, we present a miscellaneous selection of Friedrichs
models. The former is a discrete version. Here, the non-interaction
hamiltonian is a harmonic oscillators plus a bath of harmonic
oscillators. Then, an interaction is switched on between the first
oscillator and the bath. The second one has the originality of
producing an interaction between two Hamiltonians with continuous
spectrum. One of these Hamiltonians, corresponding to an internal
channel, has a bounded continuous spectrum doubly degenerated. The
second one, corresponding to the external channel, has a continuous
spectrum coinciding with the positive semiaxis and is infinitely
degenerated. An interaction intertwines both Hamiltonians. As a
consequence, the continuous spectrum of the Hamiltonian of the
internal channel is transformed in a branch cut for the resolvent of
the total Hamiltonian. This branch cut is interpreted as a type of
generalized resonance similar to the one obtained in Yukawa type
interactions \cite{N}.

We finish the paper with two Appendices. In the former, we discuss
the concept and main properties of Rigged Hilbert Spaces (RHS). RHS
are essential in the definition and presentation of properties of
Gamow vectors, or vector states for resonances. In a second
Appendix, we show how to calculate the reduced resolvent, which
gives the localization of the resonances,  for the basic Friedrichs
model.

Although we are not going to study in depth the physical
consequences of this model, a comment could be in order here. With
respect to the relation between branch points and van Hove
singularities, it has been reported  \cite{TOM} a behavior which is
contrary to which is obtained in the basic Friedrichs model: for low
dimensional systems the van Hove singularity leads to a non-analytic
dependence on the decay rate associated with the resonance on the
coupling constant.

Next, we start with the definition of resonance that we shall use in
the sequel.

\subsection{Definition of resonance: comments.}

A resonance can be defined into several ways. It can be defined
through the resolvent, or the behavior of the analytic
continuation of the $S$-matrix. Both ways can be shown to be
equivalent in specific models like the standard Friedrichs model.

Any definition of resonance in nonrelativistic Quantum Mechanics
relies in the existence of two dynamics, a {\it free} dynamics,
represented by the {\it unperturbed} or {\it free} Hamiltonian
$H_0$ and a {\it perturbed} dynamics given by the {\it total}
Hamiltonian $H=H_0+V$, where $V$ is a potential responsible of the
resonance behavior. Both Hamiltonians are densely defined on an
infinite dimensional Hilbert space $\cal H$ of pure states of the
system under study.

After the properties of the resolvent of a self adjoint operator,
we now that the resolvents

\begin{equation}
R_{H_0}(\psi,z):=\langle \psi|\frac{1}{H_0-zI}|\psi\rangle
\hskip1cm R_H(\psi,z):=\langle \psi|\frac{1}{H-zI}|\psi\rangle
\label{1}
\end{equation}
are analytic on the complex variable $z$ with a branch cut that
coincides with the continuous spectrum of $H_0$ and $H$, that we
usually assume to be equal, in both cases, to the positive
semiaxis ${\mathbb R}^+$.

Next, we make the following assumptions:

i.) There is a dense set $\cal D$ of vectors in $\cal H$ such that
both $R_{H_0}(\psi,z)$ and $R_H(\psi,z)$ admit an analytic
continuation through the cut.

ii.) For some $\psi\in\cal D$, the partial resolvent $R_H(\psi,z)$
has an isolated singular point (in general a pole) at a point
$z_0$ of analyticity of $R_{H_0}(\psi,z)$.

Then, we say that the Hamiltonian pair has a resonance at the
point $z_0$ \cite{RSIV}.

This is the definition that we shall use in the Friedrichs model.
A second definition of resonance is also very much used and is
suitable for the Friedrichs model:

Let us consider that the Hamiltonian pair $\{H_0,H\}$ satisfies
the sufficient conditions such that the $S$ operator exists. In
this case, an incoming {\it free} state $\psi$ (evolving under the
free dynamics $H_0$) will undergo the interaction given by the
potential $V:=H-H_0$ and will be released as a free state
$\varphi$. The relation between this incoming and outgoing free
states is given by the $S$-operator in the form:

$$
\varphi=S\psi
$$

In the energy representation, we can write
$\varphi(E)=S(E)\psi(E)$. Thus, the $S$-operator is a function of
the energy $E$ (and eventually of other variables). In general,
$S(E)$ is a complex analytic function with no other singularities
than a branch cut coinciding with the positive semiaxis (and we
assume that ${\mathbb R}^+$ is the continuous spectrum of both
$H_0$ and $H$ \cite{N,B}.

If the function of complex variable $S(z)$ determined by $S(E)$
has analytic continuation through the cut (and this can be shown
to be the case under very general causality conditions \cite{N})
and this continuation has poles outside the negative part of the
real axis (then these poles come in complex conjugate pairs), then
these poles are assumed to be resonance poles \cite{B}.

Thus, each resonance pole has a real part $E_R$ and a nonzero
imaginary part $-i\Gamma/2$. The energy $E_R$ corresponds to the
maximum of the bump of the cross section (whenever this
equivalence applies) or the resonance energy and $\gamma$ is the
width of this bump (related to the mean life of the resonance).

The relation between the latter definition of resonance and other
definitions (bump in the cross section, sudden change in the phase
shift, etc) is discussed in \cite{B}.

There are other definitions of resonances in quantum mechanics
that appear in the literature. Under general conditions, these
definitions are equivalent either to the resolvent or the $S$
matrix point of view. These are:

i.) Resonances are the eigenvalues of the dilated Hamiltonian. A
dilation (or dilatation) is a transformation $U(\theta)$ on
$L^2({\mathbb R}^3)$ depending on a complex parameter $\theta$,
defined as follows:

$$
U(\theta)\psi({\bf x})=e^{3\theta/2}\,\psi(e^\theta{\bf x})\,.
$$
Clearly if $\theta$ is real, $U(\theta)$ is unitary. Dilations are
used to define a large class of nonlocal potentials called
dilation analytic potentials \cite{RSIII}. If $V$ is a dilation
analytic potential, then define:

$$
H(\theta):=U(\theta)[H_0+V]U^{-1}(\theta)=e^{-2\theta}\,H_0+V(\theta)\,.
$$
Then, resonances are the solutions of the eigenvalue equation:

$$
H(\theta)\psi_R(\theta)=z_R\psi_R(\theta)\,.
$$
The value of the eigenvalue $z_R$ of $H(\theta)$ does not depend
on the parameter $\theta$ and these eigenvalues are resonances
according to the definition which makes use of the partial
resolvent. For details, see \cite{RSIII}. Dilatations really
provide a method to obtain resonances called the {\it complex
scaling method}.

ii.) When the potential is spherically symmetric, one form of
finding resonances is obtaining solutions of the Schr\"odinger
equation with complex eigenvalues, $\chi(r,z_0)$ satisfying the so
called {\it purely outgoing boundary conditions}. This means that
as $k\longmapsto\infty$, then $\chi(r,z_0)\sim e^{ikr}$, i.e.,
only outgoing wave function can exists. These complex eigenvalues
$z_0$ can be identified with resonances and are poles of the $S$
matrix.

iii.) Resonances are also defined as complex eigenvalues of a
dissipative operator in the context of the Lax-Phillips formalism.
This can be found in \cite{H,S,P}.

There are relations between these definitions of resonances. As
was mentioned above, some of them are consequences of others. For
a review in this subject, see \cite{AG} and references thereof.

\section{The standard Friedrichs model.}

We shall present in here the Friedrichs model in its most simple
form, which has been described in many places
\cite{F,HM,3,4,CG,6,SU1,SU2,TOM1,PGS}. We can introduced it in two
different languages: Hilbert space language or bra ket formalism.
The latter makes use of the rigged Hilbert space construction that
we describe in the last part of the present review. Although we
shall use of the bra-ket formalism along our presentation, we
discuss briefly the Hilbert space presentation first and then we
shall go to the second one which is more familiar to most users.
Then, in its simplest basic form, the Friedrichs model includes the
following ingredients:

\smallskip
1.- A (infinite dimensional) Hilbert space $\cal H$. If we use the
energy representation, this Hilbert space is given by:

\begin{equation}
{\cal H}:={\mathbb C}\oplus L^2({\mathbb R}^+)\,. \label{2}
\end{equation}
Here, $\mathbb C$ is the field of complex numbers, ${\mathbb R}^+$
is the positive semiaxis ${\mathbb R}^+=[0,\infty)$ and
$L^2({\mathbb R}^+)$ is the Hilbert space of square integrable
functions on ${\mathbb R}^+$. We say that we are using the energy
representation because the Hamiltonians we shall use in this
Friedrichs model will have a continuous spectrum coinciding with
${\mathbb R}^+$, then the space of the wave functions in the energy
representation is then $L^2({\mathbb R}^+)$.

After (\ref{2}), any $\psi\in\cal H$ can be written in matricial
form as:

\begin{equation}
\psi=\left( \begin{array}{c} \alpha\\[2ex] \varphi(\omega)
\end{array}\right)\,, \label{3}
\end{equation}
where $\alpha$ is an arbitrary complex number and
$\varphi(\omega)$ an arbitrary function in $L^2({\mathbb R}^+)$.
The scalar product of two vectors in $\cal H$ is given by

\begin{equation}
\left( \left( \begin{array}{c} \alpha\\[2ex] \varphi(\omega)
\end{array}\right), \left( \begin{array}{c} \beta\\[2ex] \eta(\omega)
\end{array}\right) \right)=\alpha^*\beta+ \int_0^\infty
\varphi^*(\omega)\eta(\omega)\,d\omega\,. \label{4}
\end{equation}
Here, $\beta$ is a complex number and $\eta(\omega)\in
L^2({\mathbb R}^+)$.

\smallskip
2.- In order to create resonance phenomena, we need a pair of
Hamiltonians: a {\it free} Hamiltonian  $H_0$ and a {\it total} or
{\it perturbed} Hamiltonian $H$, such that the pair $\{H_0,H\}$ is
able to create resonances\footnote{See \cite{RSIII} for examples
of pairs of resonant Hamiltonians.}. We here construct such a
pair. Let us begin with $H_0$. Its domain (the subspace of all
vectors  $\psi\in\cal H$ such that $H_0\psi\in\cal H$) is given by
the vectors in the form (\ref{3}) such that\footnote{by
definition, the energy representation is those in which the
Hamiltonian is diagonal, or for the case of the continuous
spectrum without degeneracy, the multiplication operator. This is
a consequence of the spectral theorem \cite{RSI}.}
$\omega\varphi(\omega)\in L^2({\mathbb R}^+)$. The action of $H_0$
on a vector in its domain is given by

\begin{equation}
H_0 \left( \begin{array}{c} \alpha\\[2ex] \varphi(\omega)
\end{array}\right)= \left( \begin{array}{c}\omega_0 \alpha\\[2ex] \omega \varphi(\omega)
\end{array}\right)\,,\label{5}
\end{equation}
where $\omega_0$ is a real positive number, i.e., $\omega_0>0$.
Observe that the vector

\begin{equation}
|1\rangle=\left(\begin{array}{c}
1\\[2ex]0\end{array}\right)\label{6}
\end{equation}
is an eigenvector of $H_0$ with eigenvalue $\omega_0$, i.e.,
$H_0|1\rangle=\omega_0|1\rangle$. Note that

\begin{equation}
H_0 \left( \begin{array}{c} 0\\[2ex] \varphi(\omega)
\end{array}\right)= \left( \begin{array}{c} 0\\[2ex] \omega\varphi(\omega)
\end{array}\right)\,,\label{7}
\end{equation}
which means that $H_0$ is the multiplication operator, for the
second component, on the interval ${\mathbb R}^+=[0,\infty)$. We
conclude that, by construction, $H_0$ has a non degenerate
absolutely continuous spectrum that coincides with $[0,\infty)$,
plus an eigenvalue, $\omega_0$. Since $\omega_0>0$, this unique
eigenvalue of $H_0$ is embedded in the continuous spectrum of
$H_0$.

\smallskip
3.- The total Hamiltonian has the form $H=H_0+\lambda V$, where
$\lambda$ is a coupling constant and $V$ is a potential to be
defined. In order to create resonances, $V$ should produce an
interaction between discrete and continuous parts of $H_0$. It is
proposed in the form:

\begin{equation}
V\psi= \left( \begin{array}{c} \int_0^\infty
f(\omega)\varphi(\omega)\,d\omega\\[2ex] \alpha f^*(\omega)
\end{array}\right)\,, \label{8}
\end{equation}
where $\psi$ is the arbitrary vector of $\cal H$ given by
(\ref{2}) and $f(\omega)$ is a function on ${\mathbb R}^+$ called
the {\it form factor}. Observe that $V\psi\in\cal H$ for each
$\psi\in\cal H$ if and only if $f(\omega)\in L^2({\mathbb R}^+)$.
If this were the case, the domain of $V$ is $\cal H$ (and $V$ is
bounded!) and the domain of $H$ coincides with the domain of
$H_0$. This may suggest that we cannot allow\footnote{However, the
RHS formalism allows non square integrable form factors, the
Hamiltonian would be in such case a mapping from the space of test
vectors $\bf\Phi$ into its antidual ${\bf\Phi}^\times$. See last
section for this terminology.} for $f(\omega)$ outside
$L^2({\mathbb R}^+)$. The total Hamiltonian $H$ has a continuous
nondegenerate spectrum equal to $[0,\infty)$.

So far, we have introduced the basic features of the Friedrichs
model, without a discussion of its properties. Let us translate
the above definition in terms of the bra-ket formalism.

We have already defined the ket $|1\rangle\in\cal H$. For each
$\omega$ in the (absolutely) continuous spectrum of $H_0$, there
is an (generalized\footnote{It is generalized in the sense that it
does not belong to the Hilbert space, but to a bigger space that
can be defined in the context of rigged Hilbert spaces \cite{BG}.
Nevertheless, physicists use these generalized eigenvectors in
their calculations.}) eigenvector of $H_0$, $|\omega\rangle$, with
eigenvalue $\omega$, i.e.,
$H_0|\omega\rangle=\omega|\omega\rangle$. Taking this into
account, we can use the following notation for an arbitrary vector
$\psi$ as in (\ref{3}):

\begin{equation}
\psi=\alpha|1\rangle+\int_0^\infty
\varphi(\omega)|\omega\rangle\,d\omega\,.\label{9}
\end{equation}

The expression for the free Hamiltonian  can be derived from the
spectral representation theorem \cite{RSI} (see the specific form
to translate integral spectral representations into the bra-ket
formalism in \cite{AGS} and a generalization in \cite{GG}). In our
case, it takes the following form:

\begin{equation}
H_0= \omega_0|1\rangle\langle 1|+\int_0^\infty \omega\,
|\omega\rangle\langle \omega|\, d\omega\,,\label{10}
\end{equation}
where $|1\rangle\langle 1|$ is the projection into the bound state
$|1\rangle$ of $H_0$ and $|\omega\rangle\langle\omega|$ a similar
object that can be defined for $|\omega\rangle$\footnote{Observe
that the spectral measure for $H_0$ can be written as
$dE_\omega=|\omega\rangle\langle\omega|\,d\omega$. See
\cite{AGS}.}.

Apart from the coupling constant $\lambda$, the interaction
potential $V$ can be written in the following form:

\begin{equation}
V=\int_0^\infty [f(\omega)\,|\omega\rangle\langle 1|+
f^*(\omega)|1\rangle\langle \omega|\,]\,d\omega\,, \label{11}
\end{equation}
where $f(\omega)$ is  the form factor that determines the behavior
of the interaction. We recall that the form factor should be kept
square integrable on the positive semiaxis if we want to keep the
model in the context of Hilbert space, although this restriction
may be unnecessary.  Also, the form factor could be real or
complex, it does not matter. From (\ref{10}), we clearly see that
$V$ produces an interaction between the continuous and discrete
parts of $H_0$ modelled by the form factor function. We want to
repeat again that the interaction also depends on the form factor
$\lambda$, so that $H=H_0+\lambda V$.

We want to show that even this simple Friedrichs model shows
resonance behavior. In fact, this resonance is produced when the
action of the interaction ``dissolves'' the discrete spectrum of
$H_0$ in the continuum. This process depends crucially of the form
factor $f(\omega)$. The position of the resonance will also depend
on the coupling constant $\lambda$ \cite{E}.

In order to search for resonance behavior, we need to fix a
definition of resonance and we shall use the definition proposed
in \cite{RSIII} that we have presented earlier (see (\ref{1})).
Accordingly, let us consider the projection $P:=|1\rangle\langle
1|$ into the subspace spanned by the eigenvector $|1\rangle$ of
$H_0$. Pick $\psi\in\cal H$, as in (\ref{9}). Then, $P\psi=\alpha
|1\rangle$. Then, take the operator, called the reduced resolvent:

\begin{equation}
P(H-zI)^{-1}P\,, \label{12}
\end{equation}
where $I$ is the identity operator. Then, consider the function on
the complex variable $z$, depending of the vector $\psi\in\cal H$:

\begin{equation}
R_\psi(z):=\langle\psi|P(H-zI)^{-1}P|\psi\rangle = |\alpha|^2\,
\langle 1|\frac{1}{H-zI}|1\rangle\,.\label{13}
\end{equation}
The real number $|\alpha|^2$ is irrelevant in the discussion of
the analytic properties of $R_\psi(z)$, which are only contained
in the following function independent of $\psi$:

\begin{equation}
\frac{1}{\eta(z)}:=\langle 1|\frac{1}{H-zI}|1\rangle\,. \label{14}
\end{equation}

If this function, has an analytic continuation and the
continuation has an isolated pole at $z_0=E_R-i\Gamma/2$ with
nonzero imaginary part, this pole should be associated to a
resonance, according to the definition of resonances from the
resolvent properties \cite{RSIII}. The real part of $z_0$, $E_R$
is the resonance energy (or location of the maximum of the bump of
the cross section) and $\Gamma$ the width of the bump.

By extension, (\ref{14}) is also called the reduced resolvent.

Some sufficient conditions for this analytic continuation to exist
and have the desired properties have been studied in \cite{HM,E}.
These are conditions on the form factor $|f(\omega)|^2$. Under
these conditions, we can find an explicit expression for $\eta(z)$
in (\ref{4}) \cite{E}:

\begin{equation}
\eta(z)=z-\omega_0-\lambda^2\int_0^\infty
\frac{|f(\omega)|^2}{z-\omega}\,d\omega\,. \label{15}
\end{equation}

Under the sufficient conditions mentioned earlier \cite{E},
$\eta(z)$ is an analytic function on the complex plane $\mathbb C$
with a cut coinciding with the positive semiaxis (the continuous
spectrum of both $H_0$ and $H$) and admits analytic continuations
beyond the cut from above to below and from below to above. In the
first case, the boundary values of the function on the positive
semiaxis are given by:

\begin{equation}
\eta_+(x)=x-\omega_0-\lambda^2\int_0^\infty
\frac{|f(\omega)|^2}{x-\omega+i0}\,d\omega \label{16}
\end{equation}
and in the second:

\begin{equation}
\eta_-(x)=x-\omega_0-\lambda^2\int_0^\infty
\frac{|f(\omega)|^2}{x-\omega-i0}\,d\omega\,.\label{17}
\end{equation}

These analytic continuations have respective zeroes at the points
$z_0=E_R-i\Gamma/2$ and its complex conjugate $z_0^*=E_R+i\Gamma/2$
(observe that (\ref{16}) and (\ref{17}) are complex conjugate of
each other), under the mentioned sufficient conditions \cite{E}.
Correspondingly, the analytic continuation of the inverses
$\eta_+^{-1}(x)$ and $\eta_-^{-1}(x)$ have poles at $z_0$ and
$z_0^*$ respectively.

It is important to remark that the meaning of the kets
$|\omega\rangle$, some of the above formulas like
(\ref{9}-\ref{11}) as well as the Gamow vectors that will be
introduced from the next section onwards, have a meaning in the
context of extensions of Hilbert space called {\it rigged Hilbert
spaces}. We shall add in the appendices a discussion on this
mathematical tools and their applications in the theory of
resonances.

\subsection{Simple pole resonances.}

These resonance poles may be simple or multiple. The case of the
Friedrichs model with simple pole resonances is the easiest to
study, yet containing all the nontrivial information concerning
resonance behavior. As we have seen before, these resonance poles
come into complex conjugate pairs.

The next objective is to find the Gamow vector for this resonance.
With this goal, let us construct the general state vector in the
following form:

\begin{equation}
\psi= \alpha\,|1\rangle+ \int_0^\infty
\varphi(\omega)\,|\omega\rangle\,d\omega\,.\label{18}
\end{equation}

We want to emphasize that discrete and continuous degrees of
freedom are orthogonal\footnote{Being given an observable $A$ with
both discrete and continuous spectrum, the space spanned by the
bound states is always orthogonal to the space spanned by the
scattering states. }. This idea implies the following relations:

\begin{equation}
\langle 1|1\rangle=1\,, \hskip0.4cm \langle
1|\omega\rangle=\langle \omega|1\rangle =0\,, \hskip0.4cm \langle
\omega|\omega'\rangle=\delta(\omega-\omega')\,.\label{19}
\end{equation}

The Gamow vectors are defined as the generalized eigenvectors of
the total Hamiltonian with complex eigenvalues. These eigenvalues
coincide with the resonance poles (as poles of the analytic
continuation of the resolvent). As resonance poles come into
complex conjugate pairs (in the energy representation), let us
denote by $z_0$ and $z^*_0$ the poles corresponding to a given
resonance. Then, their corresponding eigenvectors (Gamow vectors)
are denoted by $|f_0\rangle$, $|\tilde f_0\rangle$, so that

\begin{equation}
H\,|f_0\rangle =z_0\,|f_0\rangle\,; \hskip0.7cm H\,|\tilde
f_0\rangle =z_0^*|\tilde f_0\rangle \,.\label{20}
\end{equation}

Thus, in order to obtain the Gamow vectors, we have to solve the
above equations. The technique is the following: Let us write the
eigenvalue equation for all $x\ge 0$:

\begin{equation}
(H-x)\Psi(x)=0 \,.\label{21}
\end{equation}
Here $\Psi(x)$ is the (generalized) eigenvector of $H$ with
eigenvalue $x$. The set of eigenvectors $\Psi(x)$ can be written
in the following form:

\begin{equation}
\Psi(x)=\alpha(x)\,|1\rangle+\int_0^\infty
\varphi(x,\omega)\,|\omega\rangle\,d\omega\,, \label{22}
\end{equation}
for all $x>0$.

The next step is to carry (\ref{22}) into (\ref{21}). The result
is:

\begin{eqnarray}
(H-x)\Psi(x)= (\omega_0-x)\alpha(x)\,|1\rangle +\lambda
\int_0^\infty f^*(\omega)\varphi(x,\omega) \,d\omega\,\,|1\rangle
\nonumber\\[2ex] +\int_0^\infty (\omega-x)
\varphi(x,\omega)\,|\omega\rangle\, d\omega +\lambda \alpha(x)
\int_0^\infty f(\omega) \,|\omega\rangle\,d\omega=0 \,. \label{23}
\end{eqnarray}

Taking into account that the bound state $|1\rangle$ and the
continuum are linearly independent, since they are orthogonal
degrees of freedom (see relations (\ref{19})) and omitting the
vectors $|1\rangle$ in the first equation and $|\omega\rangle$ in
the second,  relation (\ref{23}) becomes the following pair of
equations:

\begin{eqnarray}
(\omega_0-x) \,\alpha(x)+\lambda \int_0^\infty
\varphi(x,\omega)\,f^*(\omega)\,d\omega=0\label{24} \\[2ex]
(\omega-x)\varphi(x,\omega)+\lambda \alpha(x)\,
f(\omega)=0\,.\label{25}
\end{eqnarray}
Equation (\ref{25}) can be written as:

\begin{equation}
\varphi(x,\omega)=\lambda \frac{\alpha(x) f(\omega)}{x-\omega\pm
i0}+ c \delta(x-\omega)\,,\label{26}
\end{equation}
where $c$ is an arbitrary constant that can be chosen equal to one
($c=1$), without any change in the essential results.  We see that
two solutions appear, one for each continuation (from above to below
and from below to above). Then, equation (\ref{26}) becomes

\begin{equation}
(\omega_0-x)\alpha(x) +\lambda^2\alpha(x)\int_0^\infty
\frac{|f(\omega)|^2}{x-\omega\pm i0}\, d\omega +\lambda
f^*(x)=0\,. \label{27}
\end{equation}
If we denote as

\begin{equation}
\eta_\pm(x):= \omega_0-x+\lambda^2 \int_0^\infty
\frac{|f(\omega)|^2}{x-\omega\pm i0}\, d\omega\,,\label{28}
\end{equation}
equation (\ref{27}) yields

\begin{equation}
\alpha(x)=-\lambda\frac{f^*(x)}{\eta_\pm (x)}\,.\label{29}
\end{equation}
If we carry this results into (\ref{23}), we finally obtain:

\begin{equation}
\Psi_\pm(x)=|x\rangle -\frac{\lambda f^*(x)}{\eta_\pm (x)}\left[
|1\rangle +\lambda \int_0^\infty \frac{f(\omega)}{x-\omega\pm
i0}\,|\omega\rangle\,d\omega\right]\,.\label{30}
\end{equation}

Let us take one of these two vectors, say $\Psi_+(x)$. This vector
is a functional on a certain space of {\it test vectors},
${\bf\Phi}_+$ as we shall discuss in the Appendix. When  we apply
$\Psi_+(x)$ to a test vector $\varphi_+\in{\bf\Phi}_+$, the result
gives a function on the variable $x$, $\langle
\Psi_+(x)|\varphi_+\rangle$ which admits an analytic continuation
on the lower half plane, $\langle \Psi_+(z)|\varphi_+\rangle$.
This continuation has a pole at the point $z_0$, which is the zero
of the function $\eta_+(z)$. We assume that this pole is simple.
Sufficient conditions for this kind of situations exists and are
discussed in \cite{E}. Then, if we omit the arbitrary
$\varphi_+\in{\bf\Phi}_+$, we can write on a neighborhood of
$z_0$:

\begin{equation}
\Psi_+(z)=\frac{C}{z-z_0}+o(z)\,,\label{31}
\end{equation}
where $o(z)$ denotes a Taylor series on $z_0$, which converges on
a neighborhood of $z_0$ (both $C$ and $o(z)$ are functionals on
${\bf\Phi}_+$).

Let us go back to (\ref{7}). As the analytic continuation
$\Psi_+(z)$ exists, the eigenvalue formula (\ref{7}) should be
also valid for any $z$ at which $\Psi_+(z)$ makes sense, so that
we can write

\begin{equation}
(H-z)\Psi_+(z)=0\,.\label{32}
\end{equation}
Now, if we carry (\ref{31}) into (\ref{32}), we get:

\begin{equation}
0=(H-z)\Psi_+(z)=\frac{1}{z-z_0}\,(H-z)\,
C+(H-z)\,o(z)\,,\label{33}
\end{equation}
which gives

\begin{equation}
(H-z_0)\,C=0\Longrightarrow HC=z_0C\,. \label{34}
\end{equation}
This shows that the residue $C$ in (\ref{31}) is precisely the
Gamow vector corresponding to the resonance pole $z_0$. In order
to obtain the explicit form for this Gamow vector, we just proceed
as one does in a standard calculus of residua. As the pole comes
from a zero of $\eta_+(z)\approx {\rm constant}(z-z_0)+o(z)$, on a
neighborhood of $z_0$, we have:

\begin{equation}
\Psi_+(z)\approx \frac{\rm constant}{(z-z_0)}\;\left[ |1\rangle
+\lambda \int_0^\infty \frac{f(\omega)}{z-\omega+
i0}\,|\omega\rangle\,d\omega\right]+o(z)\,. \label{35}
\end{equation}
Since \cite{GI}

\begin{equation}
\frac{1}{z-\omega+i0}=
\frac{1}{z_0-\omega+i0}-\frac{z-z_0}{(z_0-\omega+i0)^2}+o(z)\,,\label{36}
\end{equation}
we have that

\begin{equation}
\Psi_+(z)\approx \frac{\rm constant}{(z-z_0)}\;\left[ |1\rangle
+\lambda \int_0^\infty \frac{f(\omega)}{z_0-\omega+
i0}\,|\omega\rangle\,d\omega\right]+o(z)\,. \label{37}
\end{equation}
Therefore, save for an irrelevant constant, we conclude that

\begin{equation}
C\equiv |f_0\rangle =|1\rangle+\int_0^\infty \frac{\lambda
f(\omega)}{z_0-\omega+i0}\,|\omega\rangle\,d\omega\,. \label{38}
\end{equation}

With the vector $\Psi_-(x)$ as in (\ref{30}), we can proceed
analogously, now taking into account that the function $\eta_-(x)$
has a simple pole at $z_0^*$, the complex conjugate of $z_0$. By
repeating the same arguments, we can obtain the solution
$|\widetilde f_0\rangle$ of the eigenvalue equation $H|\widetilde
f_0\rangle=z_0^*|\widetilde f_0\rangle$ as:

\begin{equation}
|\widetilde f_0\rangle= |1\rangle+\int_0^\infty \frac{\lambda
f^*(\omega)}{z_0^*-\omega-i0}\,|\omega\rangle\,d\omega\,.
\label{39}
\end{equation}

So far the discussion of the simplest Friedrichs model. The
presentation of the model with a double pole resonance comes next.

\subsection{Double pole resonances.}

In order to show that there exists a Friedrichs model allowing for
double pole resonances, we have to construct an explicit form
factor that produces this effect. From all the above discussions,
we need to show that a given form factor produces a double zero
for $\eta_\pm(z)$. Let us choose $f(\omega)$ such that

\begin{equation}
|f(\omega)|^2=\frac{\sqrt \omega}{P(\omega)}\,,\label{40}
\end{equation}
with

\begin{equation}
P(\omega):=(\omega-\beta)(\omega-\beta^*)\,, \label{41}
\end{equation}
where $\beta$ is a complex number\footnote{Note that (\ref{40})
behaves as $\omega^{-(3/2)}$ and is therefore integrable on
$[0,\infty)$, so that $f(\omega)$ is square integrable.} (with
nonvanishing imaginary part!).  With this choice, we obtain the
following expression for the function $\eta(z)$, defined as in
(\ref{14},\ref{15}):

\begin{equation}
\eta(z)= \omega_0 -z-\pi\lambda^2\left\{
\frac{\sqrt{-z}}{P(z)}-\frac{1}{\beta-\beta^*} \left(\frac{
\sqrt{-\beta}}{ z-\beta} -\frac{\sqrt{-\beta^*}}{z-\beta^*}
\right)\right\}\,.\label{42}
\end{equation}
Next, we propose a change of variables which moves our formulas
from the energy representation to the momentum representation:

\begin{equation}
z=p^2\hskip0.6cm;\hskip0.6cm \beta=b^2 \label{43}
\end{equation}
and write $\varphi(p)=\eta(p^2)=\eta(z)$. Obviously, $p$ is a
complex variable. We have that

\begin{equation}
\varphi(p)=\omega_0-p^2+\frac{i\pi
\lambda^2}{(b-b^*)(p+b)(p-b^*)}\,. \label{44}
\end{equation}

We shall show later that $\eta(z)$ has a double zero if and only
if $\varphi(p)$ has a double zero. This is the first idea to be
considered in order to determine under which conditions our model
can have a double pole resonance. In addition, we have to fix
certain constants. First of all, we have the complex constant $b$.
This gives two undetermined real constants. But we can also play
with the real constants $\omega_0$, $\lambda$ and with the real
and imaginary parts of $p_0$, the double zero of $\varphi(p)$.
Should $\varphi(p)$ have a double zero at the point $p_0$, it
might fulfill the following conditions:

\begin{equation}
\varphi(p_0)=0 \hskip0.5cm ; \hskip0.5cm  \varphi'(p_0)=0
\hskip0.5cm ; \hskip0.5cm \varphi''(p_0)\ne 0\,, \label{45}
\end{equation}
which give four equations, given by the real and imaginary
components of the two identities in (\ref{45}), and a relation for
the six parameters mentioned in the previous paragraph:
$\omega_0,\lambda, {\rm Real}\,p_0, {\rm Im}\,p_0, {\rm Real}\,b$
and ${\rm Im}\, b$. Taking into account that we have two equations
and six parameters, we have in principle two free parameters that
can be chosen arbitrarily among these six. It seems natural to
choose the real constants $\omega_0$ and $\lambda$ as free, since
they are usually data in the Friedrichs model. Then, after not
complicated algebraic manipulations, we have

\begin{eqnarray}
b=\omega_0^{1/2}+2i\left(\frac{\pi\lambda^2}{16\omega_0}\right)^{1/3}\,,
\label{46}\\[3ex]
p_0=\left[\omega_0-\left(\frac{\pi\lambda^2}{16\omega_0}\right)^{2/3}\right]
-i\left(\frac{\pi\lambda^2}{16\omega_0}\right)^{1/3} \,.\label{47}
\end{eqnarray}
If we carry (\ref{46}) and (\ref{47}) into (\ref{44}), we obtain

\begin{equation}
\varphi(p)=-\frac{(p-p_0)^2\,(p+p_0^*)^2}{(p+\tau)\,(p-\tau^*)}\,,
\label{48}
\end{equation}
where $\tau$ is some complex number different from $p_0^*$, which is
irrelevant in our discussion because it corresponds to a simple zero
of the reduced resolvent. Equation (\ref{48}) shows that
$\varphi(p)$ has a double zero at the point $p_0$ and another double
zero at its complex conjugate $p_0^*$.

Now, we have to show that the zeroes of $\eta(z)$ and $\varphi(p)$
coincide, after the change of variables, as well as their
multiplicity. To begin with, let us assume that $\eta(z)$ has a
double zero at $z_0$. Then, there exists a function $\chi(z)$
analytic on a neighborhood of $z_0$, with $\chi(z_0)\ne 0$, and
such that on this neighborhood one has:

$$
\eta(z)=(z-z_0)^2\chi(z)\,.
$$
The function $\varphi(p)=\eta(p^2)$ and therefore,

$$
\varphi(p)=
\eta(p^2)=(p^2-p_0^2)^2\chi(p^2)=(p-p_0)^2(p+p_0)^2\chi(p^2)\,.
$$
If $\Delta(p):=(p+p_0)^2\chi(p^2)$, we have that
$\Delta(p_0)=2p_0^2\chi(p_0^2)\ne 0$ and

$$
\varphi(p)=(p-p_0)^2\Delta(p)
$$
and $\varphi(p)$ has a double pole at $p_0=\sqrt{z_0}$.
Conversely, let us assume that $\varphi(p)$ has a double pole at
$p_0$. Then, there exists an analytic function $\xi(p)$ on a
neighborhood of $p_0$, with $\xi(p_0)\ne 0$, such that

$$
\varphi(p)=(p-p_0)^2\xi(p)=\eta(z)\,.
$$
Then, $\eta(z)$ has a double pole at $z_0=p_0^2$ if and only if
the function $\frac{\eta(z)}{(z-z_0)^2}$ has a nonvanishing limit
if $z$ goes to $z_0$. Thus,

$$
\lim_{z\mapsto z_0} \frac{\eta(z)}{(z-z_0)^2}=\lim_{p\mapsto p_0}
\frac{(p-p_0)^2\xi(p)}{(p^2-p_0^2)^2}= \lim_{p\mapsto
p_0}\frac{(p-p_0)^2\xi(p)}{(p-p_0)^2(p+p_0)^2}
=\frac{\xi(p_0)}{2p_0}\ne 0\,,
$$
which proves our claim. The conclusion is that $\eta(z_0)$ has a
double zero at $z_0$ if and only if $\varphi(p)$ has a double zero
at $p_0$ with $z_0=p_0^2$.

Consequently, the analytic continuation through ${\mathbb R}^+$ of
(\ref{11}) has a double pole at the points $z_0$ and $z_0^*$ and
therefore the version of the Friedrichs model here discussed shows
a double pole resonance.

We would like to obtain the Gamow vectors for this resonance. Once
we have established the existence of a double pole resonance, on a
neighborhood of $z_0$, the vector $\Psi_+(z)$ as in (\ref{30}) has
the following behavior\footnote{As $\Psi_+(z)$ represents a
functional on a certain space of test vectors ${\bf\Phi}_+$ (to be
specified later) this behavior corresponds to
$\langle\phi_+|\Psi_+(z)\rangle$ for each
$\phi_+\in{\bf\Phi}_+$.}:

\begin{equation}\label{49}
\Psi_+(z)=\frac{C_1}{(z-z_0)^2}+\frac{C_2}{z-z_0}+o(z)\,.
\end{equation}
As in the case of a simple pole resonance, we have on a
neighborhood of $z_0$

\begin{eqnarray}
  \Psi_+(z) &\approx & \frac{{\rm constant}}{(z-z_0)^2}\left\{ |1\rangle+
  \lambda\int_0^\infty \frac{f(\omega)}{z-\omega+i0}\,|\omega\rangle
  \,d\omega\right\}\nonumber \\[2ex]
   &=& \frac{{\rm constant}}{(z-z_0)^2}\left\{ |1\rangle+
  \lambda\int_0^\infty \frac{f(\omega)}{z_0-\omega+i0}\,|\omega\rangle
  \,d\omega\right.\nonumber \\ [2ex]
   &-& \left. \lambda(z-z_0)\int_0^\infty \frac{f(\omega)}{(z_0-\omega+i0)^2}\,|\omega\rangle
  \,d\omega\right\}\,.\label{50}
\end{eqnarray}
Then,

\begin{eqnarray}
  C_1 &=& |1\rangle+\lambda\int_0^\infty \frac{f(\omega)}{z_0-\omega+i0}\,|\omega\rangle\,d\omega
  \label{51} \\[2ex]
  C_2 &=& -\lambda\int_0^\infty \frac{f(\omega)}{(z_0-\omega+i0)^2}\,|\omega\rangle
  \,d\omega\,.\label{52}
\end{eqnarray}
Now, taking into account (\ref{35}), we have:

\begin{equation}\label{53}
    (H-z)\left[\frac{C_1}{(z-z_0)^2}+\frac{C_2}{z-z_0}+o(z)\right]=0\,.
\end{equation}
If we multiply (\ref{53}) by $(z-z_0)^2$ and take the limit as
$z_0\longmapsto 0$, we obtain:

\begin{equation}\label{54}
(H-z_0)\,C_1=0 \Longleftrightarrow HC_1=z_0C_1\,,
\end{equation}
since $o(z)$ is a regular function on $z$. If now, we multiply
(\ref{50}) by $z-z_0$, we get

\begin{eqnarray}
    (H-z)\left[\frac{C_1}{z-z_0}+C_2+(z-z_0)o(z)\right]=0  \nonumber\\
    [2ex]\Longrightarrow
    \frac{z_0-z}{z-z_0}C_1+(H-z)C_2+ (z-z_0)(H-z)o(z)=0\,.\label{55}
\end{eqnarray}
Taking the limit as $z\longmapsto z_0$ in (\ref{55}), one readily
obtains

\begin{equation}\label{56}
    HC_2=z_0C_2+C_1\,.
\end{equation}
Summarizing, we obtain the following relations

\begin{equation}\label{57}
HC_1=z_0C_1 \hskip0.6cm {\rm and} \hskip0.6cm HC_2=z_0C_2+C_1\,.
\end{equation}

The vectors $C_1$ and $C_2$ are linearly independent (otherwise
$C_2$ would be an eigenvector of $H$ with eigenvalue $z_0$ in
clear contradiction of (\ref{56}). These vectors belong to the
antidual ${\bf\Phi}_+^\times$ in the rigged Hilbert space
${\bf\Phi}_+\subset{\cal H}\subset{\bf\Phi}_+^\times$ (see
Appendix), thus, $C_1$ and $C_2$ are functionals. The restriction
of $H$ on the two dimensional subspace of ${\bf\Phi}_+^\times$
with basis $\{C_1,C_2\}$ has the following form

\begin{equation}\label{58}
    \left(
\begin{array}{cc}
  z_0 & 1 \\
  0 & z_0 \\
\end{array}
\right)\,,
\end{equation}
which shows a block diagonal form. This kind of structure has been
discussed in several contexts in \cite{BRA,CD,M,M1,M2}.
Conventionally, $C_1$ and $C_2$ are called $|f_0\rangle$ and
$|f_1\rangle$ respectively \cite{AGP,CG}.

As with the Friedrichs model with a simple pole resonance, we can
proceed analogously with the vector $\Psi_-(x)$ as in (\ref{27}),
now taking into account that the function $\eta_-(x)$ has a double
pole at $z_0^*$, the complex conjugate of $z_0$. By repeating the
same arguments, we can obtain the solution $|\widetilde
f_0\rangle$ of the eigenvalue equation $H|\widetilde
f_0\rangle=z_0^*|\widetilde f_0\rangle$ as in (\ref{39}) and  a
second vector $|\widetilde f_1\rangle$, which is given by:

\begin{equation}\label{59}
|\widetilde f_1\rangle=-\lambda\int_0^\infty
\frac{f^*(\omega)}{(z_0^*-\omega-i0)^2}\,|\omega\rangle\,d\omega\,.
\end{equation}

The vectors $|\widetilde f_0\rangle$ and $|\widetilde f_1\rangle$
belong to the antidual ${\bf\Phi}_-^\times$ of a rigged Hilbert
space ${\bf\Phi}_-\subset{\cal H}\subset{\bf\Phi}_-^\times$ and
are therefore functionals on the space ${\bf\Phi}_-$ to be defined
in the appendix. As in the previous case, the total Hamiltonian
$H$ can be extended into the antidual ${\bf\Phi}_-^\times$ and we
can also prove the follwoing relations:

\begin{equation}\label{60}
    H|\widetilde f_0\rangle=z_0^*|\widetilde f_0\rangle
    \hskip0.5cm;\hskip0.5cm H|\widetilde f_1\rangle=z_0|\widetilde
    f_1\rangle+ |\widetilde f_0\rangle\,.
\end{equation}
Therefore, on the subspace spanned by the vectors $|\widetilde
f_0\rangle$ and $|\widetilde f_1\rangle$ the total Hamiltonian has
the following form:

\begin{equation}\label{61}
  H=  \left(
\begin{array}{cc}
  z_0^* & 1 \\
  0 & z_0^* \\
\end{array}
\right)\,.
\end{equation}

We may also study the time evolution of the Gamow vectors
$|f_0\rangle$, $|f_1\rangle$, $|\widetilde f_0\rangle$ and
$|\widetilde f_1\rangle$ (sometimes also called Jordan-Gamow
vectors, see \cite{BJG}). As $|f_0\rangle$ and $|f_1\rangle$ in
one side and $|\widetilde f_0\rangle$ and $|\widetilde f_1\rangle$
on the other belong to different spaces, we must study their time
evolution separately. A thoroughly discussion has been presented
elsewhere \cite{AGP,CG}. The final result is given by

\begin{eqnarray}
  e^{-itH}|f_0\rangle &=& e^{-itz_0}|f_0\rangle \label{62} \\[2ex]
  e^{-itH}|f_1\rangle &=& e^{-itz_0}|f_1\rangle-it|f_0\rangle
  \label{63}
\end{eqnarray}
and

\begin{eqnarray}
  e^{-itH}|\widetilde f_0\rangle &=& e^{-itz^*_0}|\widetilde f_0\rangle \label{64}\\[2ex]
  e^{-itH}|\widetilde f_1\rangle &=& e^{-itz^*_0}|\widetilde f_1\rangle+it|\widetilde
  f_0\rangle\,.
  \label{65}
\end{eqnarray}

As for the case of simple pole resonances, the above time
evolutions can be either valid for all values of time or just for
the values $t>0$ (in (\ref{62},\ref{63})) and $t<0$ (in
(\ref{64},\ref{65})), depending on the construction of the spaces
${\bf\Phi}_\pm$ \cite{BG,CG,AGP,CGL}.

\section{Friedrichs model with two bound states.}

In the present case, we start with an unperturbed Hamiltonian of
the following form:

\begin{eqnarray}
H=\omega_1|1\rangle\langle 1|+\omega_2|2\rangle\langle
2|+\int_0^\infty \omega|\omega\rangle\langle\omega|\,d\omega
\nonumber\\[2ex]+
\sum_{i=1}^2[f_i^*(\omega)\,|\omega\rangle\langle
i|+f_i(\omega)|i\rangle\langle \omega|\,]\,d\omega\,.\label{66}
\end{eqnarray}

The most general vector in the space of pure states should have
the following form:

\begin{equation}
\Psi=\sum_{i=1}^2\varphi_i(E)\,|i\rangle+\int_0^\infty
\phi(E,\omega)\,|\omega\rangle\,d\omega\,.\label{67}
\end{equation}
Let us carry (\ref{66}) into (\ref{67}) to obtain the eigenvalue
equation. One gets:

\begin{eqnarray}
(H-E)\Psi=\sum_{i=1}^2\varphi_i(E)(\omega_i-E)\,|i\rangle
+\int_0^\infty(\omega-E)\phi(E,\omega)\,|\omega\rangle\,d\omega\nonumber\\
[2ex]+\sum_{i=1}^2\int_0^\infty
f_i^*(\omega)\varphi_i(E)\,|\omega\rangle\,d\omega + \sum_{i=1}^2
\left[\int_0^\infty
\phi(E,\omega)f_i(\omega)\,d\omega\right]\,|i\rangle \nonumber\\
[2ex]=0 \label{68}
\end{eqnarray}

From (\ref{68}), we obtain the following pair of equations:

\begin{eqnarray}
(\omega_i-E)\varphi_i(E)+\int_0^\infty
\phi(E,\omega)f_i(\omega)\,d\omega=0\,, \hskip0.5cm
i=1,2\,.\label{69}\\[2ex]
(\omega-E)\phi(E,\omega)+\sum_{i=1}^2
f_i^*(\omega)\varphi_i(E)=0\,.\label{70}
\end{eqnarray}

It is rather straightforward to obtain the solution to this system
of equations. If we write $\phi(E,\omega)$ in terms of
$\varphi_i(E)$, we get:

\begin{equation}
\phi(E,\omega)=c\delta(\omega-E
)-\sum_{i=1}^2\frac{f_i(\omega)\varphi_i(E)}{\omega-E+i0}\,.
\label{71}
\end{equation}
If we carry (\ref{71}) into (\ref{69}), we obtain:

\begin{equation}
(\omega_i-E)\varphi_i-\sum_{i=1}^2\int_0^\infty
\frac{f_i(\omega)f_k(\omega)}{\omega-E+i0
}\,\varphi_k(E)\,d\omega=-cf_i(E)\,. \label{72}
\end{equation}
Then, (\ref{72}) give a two equation system with two undetermined
functions $\varphi_i(E)$, $i=1,2$. This system can be rewritten as
follows:

\begin{eqnarray}
(\omega_1-E)\varphi_1(E)-\left[\int_0^\infty
\frac{|f_1(\omega)|^2}{\omega-E+i0
}\,d\omega\right]\varphi_1(E)-\left[\int_0^\infty
\frac{f_1(\omega)\,f_2^*(\omega)}{\omega-E+i0}\,d\omega\right]\varphi_2(E)\nonumber\\[2ex]
=-cf_1(E)\,,\hskip0.6cm \label{73}\\[2ex]
(\omega_1-E)\varphi_2(E)-\left[\int_0^\infty\frac{|f_2(\omega)|^2}{\omega-E+i0
}\,d\omega\right] \varphi_2(E)-\left[\int_0^\infty
\frac{f_2(\omega)f_1^*(\omega)}{\omega-E+i0}\,d\omega\right]\varphi_1(E)\nonumber\\[2ex]
=-cf_2(E)\,. \hskip0.6cm\label{74}
\end{eqnarray}
Note that the determinant of the coefficients is given by:

\begin{equation}
\Delta=\left| \begin{array}{cc} \omega_1-\int_0^\infty
\frac{|f_1|^2}{\omega-E+i0}\,d\omega & -\int_0^\infty
\frac{f_1(\omega)f_2^*(\omega)}{\omega-E+i0}\,d\omega\\[3ex]
-\int_0^\infty
\frac{f_1^*(\omega)f_2(\omega)}{\omega-E+i0}\,d\omega &
\omega_2-\int_0^\infty \frac{|f_2|^2}{\omega-E+i0}\,d\omega
\end{array}  \right|\;.\label{75}
\end{equation}
 Resonance poles are complex solutions on the variable energy of the equation
 $\Delta=0$. The solutions of the above system of equations
 (\ref{73},\ref{74}) are given by

 \begin{equation}
\varphi_1(E)=-\frac c\Delta\left[ \left( \omega_2-E-\int_0^\infty
\frac{|f_2(\omega)|^2}{\omega-E+i0 }\,d\omega
\right)f_1(E)+f_2(E)\int_0^\infty
\frac{f_1(\omega)f_2^*(\omega)}{\omega-E+i0}\,d\omega
\right]\label{76}
 \end{equation}
 and

 \begin{equation}
\varphi_2(E)=-\frac c\Delta\left[ \left( \omega_1-E-\int_0^\infty
\frac{|f_1(\omega)|^2}{\omega-E+i0}\,d\omega
\right)f_2(E)+f_1(E)\int_0^\infty
\frac{f_2(\omega)f_1^*(\omega)}{\omega-E+i0}\,d\omega
\right]\,.\label{77}
 \end{equation}

As we note earlier, the search for resonance poles is the search
for solutions in the energy of the equation $\Delta=0$. At the
first sight, this equation looks difficult to solve. Therefore, we
may try for an approximation that gives us an idea on the behavior
of the resonance poles. This approximation is the following:
taking into account that for $i=1,2$

\begin{equation}
\int_0^\infty \frac{|f_1(\omega)|^2}{\omega-E+i0}\,d\omega={\rm
PV} \int_0^\infty
\frac{|f_1(\omega)|^2}{\omega-E}\,d\omega-i\pi|f_1(E)|^2\,,\label{78}
\end{equation}
where PV stands for principal value. If we assume that this
principal value is negligible, the value of the determinant
$\Delta$ is left as:

\begin{eqnarray}
\Delta\approx
(\omega_1-E-i\pi|f_1(E)|^2)(\omega_2-E-i\pi|f_2(E)|^2)\nonumber\\[2ex]
-\int_0^\infty
\left(\frac{f_1(\omega)f_2^*(\omega)}{\omega-E+i0}\,d\omega\right)
\left(\int_0^\infty
\frac{f_2(\omega)f_1^*(\omega)}{\omega-E+i0}\,d\omega\right)
\nonumber\\[2ex]
\approx
(\omega_1-E)(\omega_2-E)-i\pi(\omega_1-E)|f_1(E)|^2-i\pi(\omega_2-E)|f_2(E)|^2=0\,.\label{79}
\end{eqnarray}
Note that the last approximation in (\ref{79}) neglects the terms
of fourth order in $f_i(E)$. In the last line of (\ref{79})
appears the approximate form of the equation $\Delta=0$. The
complex zeroes of this equation gives the resonance poles, which
depend on the values of the form factors $f_i(E)$. In the simplest
case, $f_i(E)$ could be approximated by constants and equation
(\ref{79}) gives two complex solutions. However, this
approximation may not be compatible with the negligibility of the
principal value. Then, equation (\ref{79}) is in general much more
complicated than a equation of second order in the variable $E$
and may give even an infinite number of complex solutions.

\section{Friedrichs model with $N$ levels: oscillating decay.}

In this section, we shall discuss a generalization of the previous
one: Now, we have $N$ discrete levels imbedded in the continuous
spectrum of the total Hamiltonian. In the present case, the
Hamiltonian is a straightforward generalization of (\ref{66}).
Here, $H=H_0+\lambda V$ with

\begin{eqnarray}
  H_0 &=& \sum_{k=1}^N \omega_k|k\rangle\langle k|+\int_0^\infty \omega|\omega\rangle\langle
  \omega|\,d\omega \nonumber\\[2ex]
  V &=& \sum_{k=1}^N\int_0^\infty f_k(\omega) [|k\rangle\langle
  \omega|+|\omega\rangle\langle k|]\,d\omega\,, \label{n1}
\end{eqnarray}
where $|k\rangle$ represents the state with energy $\omega_k>0$,
$k=1,2,\dots,N$. We shall assume no degeneracy, i.e., $\omega_k\ne
\omega_{k'}$ if $k\ne k'$. The form factors $f_k(\omega)$ are
assumed to be square integrable, just as in the simplest case.
Here, the Hilbert space of states is given by ${\cal H}={\mathbb
C}^N\oplus L^2({\mathbb R}^+)$. Thus, relations (\ref{19}) read in
the present case as

\begin{equation}\label{n2}
    \langle k|k'\rangle=\delta_{kk'}\,, \hskip1cm \langle
    \omega|k\rangle=\langle k|\omega\rangle=0 \,, \hskip1cm
    \langle\omega|\omega'\rangle=\delta(\omega-\omega')\,,
\end{equation}
with $k=1,2,\dots,N$. The situation we can expect is that, once
the potential $\lambda V$ is switch on, the $N$ levels become
resonances \cite{E}. We want to study the eigenvalue equation
(\ref{21}), in the present situation. The form of $\Psi(x)$,
solution of (\ref{21}), is given by

\begin{equation}\label{n3}
    \Psi(x)=\sum_{k=1}^N \psi_k(x)|k\rangle+\int_0^\infty
    \psi(x,\omega)|\omega\rangle\,d\omega\,.
\end{equation}
Then, we insert (\ref{n3}) into (\ref{21}) to obtain a system of
equations, as (\ref{24}) and (\ref{25}):

\begin{eqnarray}
  (\omega_k-x)\psi_k(x)+\lambda\int_0^\infty f_k(\omega)\psi(x,\omega)\,d\omega &=&
  0\label{n4}
  \\[2ex]
  (\omega-x)\psi(x,\omega)+\lambda\sum_{k=1}^N f_k(\omega)\psi_k(x) &=&
  0\,.\label{n5}
\end{eqnarray}
We can eliminate the function $\psi(x,\omega)$ in this system to
get the following equation for $\psi_k(x)$ (note that
$\psi(x,\omega)=-[\sum_{k=1}^N f_k(\omega)\,\psi_k(x)] /
(\omega-x)-C\delta(\omega-x)$, $C$ being an arbitrary constant):

\begin{equation}\label{n6}
    \sum_{k=1}^N [G^{-1}]_{kk'}(x)\psi_k(x)=-C\lambda f_k(x)\,,
\end{equation}
with

\begin{equation}\label{n7}
[G^{-1}]_{kk'}(x)=(\omega_k-x)\delta_{kk'}-\lambda^2\int_0^\infty
\frac{f_k(\omega)f_{k'}(\omega)}{\omega-x}\,d\omega\,.
\end{equation}
In (\ref{n6}), $C$ is an arbitrary constant. We have to provide
conditions for the analyticity of the function $G_{kk'}(x)$, when
$x$ is a complex variable. The solution of (\ref{n6}) is

\begin{equation}\label{n8}
    \psi_k(x)=-C\lambda\sum_{k'=1}^N G_{kk'}(x\pm i0)f_{k'}(x)\,.
\end{equation}
When the specific conditions for the functions $[G^{-1}]_{kk'}(x)$
to have an analytic continuation are satisfied \cite{E}, one can
show that this analytic continuation includes a branch cut that
coincides with the spectrum of $H$ which is purely continuous in the
positive semiaxis ${\mathbb R}^+\equiv [0,\infty)$. They have
analytic continuation from above to below and from below to above
through the cut, that obviously cannot coincide with the values of
the function of the complex variable $x$ on the lower and upper half
planes respectively\footnote{This is the reason for defining these
analytic continuations on the second sheet of a Riemann surface.
Note that the analytic structure of the continuation is what it
matters and not the Riemann surface that we can ignore.}. The
respective boundary values on ${\mathbb R}^+$ are denoted with the
signs $-$ and $+$ respectively. This solution and (\ref{n5}) give us
the form of the function $\psi(x,\omega)$ as

\begin{equation}\label{n9}
\psi(x,\omega)=C\left[\delta(x-\omega)+\frac{\lambda^2\sum_{k'=1}^N
f_k(\omega)[G^{-1}]_{kk'}(x)f_{k'}(x)}{x-\omega\pm i0}\right]\,.
\end{equation}
Once we have obtained the solutions of equations
(\ref{n4},\ref{n5}), we have two solutions of the eigenvalue
problem $(H-E)\Psi(x)=0$ that we shall denote as $\Psi_\pm(x)$.
Alternatively and following the typical terminology of scattering
theory, these solutions are called ``in'' (for $+$) and ``out''
(for $-$), i.e., $\Psi_+(x)\equiv\Psi_{\rm in}(x)$ and
$\Psi_-(x)\equiv\Psi_{\rm out}$ \cite{AKPY}. These solutions are:

\begin{equation}\label{n10}
    \Psi_\pm(x)=|x\rangle +\lambda \sum_{k,l=1}^N
    f_l(x)G_{kl}(x\pm i0) \left\{\int_0^\infty \frac{\lambda f_k(\omega)}{\omega-x\mp i0}
    \,|\omega\rangle-|k\rangle\right\}\,.
\end{equation}
Here, $H_0|x\rangle=x|x\rangle$. The $|x\rangle$ in (\ref{n10})
comes from the delta term in (\ref{n9}). We have chosen the value
$C=1$, which gives the following normalization condition:

\begin{equation}\label{n11}
    \langle\Psi_\pm(x)|\Psi_\pm(x')\rangle=\delta(x-x')\,.
\end{equation}
The solutions of the eigenvalue problem $(H-E)\Psi(E)=0$, for
$E\in[0,\infty)$, are complete in the sense that

\begin{equation}\label{n12}
  I=  \int_0^\infty
    |\Psi_\pm(x)\rangle\langle\Psi_\pm(x)|\,dx
\end{equation}
and diagonalice the total Hamiltonian:

\begin{equation}\label{n13}
    H=\int_0^\infty
    x\,|\Psi_\pm(x)\rangle\langle\Psi_\pm(x)|\,dx\,.
\end{equation}
Note that these results are valid for any value of $N$ including
$N=1$ \cite{AKPY}. From here, we can obtain a series of formal
expressions whose validity will be checked in the next subsection.
Taking into account that, by hypothesis, the bound states
$\{|k\rangle\}$ and the kets in the continuum,
$\{|\omega\rangle\}$, form a complete set, we have that

\begin{equation}\label{n14}
    I=\sum_{k=1}^N|k\rangle\langle k|+\int_0^\infty
    |\omega\rangle\langle\omega|\,d\omega\,.
\end{equation}
Assuming that identities in (\ref{n12}) and (\ref{n14}) are the
same, we immediately find the following formal
identities\footnote{As $\Psi_\pm(x)$ and $|\omega\rangle$ are
functionals, the precise meaning of expressions like
$\langle\Psi_\pm(x)|k\rangle$ or
$\langle\Psi_\pm(x)|\omega\rangle$ should be clarified.}:

\begin{equation}\label{n15}
    |k\rangle=\int_0^\infty |\Psi_\pm(x)\rangle\langle
    \Psi_\pm(x)|k\rangle\,dx\hskip0.4cm;\hskip0.4cm
    |\omega\rangle= \int_0^\infty |\Psi_\pm(x)\rangle\langle
    \Psi_\pm(x)|\omega\rangle\,dx\,.
\end{equation}
The expressions $\langle\Psi_\pm(x)|k\rangle$ and
$\langle\Psi_\pm(x)|\omega\rangle$ should be the complex
conjugates of $\langle k|\Psi_\pm(x)\rangle$ and
$\langle\omega|\Psi_\pm(x)\rangle$ that can be determined from
(\ref{n10}) and (\ref{n2}). The result can be easily obtained to
be

\begin{eqnarray}
  \langle k|\Psi_\pm(x)\rangle &=& -\lambda\sum_{l=1}^N f_l(x)\,G_{kl}(x\pm i0)\,, \label{n16}
  \\[2ex]
  \langle\omega|\Psi_\pm(x)\rangle &=&
  \delta(\omega-x)-\sum_{k,l=1}^N \frac{\lambda^2\,f_k(\omega)\,f_l(x)\,G_{k,l}(x)}{
  x-\omega\mp i0}\,. \label{n17}
\end{eqnarray}
In order to obtain explicit expressions of $|k\rangle$ and
$|\omega\rangle$ in terms of $|\Psi_\pm(x)\rangle$, we have to
carry (\ref{n16}) and (\ref{n17}) into the first and second
equations in (\ref{n15}) respectively. The results are

\begin{eqnarray}
  |k\rangle &=& -\lambda\sum_{l=1}^N\int_0^\infty dx\, f_l(x)\, G_{kl}(x\pm i0)\, |\Psi_\mp(x)\rangle \label{n18}
  \\[2ex]
 |\omega\rangle &=& |\Psi_+(\omega)\rangle -\lambda \sum_{k,l=1}^N
 f_k(\omega)\int_0^\infty
 dx\,\frac{\lambda\,f_l(x)\,G_{k,l}(x)}{x-\omega\pm i0}\,|\Psi_\pm(x)\rangle\,.
 \label{n19}
\end{eqnarray}
Note that $G_{k,l}^*(x+i0)=G_{k,l}(x-i0)$, where the star denotes
complex conjugation. These inverse relations will be used below.

\subsection{Survival probability.}

We know that the vectors $|\Psi_\pm(x)\rangle$ are eigenvectors of
the total Hamiltonian $H$ with eigenvalue given by $\omega$. Then,
time evolution for $|\Psi_+(x)\rangle$ is given by

\begin{equation}\label{n20}
    e^{-itH}\,|\Psi_+(x)\rangle=e^{-it\omega}\,|\Psi_+(x)\rangle\,.
\end{equation}
This allows us to obtain the time evolution, with respect to the
total evolution $H$, of the eigenvectors of the free Hamiltonian
$H_0$, $|k\rangle$. Let us apply the time evolution operator
$e^{-itH}$ to (\ref{n18}), with sign $+$. We readily obtain the
following expression:

\begin{equation}\label{n21}
    |k\rangle_t:= e^{-itH}|k\rangle=
    -\lambda\sum_{l=1}^N\int_0^\infty d\omega\,e^{-it\omega}
    \,f_l(\omega)\,G_{kl}(\omega-i0)\,|\Psi_+(x)\rangle\,.
\end{equation}
This vectors $|k\rangle_t$ must have a general similar to
(\ref{n3}). Also, inserting (\ref{n10}) into (\ref{n21}), we
obtain that this form should be

\begin{equation}\label{n22}
|k\rangle_t= \sum_{l=1}^N A_{kl}(t)\,|l\rangle+
\lambda\sum_{l=1}^N \int_0^\infty
d\omega\,f_l(\omega)\,g_{kl}(\omega,t)\,|\omega\rangle\,,
\end{equation}
where $|l\rangle$, $l=0,1,\dots,N$ are the eigenvectors of the
free Hamiltonian $H_0$. The explicit form of the functions
$A_{kl}(t)$ and $g_{kl}(\omega,t)$ is obtained through a
cumbersome but straightforward calculation, for which the main
steps are explicitly given in \cite{AKPY}. The final results are
for $ A_{kl}(t)$

\begin{equation}\label{n23}
    A_{kl}(t)= \frac{1}{2\pi i}\int_C
    d\omega\,e^{-it\omega}\,G_{kl}(\omega)
\end{equation}
and $C$ is the contour depicted in Figure 1. For $g_{kl}(\omega,t)$,
we have

\begin{figure*}
\begin{center}
\includegraphics[width=10cm]{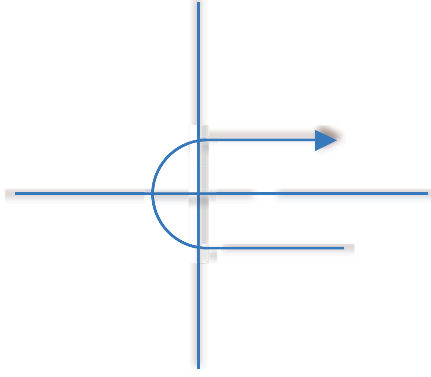}
\end{center}
\caption{Contour of integration $C$.} \label{fig:1}
\end{figure*}

\begin{equation}\label{n24}
g_{kl}(\omega,t)= \frac{1}{2\pi i}\int_C
d\omega'\,G_{kl}(\omega')\,\frac{e^{-it\omega'}}{\omega'-\omega}\,,
\end{equation}
where the contour $C$ is again the same as in (\ref{n23}). There
exists the following relation between $ A_{kl}(t)$ and
$g_{kl}(\omega,t)$:

\begin{equation}\label{n25}
 A_{kl}(t)=\left( i\frac{d}{dt}-\omega
 \right)\,g_{kl}(\omega,t)\,,
\end{equation}
as it can be easily derived from (\ref{n23}) and (\ref{n24}).

In order to obtain the survival probability for the bound states
of $H_0$, we just calculate the survival amplitude for these
states. Let $|{\Phi}\rangle$ be the most general linear
superposition of the eigenvectors of $H_0$, it has the following
form:

\begin{equation}\label{n26}
    |{\Phi}\rangle=\sum_{k=1}^N a_k\,|k\rangle\,.
\end{equation}
Obviously, time evolution for (\ref{n26}) must have the following
form

\begin{equation}\label{n27}
    |{\Phi}(t)\rangle=\sum_{k=1}^N a_k\,|k\rangle_t\,.
\end{equation}
We are now in the position of computing the survival amplitude
$A(t)$ and hence the survival probability $|A(t)|^2$. For the
survival amplitude, we have

\begin{equation}\label{n28}
    A(t)=\langle \Phi|\Phi(t)\rangle=\sum_{k,k'=1}^N
    a_ka^*_{k'}\,\langle k|k'\rangle_t= \sum_{k,k'=1}^N
    a_ka^*_{k'}\,A_{kk'}(t)\,.
\end{equation}
There is a form to write (\ref{n28}) as a sum of the contributions
of the resonance poles plus the background. We need to change the
contour $C$ into $C_1$ as depicted in Figure   2. Note that for this
contour transformation we are using the idea that the analytic
continuation from above to below is produced in the second sheet of
a Riemann surface. With these ideas in mind, we obtain the following
expression for $A_{kk'}(t)$:

\begin{equation}\label{n29}
A_{kk'}(t)= -\sum_j r^j_{kk'}\,e^{-iz_jt}+\frac{1}{2\pi i}
\int_{C_1}d\omega\,e^{-i\omega t}\, G_{kk'}(\omega)\,.
\end{equation}
Here the sum in $j$ runs out the poles of $G_{kk'}(\omega)$ in the
forth quadrant of the complex plane (see Figure  .). The term
$r^j_{kk'}$ is the residue of $G_{kk'}(\omega)$ at the pole $z_j$:

\begin{figure*}
\begin{center}
\includegraphics[width=10cm]{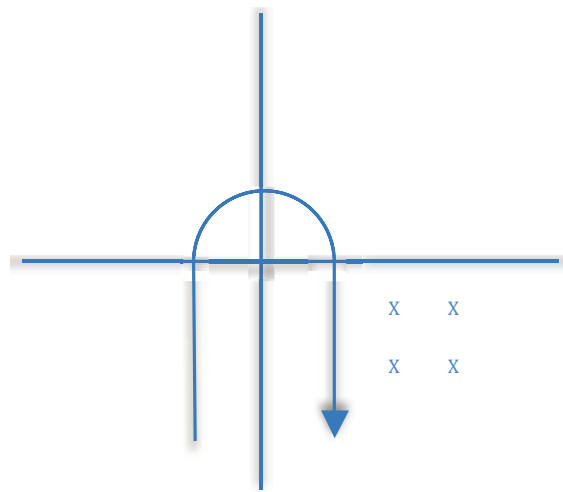}
\end{center}
\caption{Contour of integration $C_1$. By x we represent the
location of the resonance poles. } \label{fig:1}
\end{figure*}

\begin{equation}\label{n30}
r^j_{kk'}={\rm res\;}G_{kk'}(\omega)\Big|_{\omega=z_j}\,.
\end{equation}
Formula (\ref{n29}) can be written in a more explicit form,
provided that we are assuming that the form factors $f_k(\omega)$
are analytically continuable into the lower half plane. After some
calculations (given in \cite{AKPY}\,), one obtains:

\begin{equation}\label{n31}
A_{kk'}(t)= \sum_jN_j^2\,e^{-iz_j t}\sum_{k,l,k',l'=1}^N \lambda^2
f_l(z_j)\,f_{l'}(z_j)\,r^j_{kl}\,r^j_{k'l'}+ \frac{1}{2\pi i}
\int_{C_1}d\omega\,e^{-i\omega t}\, G_{kk'}(\omega)\,,
\end{equation}
where the $N_j$ are certain normalization constants \cite{AKPY}.

The form of the terms $A_{kk'}(t)$ and hence the form of the
survival amplitude is particularly interesting. It has an
oscillating term plus another term called the background and that
is represented by the integral in (\ref{n29}) or (\ref{n31}). The
background plays an essential role in the behavior of the decay
phenomena at very short and very long values of time $t$. This
behavior has been extensively studied \cite{K,FP,A,FI,KK,NA}.

In the simplest Fridrichs model, $N=1$ and there is a simple pole
resonance only located at $z_0$. Then, the form of $A(t)$ is
simply

\begin{equation}\label{n32}
A(t)=-r\,e^{-itz_0}+ \frac{1}{2\pi i}
\int_{C_1}d\omega\,e^{-i\omega t}\, G(\omega)\,.
\end{equation}
In this case, the survival probability $|A(t)|^2$ has a term that
decays exponentially as $e^{-t \,{\rm Im\,}z_0}$ and that
corresponds to the behavior of resonance phenomena for
intermediate times, i. e., neither too short or two large
\cite{FGR}. However, if $N\ge 2$, the survival probability has an
oscillating behavior for which the maximus decay as $t$ goes to
infinite.

\subsection{Two level model.}

The simplest case of the $N$ level model is of course $N=2$. Also
for simplicity, we have chosen the coupling constant $\lambda$ in
$H=H_0+V$ equal to one, $\lambda=1$ This situation has already
several interesting features and for this reason and also for its
simplicity it is an example of obligatory study.

First of all, it is important to choose the form factors
adequately. The following choice was proposed in the context of a
study of decay for elementary particles \cite{LP}:

\begin{equation}\label{n33}
f_k(\omega)=\frac{\omega^{1/4}}{\omega+\rho_k^2}\;, \hskip2cm
k=1,2\,,
\end{equation}
where $\rho_k$ are positive numbers. These form factors permit to
calculate the matrix elements $G^{-1}_{kk'}(\omega)$ from
(\ref{n6}) as

\begin{equation}\label{n34}
G^{-1}_{kk'}(\omega)=(\omega_k-\omega)\delta_{kk'}+\frac{\pi\lambda^2}{\rho_k+\rho_{k'}}
\;\frac{1}{(\sqrt\omega+i\rho_k)(\sqrt\omega+i\rho_{k'})}\,,
\end{equation}
with $k,k'=1,2$. The branch square root is defined so that a
positive real number had a positive square root.

From (\ref{n34}) obtaining the $2\times 2$ matrix $G(\omega)$ is
just a straightforward manipulation. In order to obtain the
resonances as poles of $G(\omega)$ the technique will be to obtain
these resonances as zeroes of the analytic continuation of the
inverse of the determinant of the $2\times 2$ matrix $G(\omega)$.
If we make the simple change of variables given by
$\sqrt\omega=ix$, we need to solve the following equation in $x$:

\begin{eqnarray}
(\det G(\omega))^{-1}= \nonumber
\end{eqnarray}
\begin{eqnarray}
 \left[(\omega_1+x^2)(x+\rho_1)^2-\frac{\pi\lambda^2}{2\rho_1}\right]
 \times \left[(\omega_2+x^2)(x+\rho_2)^2-\frac{\pi\lambda^2}{2\rho_2}
 \right]-\left(\frac{\pi\lambda^2}{\rho_1+\rho_2}
 \right)^2 \nonumber
\end{eqnarray}
\begin{equation}\label{n35}
    =0\,.
\end{equation}
This is an algebraic equation of degree 8 with real coefficients.
Therefore, all its roots are either real or appear in complex
conjugate pairs. If certain conditions on $\omega_i$ and $\rho_i$
are fulfilled \cite{AKPY}, $G(\omega)$ has no singularities except
for a brunch cut coinciding with the positive semiaxis. Then, all
the resonance poles are poles of the analytic continuation, or in
the language of Riemann surfaces, they are located in the second
Riemann sheet. Real singularities correspond to virtual poles and
lie in the negative semiaxis (second sheet).

In the case of weak coupling, i. e., when the coupling constant
$\lambda$ is very small,  two pairs of complex conjugate poles can
be evaluated perturbatively. The result is given by:

\begin{eqnarray}
  z_j &=& \omega_j+\frac{\pi \lambda^2}{2\rho_j}\;
  \frac{(\sqrt{\omega_j} -i\rho_j)^2}{(\omega_j+\rho_j^2)^2} +
  \frac{\pi^2\lambda^4}{(\sqrt{\omega_j}+i\rho_j)^2}\nonumber
  \\[2ex]
   &\times & \left(\frac{1}{(\omega_j-\omega_k)(\rho_1+\rho_2)^2(\sqrt{\omega_j}+ i\rho_k)^2}
  -\frac{1}{4\rho_j^2\,\sqrt{\omega_j}\,(\sqrt{\omega_j}+i\rho_j)^2} \right)
  \nonumber\\[2ex]
   &+& o(\lambda^6)\,,\quad j=1,2\,, \quad k\ne j\,. \label{n36}
\end{eqnarray}
In this case, we can obtain approximate expressions for the real
and imaginary parts of $z_j$ and the result is given by

\begin{eqnarray}
   && \widetilde\omega:= {\rm Real}\,z_j=  \omega_j+\frac{\pi\lambda^2}{2\rho_j}
   \; \frac{\omega_j-\rho_j^2}{(\omega_j+\rho_j^2)^2}
   +o(\lambda^4)\,,
   \quad j=1,2\,, \nonumber \\[2ex]
  && \gamma_j:=-{\rm Im}\,z_j= \frac{\pi\lambda^2\,\sqrt{\omega_j}}{(\omega_j+\rho_j^2)^2}
  +o(\lambda^4)\,,\quad j=1,2\,.
\label{n37}
\end{eqnarray}

Note that (\ref{n35}) is of eight degree, so that it must have eight
solutions. We have already calculated four of them. The other four
are real and correspond to virtual states \cite{AKPY}, also called
anti-bound states.

Then, (\ref{n37}) permits us to write an approximate expression
for the survival amplitud $A(t)$. If we neglect the background
term, (\ref{n37}) yields \cite{AKPY}

\begin{eqnarray}
  A(t) &\approx & \sum_{k,k'=1,2} a_k a_{k'}^*\sum_{j=1,2}
  e^{-\gamma_j t}\, e^{-i\widehat\omega t}\,r_{kk'}^j \nonumber
  \\[2ex]
   &=& \sum_{j=1}^2 |a_j|^2\,e^{-iz_jt}-\lambda^2 \sum_{j=1}^2
   \left(\frac{i\pi|a_j|^2\,e^{-iz_jt}}{2\rho_j(\rho_j-i\sqrt{\omega_j})^3\,\sqrt{\omega_j}}
   \right.
   \nonumber \\[2ex]
   &+&\left. \frac{2\pi\,{\rm Real}\, (a_1a_2^*)\,e^{-iz_jt}}
   {(\rho_1+\rho_2)(\rho_j-i\sqrt{\omega_j})(\rho_l-i\sqrt{\omega_j})(\omega_j-\omega_l)}
   \right) +o(\lambda^4)\, \quad l\ne j\,.\nonumber\\[2ex]
   \label{n38}
\end{eqnarray}
It is important to remark that both in (\ref{n36}) and (\ref{n38})
the term $1/(\omega_k-\omega_l)$ appears. Then, these expressions
cannot be used in the case of the existence of degeneracy levels
for $H_0$.

Note that, although for some initial conditions like $a_1=1$,
$a_2=0$, the survival amplitude (\ref{n38}) does not show
oscillations, these oscillations must be present in the terms of
the order $\lambda^4$. For the lowest order $\lambda^2$, the
survival probability can be written in the following form

\begin{equation}\label{n39}
|A(t)|^2= |\,|a_1|^2\,e^{-\gamma_1 t}+|a_2|^2 \,e^{-\gamma_2 t}\,
e^{-2i\nu t}|^2\,,
\end{equation}
with

\begin{equation}\label{n40}
    \nu:=\frac{\widetilde\omega_1-\widetilde\omega_2}{2}\;.
\end{equation}
From (\ref{n39}), we can see the dependence of the survival
probability with respect to the initial conditions. For $a_1=0$ or
$a_2=0$, the survival probability behaves exponentially without
oscillations, at this order of approximation. If $a_1\approx a_2$
the survival probability shows considerable oscillations.

\subsection{The $N$-level model.}

An analysis can be carried out for the $N$ level model in the weak
coupling regime. We can use arbitrary form factors $f_k(\omega)$
in (\ref{n7}). Then, we can get

\begin{equation}\label{n41}
    (\det G(\omega))^{-1} =\prod_{k=1}^N (\omega_k-\omega)-
    \lambda^2\sum_{k=1}^N I_{kk}(\omega)\prod_{m\ne k}^N
    (\omega_m-\omega) +o(\lambda^4)\,,
\end{equation}
where

\begin{equation}\label{n42}
    I_{kl}(\omega)=\int_0^\infty d\omega'\,\frac{f_k(\omega')\,f_l(\omega')}
    {\omega'-\omega-i0}\;.
\end{equation}
The zeroes of (\ref{n41}) give the position of the resonances:

\begin{equation}\label{n43}
z_k=\omega_k-\lambda^2
I_{kk}(\omega_k)+o(\lambda^4)=\widetilde\omega_k-i\gamma_k\,,
\quad j=1,2,\dots,N\,,
\end{equation}
where, again, $\widetilde\omega_k={\rm Real}\,z_k$ and
$\gamma_k={\rm Im}\,z_k$.

Then, in the first nontrivial order, which is $\lambda^2$, we have
for the real and imaginary part of the resonance pole located at
$z_k$ the following expression:

\begin{equation}\label{n44}
    \widetilde \omega_k=\omega_k\;, \hskip2cm
    \gamma_k=\pi\lambda^2\,f_k^2(\omega_k)\;.
\end{equation}
It is not a surprise that, at first order, real parts of $z_k$ do
not depend on $\lambda$, while imaginary parts depend on
$\lambda^2$. This is exactly what happens in the two level model
studied in the previous subsection, see (\ref{n36}). At this order
of approximation, we can also calculate the partial resolvent $G$
\cite{AKPY}

\begin{equation}\label{n45}
G_{kk'}(\omega)=(\omega_k-\omega-\lambda^2I_{kk'}(\omega)\,)^{-1}\,\delta_{kk'}
+o(\lambda^2)\,,
\end{equation}
the residues

\begin{equation}\label{n46}
    r_{kk'}^j=-\delta_{kk'}\delta_{kj}+o(\lambda^2)
\end{equation}
and the survival amplitude

\begin{equation}\label{n47}
    A(t)= \sum_{k=1}^N |a_k|^2\,e^{-i\omega_k t}\,e^{-\pi\lambda^2
    f_k^2(\omega_k)t}\;.
\end{equation}
The conclusion that one arrives is that, although the survival
probability decays, the decaying mode can be strongly oscillatory,
depending on the values of the parameters $a_k$, $k=1,2,\dots,N$
(initial conditions). One study of particular forms of this decay
can be seen in \cite{AKPY}.

\section{Some types of Friedrichs-like models and their applications to Quantum Field Theory.}

More sophisticated versions of the  Friedrichs model have been
used to introduced models for unstable relativistic quantum fields
exactly solvable on a sector or under a well determined
simplification or approximation. We want to discuss a few models
that have been introduced elsewhere
\cite{AGPP,AGP,KPPP,AGKPP,AGMP} that are very useful to give us an
idea on the construction and behavior of relativistic resonances,
not making use of the $S$ matrix formalism (for a study on
relativistic resonances in the context of the $S$ matrix see
\cite{BK,BKW,BS}).

Along this section, we are going to present four different
situations, which are somehow related with unstable quantum field
theory. In the first one, we consider two local fields with a cubic
interaction. We show that, although the model is in principle not
exactly solvable, it can be solved in a sector in which a particle
of one of the spices considered decays into two of the other spice.
In the second one, the interaction is between a local quantum field
and a bilocal quantum field with a continuous and bounded mass
spectrum. We choose a quadratic interaction between these two
fields, which makes the exact solution tractable through a rather
complicated generalization of the Friedrichs model. Due to the
complexity of this example, we shall limit ourselves to give here a
summary of it. Readers interested in a deeper understanding of this
example can go to the original paper \cite{AGPP}. Finally, we
propose an example of ``second quantized Friedrichs model'', the
Friedrichs model with virtual transitions, which is probably the
simplest case of these second quantized models.

\subsection{First model.}

Assume that we have two real scalar relativistic quantum fields
$\varphi(x)$ and $\psi(x)$, where  $x$ is the fourth component
space time, with respective masses given by $m$ and $M$, coupled
with the simplest cubic interaction. The Hamiltonian of the system
is given by

\begin{equation}
H=H_m+H_M+V\,,
\end{equation}
where

\begin{eqnarray}
  H_m &=& \int
  dx\,(\dot\varphi^2+(\nabla\varphi)^2+m^2\varphi^2)\label{Hm}
  \\[2ex]
  H_M &=& \int dx\,(\dot\psi^2+(\nabla\psi)^2+M^2\psi^2) \label{HM} \\[2ex]
  V &=& \lambda\int dx\psi(x)\varphi^2(x) \label{V}\,.
\end{eqnarray}
The dot means first derivative with respect to time. By boldface
letters we  denote three-dimensional vectors. Four-dimensional
vectors in Minkowski space are denoted by roman style letters. The
products of two four vectors as well as the scalar products of two
three vectors are denoted by a dot. We use the standard metric
$(+,-,-,-)$ of Minkowski space. For example: $k\cdot x=k_0x_0-{\bf
k}\cdot{\bf x}$.

In terms of the creation and annihilation operators, the above
formulae can be written in the interaction picture as:

\begin{eqnarray}
  H_m &=& \int d\widetilde {\bf q}\,\rho({\bf q})b^\dagger({\bf q}) b ({\bf
  q})\label{a}\\[2ex]
  H_M &=& \int d\widetilde{\bf k}\,\omega({\bf k}) a^\dagger({\bf k}) a ({\bf k})
  \label{b}\\[2ex]
  V_I(t) &=& \lambda\int d\widetilde{\bf k}\,d\widetilde {\bf q}_1\, d\widetilde {\bf q}_2\,
 \int d^3{\bf x}\,[a^\dagger ({\bf k})\, \exp(ik\cdot x)+a({\bf k})\,\exp(-ik\cdot
  x)]\nonumber\\[2ex]
   & \times& [b^\dagger ({\bf q}_1)\, \exp(iq_1\cdot x)+b({\bf q}_1)\,\exp(-iq_1\cdot x)] \nonumber
   \\[2ex]
   &\times & [b^\dagger ({\bf q}_2)\, \exp(iq_2\cdot x)+b({\bf q}_2)\,\exp(-iq_2\cdot
   x)]\,.\label{c}
\end{eqnarray}
The quantum field operators in (\ref{Hm}-\ref{V}) are given by

\begin{eqnarray}
  \psi({\bf x},t) &=& \int d\widetilde {\bf k}\,[a^\dagger ({\bf k})\, \exp(ik\cdot x)+a({\bf k})\,\exp(-ik\cdot
  x)] \nonumber \\
\varphi({\bf x},t)   &=& \int d\widetilde {\bf q} \, [b^\dagger
({\bf q})\, \exp(iq\cdot x)+b({\bf q})\,\exp(-iq\cdot
   x)]
\end{eqnarray}
with the Lorentz invariant measure

\begin{eqnarray}
  d\widetilde{\bf k} &=& \frac{d^3{\bf k}}{(2\pi)^3 2\omega({\bf k})}\,,
  \hskip1cm \omega({\bf k}) =({\bf k}^2+M^2)^{1/2} \nonumber
  \\[2ex]
  d\widetilde{\bf q} &=&  \frac{d^3{\bf q}}{(2\pi)^3 2\rho({\bf q})}\,,
  \hskip1cm \rho({\bf q}) =({\bf q}^2+m^2)^{1/2}\;.\label{e}
\end{eqnarray}
The creation and annihilation operators in (\ref{a}-\ref{c})
satisfy the usual commutation relations:

\begin{eqnarray}
  [a({\bf k}),a^\dagger ({\bf k}')] &=& (2\pi)^3 2\omega({\bf k})\delta({\bf k}-{\bf k}') \nonumber
  \\[2ex]
[b({\bf q}), b^\dagger ({\bf q})] &=& (2\pi)^3 2\rho({\bf
q})\delta({\bf q}-{\bf q}')\,.\label{f}
\end{eqnarray}

The first goal is to obtain an appropriate asymptotic dynamics for
this situation. Note that scattering theory applies if asymptotic
completeness holds true between the free and the interacting fields
\cite{BLT}. If this assumption is not valid, one tries to find
another solvable evolution which satisfies the asymptotic condition
and reestablish scattering theory as a comparison between the
interaction field  and the redefined asymptotic field. A typical
example is the Faddeev-Kulish \cite{FK} removal of infrared
divergencies.

After integration over the three dimensional space, i.e., with
respect to $d^3{\bf x}$, on the right hand side of (\ref{c}), we
obtain eight terms of products of creation and annihilation
operators with $t$-dependent exponents accomplished with three
dimensional $\delta$ functions of momentum conservation. According
to the Riemann-Lebuesgue lemma \cite{RSII}, the asymptotic
behavior of $V_I(t)$ is defined by the behavior of these
$t$-dependent exponents in the integration domain. For example,
one of the terms on the right hand side of (\ref{c}) is

\begin{equation}\label{g}
\int d\widetilde{\bf k}\,d\widetilde {\bf q}_1\, d\widetilde {\bf
    q}_2 \,a^\dagger ({\bf k})\, b({\bf q}_1)\, b({\bf
    q}_2)\,\exp(i[\omega({\bf k})-\rho({\bf q}_1)-\rho({\bf
    q}_2)]t) (2\pi)^3 \,\delta({\bf k}+{\bf q}_1+{\bf q}_2)\,.
\end{equation}
The asymptotic behavior of the integral (\ref{g}) as
$t\longmapsto\pm\infty$ is determined by the term $[\omega({\bf
k})-\rho({\bf q}_1)-\rho({\bf q}_2)]$. Notice that

\begin{equation}\label{h}
    \omega ({\bf k})+\rho ({\bf q}_1)+\rho ({\bf q}_2)\bigg|_{{\bf
k}+{\bf q}_1+{\bf q}_2={\bf 0}} \ge M+2m>0 \,.
\end{equation}
Therefore, the integral (\ref{g}) goes rapidly to zero due to the
fast oscillations (indeed to the Riemann-Lebesgue lemma). Let us
consider another term

\begin{equation}\label{i}
    \int d\widetilde{\bf k} d\widetilde {\bf q}_1 d\widetilde {\bf q}_2\; a^\dagger({\bf k})
b({\bf q}_1)b({\bf q}_2)e^{i[\omega ({\bf k})-\rho ({\bf
q}_1)-\rho ({\bf q}_2)]t}\, (2\pi)^3\,\delta ({\bf k}-{\bf
q}_1-{\bf q}_2)\,.
\end{equation}
As for the case of the previous integral (\ref{g}), the asymptotic
behavior of (\ref {i}) is defined by the quantity

\begin{equation}\label{j}
    \Delta ({\bf k}, {\bf q}_1)=\omega ({\bf k})-\rho
({\bf q}_1) -\rho ({\bf q}_2)\bigg |_{{\bf k}-{\bf q}_1-{\bf
q}_2=0}\,.
\end{equation}
The values of $\Delta ({\bf k}, {\bf q}_1)$ depend on the masses
of the fields $\varphi$ and $\psi$. If $M<2m$, then $\Delta ({\bf
k},{\bf q}_1)$ is strictly negative:

\begin{equation}\label{k}
    \Delta ({\bf k},{\bf
q}_1)\le  2(m^2+\frac{{\bf k}^2}{4})^{1/2} -(M^2+{\bf
k}^2)^{1/2}<0\,.
\end{equation}
To show (\ref{k}), we obtain the maximum of $\Delta ({\bf k}, {\bf
q}_1)$ on ${\bf q}_1$ for each $\bf k$. This maximum is at ${\bf
q}_1={\bf k}/2$ (it is certainly a maximum) and the value of
$\Delta ({\bf k},{\bf q}_1)$ at the maximum is

\begin{equation}\label{l}
    2(m^2+\frac{{\bf k}^2}{4})^{1/2}-(M^2+{\bf k}^2)^{1/2}\,.
\end{equation}
Note that if $M<2m$, (\ref{l}) is indeed smaller than zero. Thus,
$\Delta ({\bf k},{\bf q}_1)$ has constant sign and therefore the
Riemann-Lebesgue lemma applies. As a consequence, the integral
(\ref{i}) decreases fast as $t\longmapsto\pm\infty$.

In contrast, let us assume that $M>2m$. Then the maximum (\ref{i})
is bigger than zero. However, for fixed $\bf k$ the term

\begin{equation}\label{m}
\Delta ({\bf k},{\bf q}_1)=(M^2+{\bf k}^2)^{1/2}- (m^2+{\bf
q}^2_1) - (m^2+({\bf q}_1-{\bf k})^2)^{1/2}
\end{equation}
is obviously smaller than zero for high values of $|{\bf q}_1|$.
Let us consider the manifold for which

\begin{equation}\label{n}
\Delta ({\bf k},{\bf q}_1)=0\,.
\end{equation}

The integral (\ref{i}) over this region does not vanish
asymptotically and the same behavior can be observed on its
complex conjugate. In other words, by relinquishing the stability
condition $M<2m$, we obtain an unstable field theory where the
asymptotic condition for scattering theory fails. Therefore,
following a standard procedure in field theory, we redefine the
asymptotic evolution in the interaction picture as follows:

\begin{equation}\label{o}
h_{as}=H_0+V_{as}(t)\;,
\end{equation}
where $V_{as}$ includes the slowly decreasing part of $V_I$. There
is an ambiguity in the definition of $V_{as}$ and we shall take it
as the sum of (\ref{i}) and its complex conjugate, so that
$V_{as}$ is Hermitian. This is the simplest choice for $V_{as}$:

\begin{eqnarray}
  V_{as}(t) &=& \lambda\int d\widetilde{\bf k}\, d\widetilde{\bf q}_1\,
   d\widetilde{\bf q}_2\;
[a^\dagger({\bf k})b({\bf q}_1) b({\bf q}_2) e^{i\Delta ({\bf
k},{\bf q}_1)t}+\nonumber \\
[2ex] &+&a({\bf k}) b^\dagger({\bf q}_1)b^\dagger({\bf q}_2)
e^{-i\Delta ({\bf k},{\bf q}_1)t}]
(2\pi)^3\,\delta ({\bf k}-{\bf q}_1-{\bf q}_2)\nonumber \\
[2ex] &=&\lambda\int d^3{\bf x} [\psi^{(+)}(x) \varphi^{(-)2}
(x)+\psi^{(-)}(x) \varphi^{(+)2} (x)] \,,\label{p}
\end{eqnarray}
where $\Delta$ was defined in (\ref{j}) and $\varphi^{(+)},
\varphi^{(-)}$ denote the positive and negative frequency parts of
$\varphi$. Same for $\psi$.

\bigskip\begin{itemize}
    \item The above comments have an obvious physical interpretation:
The decay process given by

$$
\psi\longmapsto \varphi+\varphi
$$
is only possible if the mass $M$ of the $\psi$-particle is bigger
than the mass $2m$ corresponding to two $\varphi$-particles.

    \item {\it Renormalization}. As we shall see below the $h_{as}$
produces a ultraviolet divergence in the equation for its
eigenstates. In order to remove this divergence we add to $h_{as}$
the appropriate counterterm

\begin{equation}\label{q}
H_{c.t.}=\frac{1}{2}\int d^3x\,\delta M^2
\varphi^{(+)}(x)\varphi^{(-)}(x)=\int d{\bf k}\;\frac{\delta
M^2}{\omega({\bf k})}\;a^\dagger({\bf k})\,a({\bf k})\,,
\end{equation}
where the mass renormalization $\delta M^2$ is of order
$\lambda^2$. The appearance of  ultraviolet counterterms is due to
our choice of the asymptotic interaction. We can of course
introduce some smooth cut off in $V_{as}$, but it will involve
additional parameters to the asymptotic states and it will
generally break the relativistic invariance of the system.
\end{itemize}

In order to obtain the desired formula for the asymptotic
Hamiltonian, we add $h_{as}$ and $H_{c.t.}$:

\begin{equation}\label{r}
    H_{as}(t)=h_{as}(t)+H_{c.t.}\,.
\end{equation}

In the Schr\"odinger picture,  the time dependent exponents
$e^{\pm i\Delta (k,q_1)t}$ in equation (\ref{p}) disappear. Now,
we shall consider the eigenstates of $H_{as}=H_{as}(0)$, which is
a Hermitian operator in the Fock space corresponding to the
creation and annihilation operators satisfying (\ref{f}).

From now on, we shall work in the Schr\"odinger picture, in order
to make calculations easier. In the Schr\"odinger picture the
total Hamiltonian $H$ will be the following:

\begin{eqnarray}
 H_{as}=H_M+H_m+ H_{c.t.} +\lambda\int d\widetilde{\bf k}\, d\widetilde{\bf q}_1\,
d\widetilde{\bf q}_2\; [a^\dagger({\bf k})b({\bf q}_1) b({\bf
q}_2)\nonumber\\[2ex]
+a({\bf k}) b^\dagger({\bf q}_1)b^\dagger({\bf
q}_2)]\,(2\pi)^3\,\delta ({\bf k}-{\bf q}_1-{\bf q}_2) \,.
\label{s}
\end{eqnarray}

We shall show how the above problem of unstable quantum field can
be solved exactly on a sector. Of course, as the interaction is
not quadratic, no exact solution is know for all sectors. This
sector corresponds to the decay of one $\psi$-particle into two
$\varphi$-particles. The Hilbert space of this system is given by

\begin{equation}\label{t}
    {\cal H}:= {\cal H}_\varphi\otimes{\cal H}_\varphi\oplus{\cal
    H}_\psi\,.
\end{equation}
The eigenspaces of the operator

\begin{equation}\label{u}
    N=N_\psi +2N_\varphi
\end{equation}
are invariant subspaces of $H_{as}$. The operators $N_\psi$ and
$N_\varphi$ are the number operators of $\psi$ and
$\varphi$-particles. The asymptotic interaction does not affect
the vacuum state and the one $\psi$-particle state, which are the
only nondegenerate eigenspaces of $N$. The first degenerate case
is the eigenspace of $N$ with eigenvalue $2$, which is precisely
$\cal H$ as given in (\ref{s}).  In the subspace (\ref{s}) the
term $V_{as}$ in $H_{as}$ produces transitions of two
$\varphi$-particles state to one $\psi$-particle and vice versa.
We should mention that not every two $\varphi$-particle state will
mix with one $\psi$-particle state, but only with those in which
the two $\varphi$-particles are in the  $S$-wave state.

The most general linear combination of two $\varphi$-particles in
the $S$-wave state with one $\psi$-particle, can be written as:

\begin{equation}\label{v}
    \Phi (E,{\bf k})=f (E,{\bf k})|{\bf k}\rangle + \int dE'\,f
(E,E',{\bf k})|E', {\bf k}\rangle\,,
\end{equation}
where $|E', {\bf k}\rangle$ is the two $\varphi-$particle $S-$wave
and $|{\bf k}\rangle$ is the one $\psi$ particle state. The
objective is to solve the eigenvalue equation

\begin{equation}\label{w}
    (H_{as}-E)\Phi (E,{\bf k})=0\,,
\end{equation}
where $H_{as}$ was given in (\ref{s}). The method to solve this
equation consists in replacing (\ref{v}) into (\ref{w}). Thus, we
can obtain the coefficients $f (E,{\bf k})$ and $f (E,E',{\bf k})$.

We are interested in obtaining these coefficients in the unstable
case characterized by the condition $M>2m$, where $M$ and $m$ are
the respective masses of the $\psi$ and $\varphi$ particles. The
method is tricky as it requieres the use of ``counterterms'' and a
particular choice of a normalization, but it is explained in detail
in \cite{AGP}. The final result is given by

\begin{equation}\label{y}
    f(E,{\bf k})=-A\,\frac{1}{\eta (E,{\bf
k})}\,\frac{\lambda\sqrt{2}}{2\omega ({\bf k})}\, \tau (E^2-{\bf
k}^2)
\end{equation}
and

\begin{equation}\label{z}
f(E,E',{\bf k})=A\left \{ \delta (E'-E)+
\frac{2\lambda^2(2\pi)^3}{2\omega ({\bf k})}\, \frac{\tau(E^2-{\bf
k}^2)}{\eta (E,{\bf k})}\, \frac{1}{E'-E}\right \}\;,
\end{equation}
where

\begin{equation}\label{ab}
    \langle{\bf
k}|\frac{1}{H_{as}-E}|{\bf k}\rangle=\frac{1}{\eta(E,{\bf k})}
\end{equation}
is the partial resolvent of the Hamiltonian $H_{as}$. This partial
resolvent has an explicit expression \cite{AGMP}. The function
$\tau$ is defined on the halfline $E> (4m^2+{\bf k}^2)^{1/2}$ by

\begin{eqnarray}
 \tau
(E^2-{\bf k}^2) &=& \int d\widetilde{\bf q}_1 d \widetilde{\bf
q}_2 \delta ({\bf k}-{\bf q}_1-{\bf q}_2) \delta (E-\rho ({\bf
q}_1)-\rho({\bf q}_2)) \nonumber
\\[2ex]
   &=& \frac{1}{4(2\pi)^5} \left (1-\frac{4m^2}{E^2-{\bf
k}^2}\right)^{1/2} \theta (E-(4m^2+{\bf
k}^2)^{1/2})\;,\nonumber\\[2ex]\label{ac}
\end{eqnarray}
where $\theta (x)$ is the usual Heaviside step function,
$d\widetilde{\bf q}_i$ and $\omega({\bf k})$ were defined in
(\ref{e}) and $A$ is an arbitrary constant.

In (\ref{z}) there is an ambiguity which arises due to the
singular denominator $(E'-E)$. This ambiguity also exists in
(\ref{y}) although it is there hidden in the partial resolvent
(\ref{ab}). Consequently, we shall define two solutions of
eigenvalue problem (\ref{w}) that correspond to the incoming and
the outgoing solution. The incoming and outgoing solutions are
defined as boundary
 functions of analytic functions from above
$(E+i0)$ and from below $(E-i0)$  the real axis respectively:

\begin{eqnarray}
  \Phi^{in\atop out} (E,{\bf k}) &=& A\left \{| E,{\bf k}\rangle+
\frac{\lambda\sqrt{2}\cdot\tau (E^2-{\bf k})}{2\omega ({\bf
k})\eta (E\pm i0,{\bf k})} \times\right. \nonumber\\[2ex]
   &=& \left.\times [\lambda\sqrt{2}\cdot (2\pi)^3 \int_{E_0}^\infty
dE'\frac{1}{E'-E\mp i0} |E',{\bf k}\rangle-|{\bf
k}\rangle]\right\}.\nonumber\\[2ex]\label{ad}
\end{eqnarray}

This formula explicitly demonstrates that the energy spectrum of
the asymptotic states in the eigenspace of $N$, as defined in
(\ref{u}), corresponding to the eigenvalue 2, is absolutely
continuous over the interval $[E_0,\infty)$ where $E_0=(4m^2+{\bf
k}^2)^{1/2}$. The isolated eigenvalue associated to the one
$\psi$-particle state $|{\bf k}\rangle$ has been dissolved to the
continuum due to the interaction. This reproduces somehow the
situation that appears in the standard Friedrichs model discussed
in Section 2.

\subsubsection{The Gamow vectors.}

In \cite{AGMP}, it has been shown that the inverse partial resolvent
$\eta(E,{\bf k})$ admits a continuation on the variable $E$,
$\eta_+(E,{\bf k})$, from above to below through the branch cut
$[E_0,\infty)$ and has a complex zero $z_R$ on the lower halfplane.
In order to find resonances, we have to solve the equation
$\eta(E,{\bf k})=0$. The explicit form of $\eta(E,{\bf k})$ is

\begin{equation}\label{5.33}
  \eta(E,{\bf k})=\omega({\bf
  k})-E-\lambda^2(2\pi)^3(E-a)\int_{E_0}^\infty dE'\,\frac{\tau({E'}^2-{\bf
  k}^2)}{(E'-E)(E'-a)}\,,
\end{equation}
where $\tau({E'}^2-{\bf  k}^2)$ was given in (\ref{ac}).

Note that the interaction $V_{as}$ in (\ref{p}) depends on a
multiplicative coupling constant $\lambda$. When $\lambda=0$, the
interaction is switched off, exactly as it happens in the ordinary
Friedrichs model. It has been shown in \cite{AGMP} that in absence
of interaction ($\lambda=0$), $E=\omega({\bf k})$ and that for
small values of $\lambda$ one has that the solution of the
equation $\eta(E,{\bf k})=0$ corresponding to the lower half plane
should have the following form

\begin{equation}\label{ae}
    z_R=\omega({\bf k})+\lambda^2 z_1+O(\lambda^4)\,,
\end{equation}
where

\begin{equation}\label{af}
    z_1= -\frac{i}{8\pi}\left(1-\frac{4m^2}{M^2}
\right)^{1/2}\,.
\end{equation}
Note that (\ref{e}) shows that $M^2=\omega^2({\bf k})-{\bf k}^2$,
where $M$ is the mass of the $\psi-$particle. Then, (\ref{ae}) and
(\ref{af}) gives

\begin{equation}\label{ag}
    z_R=\omega ({\bf k})- i\frac{M\Gamma}{2\omega ({\bf k})}\simeq
({\bf k}^2 +M^2-iM\Gamma)^{1/2}\,,
\end{equation}
where we have denoted

\begin{equation}\label{ah}
    M\Gamma =\frac{\lambda^2}{8\pi} \left (1-\frac{4m^2}{M^2}\right
)^{1/2}\,.
\end{equation}

This zero has a complex conjugate on the analytic continuation
from above to below. Equation (\ref{ag}) demonstrates that the
zero of $\eta (E,{\bf k})$ corresponds to a particle with complex
square mass

\begin{equation}\label{ai}
    M^2_c=M^2-iM\Gamma\,,
\end{equation}
which is the remnant of a $\psi$-particle. The eigenstate
(\ref{ad}) as a function of $E$ has a pole at the point $E=z_R$
and its residue in this pole is the eigenstate with complex
energy. Up to an irrelevant normalization constant this state is:

\begin{equation}\label{aj}
    \Phi^G(z_R,{\bf
k})=\lambda\sqrt{2}\cdot (2\pi)^3\int_{E_0}^\infty
dE'\frac{1}{E'-z_R}|E',{\bf k}\rangle- |{\bf k}\rangle\,.
\end{equation}

This Gamow vector has a clear meaning as an antilinear functional
on a suitably chosen space of test functions $\bf\Phi$. The test
function space should be dense on the Hilbert space (\ref{t}) of
the states of two $\varphi$-particles and one $\psi$-particle,
which is contained in the Fock space of the states of $\varphi$
particles and $\psi$-particles. Following the spirit of the
construction for simple non relativistic resonances, a simple
choice for the test function space, dense in the Hilbert space
(\ref{t}), is given by

\begin{equation}\label{ak}
    {\bf \Phi}:= [({\cal H}^2_-\cap S)\otimes S({\mathbb R}^3)]\oplus
    S({\mathbb R}^3)\,.
\end{equation}

We recall that

\begin{itemize}

\item{${\cal H}^2_-$ is the space of Hardy functions on the lower
half plane.}

\item{$S$ is the space of all real valued complex functions that
are differentiable at all orders such that they as well as their
derivatives vanish at the infinity faster than the inverse of any
polynomial (the one dimensional Schwartz space).}

\item{$S({\mathbb R}^3)$ represents the space of all complex
functions on the three dimensional real space ${\mathbb R}^3$
having the same properties as the Schwartz space $S$.}

\end{itemize}

Then, the rigged Hilbert space is a triplet of the form ${\bf
\Phi}\subset{\cal H}\subset {\bf\Phi}^\times$, where
${\bf\Phi}^\times$ is the dual space of $\bf\Phi$. Then, Gamow
vectors belong to this dual ${\bf\Phi}^\times$.

 A typical function in $\bf\Phi$ has the form $F(E,{\bf k})+g({\bf
k'})$ where $F(E,{\bf k})\in ({\cal H}^2_-\cap S)\otimes
S({\mathbb R}^3)$ and $g({\bf k'})\in S({\mathbb R}^3)$. The
variables $E$ and $\bf k$ in $F(E,{\bf k})$ represents
respectively the total energy and the sum of the momenta for the
two $\varphi$-particles. The variable $\bf k'$ is the momentum of
the $\psi$-particle.

The action of the functional (\ref{aj}) on the function $K(E,{\bf
k''},{\bf k'}):=F(E,{\bf k''})$ $+g({\bf k'})$ is given by

\begin{eqnarray}
  \langle K(E,{\bf k''},{\bf k'})|\Phi^G (z_R,{\bf
k})\rangle &=& \lambda\sqrt{2}\cdot (2\pi)^3\,\int_{E_0}^\infty
dE'\frac{1}{E'-z_R}\langle F(E,{\bf k''})|E',{\bf k}\rangle
\nonumber
\\[2ex]
 - \langle g({\bf k'})|{\bf k}\rangle  &=& \lambda\sqrt{2}\cdot (2\pi)^3\,\int_{E_0}^\infty
dE'\,\frac{F(E,{\bf k})}{E'-z_R}+g({\bf k})\;. \label{al}
\end{eqnarray}
For the fixed variable $\bf k$, the function $F(E,{\bf k})$ is a
Hardy function and, therefore, the integral term of (\ref{al})
converges. In the usual Fock space notation, a typical element of
${\bf\Phi}^\times$ can be written  as

\begin{equation}\label{am}
    F(E,{\bf k''})\,|E,{\bf k}''\rangle+g({\bf k'})|{\bf
    k}'\rangle\,.
\end{equation}

The functional $\Phi^G(z_R,{\bf k})$ has the following property:

\begin{equation}\label{an}
    H_{as}\,\Phi^G(z_R,{\bf k})=z_R\,\Phi^G(z_R,{\bf k})\,.
\end{equation}

It may be interesting to discuss the proof of this latter
statement. We have seen  that $\Phi(E,{\bf k})$ has a meromorphic
extension from above to below on the variable $E$ with a pole at
the point $z_R$. Lets us call this extension $\Phi_C(z,{\bf k})$.
The residue of $\Phi_C(z,{\bf k})$ at $z_R$ gives the Gamow vector
$\Phi^G(z_R,{\bf k})$. On a neighborhood of $z_R$ one has the
following expansion in terms of the complex energy $z$:

\begin{equation}\label{an}
    \Phi_C(z,{\bf k})=\frac{1}{z-z_R}\Phi^G(z_R,{\bf k})+{\rm
regular}\;\; {\rm terms}\,.
\end{equation}
The meromorphic extension of $\Phi(E,{\bf k})$ allows us to extend
(\ref{w}), so that, we have:

\begin{equation}\label{ao}
    H_{as}\Phi_C(z,{\bf k})=z_R\,\Phi_C(z,{\bf k})\,,
\end{equation}
for values of the complex energy $z$ in the lower half plane. If
we bring (\ref{ao})  into (\ref{an}), we get:

\begin{equation}\label{ap}
    \frac{H_{as}\Phi^G(z_R,{\bf k})}{z-z_R}+{\rm regular}\;\; {\rm
terms}=(z-z_R+z_R)\,\frac{\Phi^G(z_R,{\bf k})}{z-z_R}+{\rm
regular}\;\; {\rm terms}\,.
\end{equation}
If we identify the pole terms in the left and right hand sides of
(\ref{ap}), we finally arrive to (\ref{am}). This concludes the
presentation of the model.

\subsection{Resonances in the interaction between a local and a bilocal boson field.}

In this example, developed in \cite{AGPP}, we shall consider an
exactly solvable model for relativistic unstable quantum fields.
This model considers a quadratic interaction between a local
scalar field $\varphi$ with a mass $M$ and a bilocal scalar field
$\psi$ with a continuous mass spectrum (we shall explain this
terminology soon). The interaction makes the local scalar field
unstable. The interaction is exactly solvable and it can be worked
out as a generalization of the Friedrichs model.

It is very convenient to write the local scalar field $\varphi(x)$
with mass $M$ in terms of the creation and annihilation operators
as follows:

\begin{equation}\label{f1}
    \varphi({\bf x},t)=\int d\widetilde{\bf k}\,[a^\dagger({\bf
    k})\,e^{ik\cdot x}+a({\bf k})\,e^{-ik\cdot x}\,]\,.
\end{equation}
Note that $k=(k_0,{\bf k})$ and $x=(x_0,{\bf x})$. By boldface
letters we always denote three dimensional vectors. Four
dimensional vectors in in Minkowski space are denoted by Roman
style letters. The products of two vectors in Minkowski space are
characterized by a dot. In Minkowski space we use the metric
$+---$. For example, $k\cdot x=k_0x_0-{\bf kx}$. The Lorentz
invariant measure $d\widetilde{\bf k}$ was defined in (\ref{e}).

The creation and annihilation operators in (\ref{f1}) satisfy the
following commutation relations:

\begin{equation}\label{f2}
    [a({\bf k}),a^\dagger({\bf k})]=(2\pi)^3\,2\,\omega(({\bf
    k})\,\delta({\bf k}-{\bf k'})\,.
\end{equation}
The Hamiltonian for the field $\varphi(x)$ is given by

\begin{equation}\label{f3}
    H_m=\int d\widetilde{\bf k} \,\omega(({\bf
    k})\,a^\dagger({\bf k})\,a({\bf k})\,.
\end{equation}
Therefore, the field $\varphi(x)$ is a free Klein-Gordon field.

Next, we consider a bilocal scalar field $\psi(x_\mu,q)$ with
continuous mass spectrum and we shall describe this field as
follows. Apart from the dependence of $\psi$ on the four vector
$x$ with components $x_\mu$, $\mu=0,1,2,3$, it depends on an
additional real variable $q$ representing an internal degree of
freedom.

For simplicity in our calculations and because it does not
interfere in our discussion, we shall assume that $\psi(x_\mu,q)$
is an even function of $q$. This means that

\begin{equation}\label{f4}
    \psi(x_\mu,q)=\psi(x_\mu,-q)\,.
\end{equation}
The mass operator affects to this internal variable and in its
simplest form has the form

\begin{equation}\label{f5}
    {\bf M}^2=4m-\frac{\partial^2}{\partial q^2}\,.
\end{equation}
Then, the field $\psi(x_\mu,q)$ satisfies the following
generalized Klein-Gordon equation:

\begin{equation}\label{f6}
    (\Box -{\bf M}^2)\,\psi(x_\mu,q)=0\,,
\end{equation}
where $\Box$ is the usual D'Alembert operator.

Bilocal fields were introduced by Yukawa \cite{Y} and Markov
\cite{MA} in connection with extended particles. The motivation of
its use here is described in \cite{AGMP} in which the solution
$\psi(x_\mu,q)$ of (\ref{f6}) can be seen as the state of two
particles. The fact that the second field is bilocal with a mass
operator with continuous spectrum is causes a quadratic
interaction between these two fields to have non trivial features
such as instabilities.

The solution of equation (\ref{f6}) with condition (\ref{f4}) can
be written in the following form

\begin{equation}\label{f7}
    \psi(x_\mu,q)=\int d\kappa\int \frac{d^3{\bf k}\cos \kappa
q}{(2\pi)^4\,2E({\bf k},\kappa)}\,(B^*({\bf k},\kappa)\,e^{ix\cdot
k} + B({\bf k},\kappa)\,e^{-ix\cdot k})\,,
\end{equation}
where

\begin{equation}\label{f8}
    \kappa=(E^2-{\bf k}^2-4m^2)^{1/2}= \kappa(k_\mu)\,.
\end{equation}
We change the variables in (\ref{f7}), in order to use $E$ as a
new independent variable instead of $\kappa$:

\begin{equation}\label{f9}
    \frac{d\kappa}{E}= \frac{dE}{\kappa}\,.
\end{equation}
After (\ref{f9}), equation (\ref{f7}) reads

\begin{equation}\label{f10}
    \psi(x_\mu,q)=\int_0^\infty dE\int \frac{d^3{\bf k} \cos
\kappa(k_\mu)q}{(3\pi)^4\,\kappa(k_\mu)}\,(B^*({\bf
k},E)\,e^{ix\cdot k} + B({\bf k},E)\,e^{-ix\cdot k})\,.
\end{equation}

Following the routine of second quantization, we replace the
function $B({\bf k},E)$ by an operator, that we also will call
$B({\bf k},E)$ for simplicity, and its complex conjugate $B^*({\bf
k},E)$ by the adjoint operator $B^\dagger({\bf k},E)$, satisfying
the following commutation relations:

\begin{equation}\label{f11}
[B({\bf k},E),B^\dagger({\bf
k},E)]=(2\pi)^4\,\kappa(k_\mu)\,\delta^4(k_\mu-k'_\mu)\,.
\end{equation}
Then, the function $\psi(x_\mu,q)$ becomes the operator

\begin{equation}\label{f12}
    \psi(x_\mu,q)=\int_0^\infty dE\int \frac{d^3{\bf k} \cos
\kappa(k_\mu)q}{(3\pi)^4\,\kappa(k_\mu)}\,(B^\dagger({\bf
k},E)\,e^{ix\cdot k} + B({\bf k},E)\,e^{-ix\cdot k})\,.
\end{equation}
Then, we have constructed the quantum bilocal field. Recall that
the solution $\psi(x_\mu,q)$ to our bilocal field, represents
somehow the state of two particles.

Now, let us assume that the fields (\ref{f12}) and (\ref{f1})
interact and that the interaction is given by

\begin{equation}\label{f13}
    H_{int}= -\lambda\int d^3{\bf x}\int _{-\infty}^\infty
dq\,\psi(x_\mu,q)\, f(q)\,\varphi(x_\mu)\,.
\end{equation}
The function $f(q)$ is called the form factor and we choose it to
be a smooth even function. As always, the coupling constant
$\lambda$ is assumed real and positive. As we mentioned already,
the bilocal field results as an approximation of the behavior of
two relativistic interacting particles. The interaction of the
bilocal field with the field $\varphi(x)$ represents the
interaction of the two former particles with a third of fixed mass
$M$. This is therefore, a model that approaches the behavior of
three interacting particles, just like the other one described in
the precedent section.

If $\alpha(y)$ is the Fourier transform of $f(q)$ and $a({\bf k})$,
$a^\dagger({\bf k})$ are the respective annihilation and creation
operators for the local field $\varphi(x)$, the total Hamiltonian is
given by

\begin{eqnarray}
 P_0= \int \frac{d^3{\bf k} dE}{(2\pi)^4\kappa ({\bf
k},E)}\,E B^\dagger({\bf k},E)B({\bf k},E)+\int \frac{d^3{\bf
k}}{(2\pi)^32\omega({\bf k})}\, \omega({\bf k}) a^\dagger({\bf
k})a({\bf k})
\nonumber \\[2ex]
  + \int \frac{d^3{\bf
k}dE}{(2\pi)^32\omega} \frac{\lambda\alpha (\kappa ({\bf
k},E))}{\kappa ({\bf k},E)} (a({\bf k})+a^\dagger(-{\bf k}))
(B^\dagger({\bf k},E)+B(-{\bf k},E))\,,\nonumber\\[2ex]\label{f14}
\end{eqnarray}
and the three momentum is:

\begin{eqnarray}
 {\bf P}=\int \frac{d^3{\bf k}dE}{(2\pi)^4\kappa({\bf k},E)}{\bf k}
B^\dagger({\bf k},E)B({\bf k},E)+ \int \frac{d^3{\bf k}\,{\bf
k}}{(2\pi)^32\omega({\bf k})}\, a^\dagger({\bf k}) a({\bf k})
\,,\nonumber \\[2ex]
  \label{f15}
\end{eqnarray}
where $\omega({\bf k})$ was given in (\ref{e}).

The next step is to {\it diagonalize} the four momentum
(\ref{f14}) and (\ref{f15}). This means that we are looking for
creation, $b^\dagger (E,{\bf k})$, and annihilation, $b (E,{\bf
k})$, operators such that the four momentum components $P_\mu$ can
be written as:

\begin{equation}\label{f16}
    P_\mu=\int\frac{d^3{\bf k}\,dE}{(2\pi)^4\, \kappa (E,{\bf
k})}\,k_\mu\,b^\dagger (E,{\bf k})\, b (E,{\bf k})\,,
\end{equation}
with $\kappa (E,{\bf k})=\kappa(k_\mu)$ as given in (\ref{f8}). To
get this diagonalization, we pose the eigenvalue equation

\begin{equation}\label{f17}
    [P_\mu, b^\dagger(E,{\bf k})]=k_\mu b^\dagger(E,{\bf
k})\,,
\end{equation}
where $k_\mu=(E,{\bf k})$ and  the $P_\mu$ are given by
(\ref{f14}), (\ref{f15}). To solve (\ref{f17}), i.e., to obtain
the creation operators $b^\dagger(E,{\bf k})$, we make the
following Ansatz:

\begin{eqnarray}
  b^\dagger(E,{\bf k})=\int dE'\left(T(E,E',{\bf k})
B^\dagger(E',{\bf k})+R(E,E',{\bf k})B(E',-{\bf k})\right) \nonumber
\\[2ex]
 +t(E,{\bf k})a^\dagger({\bf k})+r(E,{\bf
k})a(-{\bf k})\,,\label{f18}
\end{eqnarray}
which obviously means that we are assuming that $b^\dagger(E,{\bf
k})$ is a linear combination of $B^\dagger(E,{\bf k}), B(E,{\bf
k}); a^\dagger({\bf k})$ and $a({\bf k})$. This problem was solved
in \cite{AGPP}. The coefficients $T(E,E',{\bf k})$, $R(E,E',{\bf
k})$, $t(E,{\bf k})$ and $r(E,{\bf k})$ depend on the form factor
$f(q)$ in (\ref{f13}) and can be written in terms of the Green
function \cite{AGPP}:

\begin{equation}\label{f19}
   G(E,{\bf k})=\frac{1}{\omega^2-E^2-\Pi (E,{\bf k})}\,,
\end{equation}
with

\begin{equation}\label{f20}
    \Pi (E,{\bf k})=\int\limits^{\infty}_{E_0} dE'2E'\,\frac{\rho
(E',{\bf k})}{E'^2-E^2}
\end{equation}
and $E_0=(4m^2+{\bf k}^2)^{1/2}$. Therefore,  $G(E,{\bf k})$
depends on

\begin{equation}\label{f21}
    \rho (E,{\bf k})=2\pi \frac{\lambda^2\alpha^2(\kappa(E,{\bf
k}))}{\kappa(E,{\bf k})}
\end{equation}
through $\Pi (E,{\bf k})$, since $\rho (E,{\bf k})$ is given by
equation (\ref{f21}). Note that $\rho (E,{\bf k})$ depends on
$\alpha(y)$ which is the Fourier transform of the form factor
$f(q)$.

From (\ref{f21}), we see that we cannot choose arbitrarily the
form factor $f(q)$. For instance, if we fix $f(q)\equiv 1$, its
Fourier transform $\alpha(y)$ is a Dirac delta that cannot be
squared thus making (\ref{f21}) meaningless. To be in the safe
side, we may choose the form factor $f(q)$ to be a smooth (i.e.,
Schwartz) function.

For $\bf k$ fixed, the function $G(E,{\bf k})$ is a function of
the complex variable $E$ with a cut on the real semiaxis
$[E_0,\infty)$. The boundary values of the complex variable
function $G(E,{\bf k})$ from above to below and from below to
above are respectively given by $G_+(E,{\bf k})$ and $G_-(E,{\bf
k})$.

The coefficients $T(E,E',{\bf k})$, $R(E,E',{\bf k})$, $t(E,{\bf
k})$ and $r(E,{\bf k})$ in (\ref{f18}) depend on $G(E,{\bf k})$
and hence on the operators $b^\dagger(E,{\bf k})$ and its adjoint
$b(E,{\bf k})$. If we use $G_+(E,{\bf k})$ instead of $G(E,{\bf
k})$, we are using the so called {\it incoming boundary
conditions}. Henceforth, we shall denote by $b^\dagger_{\rm
in}(E,{\bf k})$ and $b_{\rm in}(E,{\bf k})$ to the solutions of
(\ref{f18}) and its adjoint equation, where we have used
$G_+(E,{\bf k})$ instead of $G(E,{\bf k})$ (If instead of
$G_+(E,{\bf k})$, we use $G_-(E,{\bf k})$, we obtain the new
solution $b^\dagger_{\rm out}(E,{\bf k})$ and $b_{\rm out}(E,{\bf
k})$ corresponding to the {\it outgoing boundary conditions}).

Now, there are two possible situations. If $M<2m$, the functions
$G_+(E,{\bf k})$ and $G_-(E,{\bf k})$ are analytic with no
singularities on the upper and lower half plane respectively
\cite{AGPP}. On the other hand, if $M\ge 2m$, $G_+(E,{\bf k})$ has
a pole at $E^2={\bf k}^2+\mu_c^2$ and $G_-(E,{\bf k})$ has a pole
at $E^2={\bf k}^2+\mu_c^{*2}$, where the star denotes complex
conjugation, and

\begin{equation}\label{f22}
    \mu^2_c=\mu^2-i\mu\Gamma\,.
\end{equation}
The real and positive numbers $\mu$ and $\Gamma$ depend on the
form factor $f(q)$.

The existence of a pole of this kind implies the presence of a
metastable state (resonance) of the system \cite{AGPP}. Carrying
the analogy with the previous model further, the metastable state
appears when the particle of mass $M$ can decay into the two
particle system described by the bilocal field. The condition
$M\ge 2m$ is of course necessary for this process to take place.

The equation for the complex pole $\mu_c$ is \cite{AGPP}

\begin{equation}\label{f23}
    \omega^2({\bf k})-E^2-\int dE'^2\;\frac{\rho(E',{\bf
k})}{E'^2-E^2}=0\,.
\end{equation}
For small values of the coupling constant $\lambda$, we find
\cite{AGPP}

\begin{equation}\label{f24}
    \mu^2_c=
M^2-2i\pi^2\lambda^2\,\frac{[\alpha(\sqrt{M^2-4m^2})]^2}
{\sqrt{M^2-4m^2}}\,.
\end{equation}

In order to obtain the Gamow vectors for this resonance, we have
to obtain first the vacuum $\Omega$ being annihilated by the
operators $b(E,{\bf k})$. This can be obtained by the vacuum
$|0\rangle$, annihilated by $B(E,{\bf k})$ and $a({\bf k})$, i.e.,
the initial vacuum state defined by

\begin{equation}\label{f25}
    B(E,{\bf k})\,|0\rangle=0\,, \hskip0.6cm a({\bf
k})\,|0\rangle=0\,.
\end{equation}
This vacuum is not annihilated by $b(E,{\bf k})$, i.e., $b(E,{\bf
k})|0\rangle\ne 0$. As a function of the energy $E$, $b(E,{\bf k})$
is analytic with a branch cut in $[E_0,\infty)$ \cite{AGPP}.  The
new vacuum $\Omega$ is required to have the following property

\begin{equation}\label{f27}
    b_{\rm in}(E,{\bf k})\,\Omega=0\,,
\end{equation}
where $b_{\rm in}(E,{\bf k})$ represents the function of the
boundary values of $b(E,{\bf k})$ on the cut from above to below. We
obtain this new vacuum making use of the standard theory of
Bogolyubov transformations \cite{BLT}. It should be a superposition
of states with an arbitrary number of particles of $B$ and $a$ types
and is obtained from the old vacuum $|0\rangle$ by a Bogolubov
transformation \cite{AGPP}:

\begin{equation}\label{f26}
    \Omega=e^V\,|0\rangle \,,
\end{equation}
where $V$ depends on the creation operators $B^\dagger(E,{\bf k})$
and $a^\dagger({\bf k})$ and on three functions depending on the
energy and the three momentum. Its explicit form is rather
complicate and not strictly necessary to understand the sequel. See
details in \cite{AGPP}. The procedure requires a rather cumbersome
calculation that we prefer to omit here, see \cite{AGPP}. Now, let
us define

\begin{equation}\label{f28}
    \Phi_{\rm in}(E,{\bf k})=b^\dagger_{\rm in}(E,{\bf
k})|\Omega\rangle\,,
\end{equation}
which, as a function of the complex variable $E$, has a pole in
the analytic continuation from above to below. This pole is
located at the point $z_R=({\bf k}^2+\mu^2-i\mu\Gamma)^{1/2}$
\cite{AGPP}. On a neighborhood of $z_R$, the vector $\Phi_{\rm
in}(E,{\bf k})$ has the form:

\begin{equation}\label{f29}
    \Phi_{\rm in}(E,{\bf k})= \frac{1}{E-z_R}\,\varphi_{\rm
in}^G({\bf k})+ {\rm regular\; part}\,.
\end{equation}
By construction, $\varphi_{\rm in}^G({\bf k})$ is the Gamow vector
associated to the resonance pole $z_R$. Mathematically,
$\varphi_{\rm in}^G({\bf k})$ can be rigorously defined as a
functional on a space of test vectors, dense in the Fock space
\cite{AGPP}, so that the following properties hold:

\begin{itemize}

\item{The Gamow vector $\varphi_{\rm in}^G({\bf k})$ is a
eigenvector of the Hamiltonian with eigenvalue $z_R$, i.e.,

\begin{equation}\label{f30}
    P_0\,\varphi_{\rm in}^G({\bf k})=z_R\,\varphi_{\rm in}^G({\bf
k})\,.
\end{equation}}

\item{It decays exponentially (in a weak sense, note that
$\varphi_{\rm in}^G({\bf k})$ is not in the Fock space), so that
if $t\ge 0$,

\begin{equation}\label{f31}
    e^{-itP_0}\,\varphi_{\rm in}^G({\bf k})= e^{-itz_R}\,
\varphi_{\rm in}^G({\bf k})\,.
\end{equation}}

\end{itemize}

It is important to remark that equation (\ref{f31}) implies that
the Gamow vector $\varphi_{\rm in}^G({\bf k})$ decays
exponentially. In fact, we can decompose $z_R=({\bf
k}^2+\mu^2-i\mu\Gamma)^{1/2}$ in its real and imaginary part:
$z_R=A({\bf k})-iB({\bf k})$. The form of the imaginary part is
given by

\begin{equation}\label{f32}
    B({\bf k})= \left\{\frac{[({\bf k}^2+\mu^2)^2+\mu^2\Gamma^2 ]^{1/2} -({\bf k}^2+\mu^2)}{2}
    \right\}^{1/2}\,.
\end{equation}
For small values of $\Gamma$, we obviously have using the Taylor
theorem

\begin{equation}\label{f33}
B({\bf k})\approx \frac{\mu\Gamma}{2({\bf k}^2+\mu^2)^{1/2}}\,.
\end{equation}
From (\ref{f33}) and (\ref{f31}) the exponential decay of the
Gamow vector $\varphi_{\rm in}^G({\bf k})$ follows.

\subsubsection{Construction of the rigged Hilbert space.}

Since the Gamow vector $\varphi_{\rm in}^G({\bf k})$ decays
exponentially, it cannot belong to a Hilbert space in which the
time evolution is unitary.  There is one possibility for
$\varphi_{\rm in}^G({\bf k})$ to decay under time evolution which
is to extend this time evolution into the dual space of a rigged
Hilbert space. This dual space will include $\varphi_{\rm
in}^G({\bf k})$. The procedure is identical to what we have done
in the case of the standard Friedrichs model although the
construction should be obviously more complicated.

We begin with the following space

\begin{equation}\label{f34}
    {\bf\Psi}=({\cal H}_-^2\cap{\cal S})\otimes {\cal S}({\mathbb
    R}^3)\,,
\end{equation}
for which ${\cal H}_-^2$, $\cal S$ and ${\cal S}({\mathbb
    R}^3)$ have been defined in the previous section right after
    (\ref{ak}). A typical function in $({\cal H}_-^2\cap{\cal S})\otimes {\cal S}({\mathbb
    R}^3)$ is of the form $\varphi(E,{\bf k})$. For a fixed value
    of the momentum $\bf k$, this is a function of the energy
    which belongs to ${\cal H}_-^2\cap{\cal S}$. For a fixed value
    of the energy $E$, $\varphi(E,{\bf k})\in {\cal S}({\mathbb
    R}^3)$.

Then, if $\varphi(E,{\bf k})$ is in ${\bf\Psi}=({\cal
H}_-^2\cap{\cal S})\otimes {\cal S}({\mathbb R}^3)$, we define the
following vectors, written in bra form:

\begin{equation}\label{f35}
    \langle\varphi|=\int_{E_0}^\infty \frac{d^3{\bf k} dE}{(2\pi)^4\kappa ({\bf
k},E)} \,\langle\Omega|\,b_{out}(E,{\bf k})\,\varphi(E,{\bf k})\,,
\end{equation}
where $|\Omega\rangle$ denotes the new vacuum introduced in
(\ref{f26}). Note that the above integral goes from $E_0$ to
$\infty$, which means that the energy is always positive. Also,
its is important to point out that for fixed $\bf k$, the function
$\varphi(E,{\bf k})$ is a Hardy function on the lower half plane
that is completely determined by its values on the interval
$[E_0,\infty)$ \cite{AGPP}.

By construction, the set of the abstract vectors of the form
(\ref{f35}) is a vector space $\bf\Phi$. The mapping

\begin{equation}\label{f36}
    f: \;\varphi(E,{\bf k})\longmapsto |\varphi\rangle
\end{equation}
defines a one to one correspondence from ${\bf\Psi}=({\cal
H}_-^2\cap{\cal S})\otimes {\cal S}({\mathbb R}^3)$ onto
$\bf\Phi$. We do not have defined a topology in $\bf\Phi$ yet, but
we have an isomorphism $g$ from $\bf\Psi$ onto $\bf\Phi$ and a
topology on $\bf\Psi$, which is a tensor product of two Schwartz
topologies \cite{SC}. Then, we use $g$ to transport the topology
from $\bf\Psi$ to $\bf\Phi$. This is a usual procedure in the
theory of locally convex spaces \cite{SC}.

Then, we want to outline in here the construction for the rigged
Hilbert space in which are well defined the Gamow vector
$\varphi_{\rm in}^G({\bf k})$ and its time evolution. We shall
skip some technical details that go beyond the scope of the
present work.

First of all, we construct tensor products of the form
${\bf\Phi}\otimes{\bf\Phi}$, ${\bf\Phi}\otimes
{\bf\Phi}\otimes{\bf\Phi}$, etc. The algebraic tensor products
give topological vector spaces that have to be completed with
respect to some topology (which in particular assures the
convergence of Cauchy sequences) \cite{J}. Then, let us construct
the following sequence:

\begin{eqnarray}
  {\bf\Phi}_0&=& {\mathbb C} \nonumber\\[2ex]
  {\bf\Phi}_1 &=& {\mathbb C}\oplus{\bf\Phi} \nonumber \\[2ex]
  {\bf\Phi}_2 &=& {\mathbb C}\oplus{\bf\Phi}\oplus {\bf\Phi}\otimes{\bf\Phi}
  \nonumber\\[2ex]
  \dots & & \dots \dots\dots \nonumber \\[2ex]
  {\bf\Phi}_n &=& {\mathbb C}\oplus{\bf\Phi}\oplus {\bf\Phi}\otimes{\bf\Phi}
  \oplus\dots\oplus
  {\bf\Phi}\otimes \dots \otimes {\bf\Phi}  \nonumber\\[2ex]
  \dots & & \dots \dots\dots \;.\label{f36}
\end{eqnarray}
Note that, for all values of $n$, ${\bf
\Phi}_n\subset{\bf\Phi}_{n+1}$. It is also true that the identity
mapping from ${\bf \Phi}_n$ into ${\bf\Phi}_{n+1}$ is continuous.
Then, let us construct the following space:

\begin{equation}\label{f37}
    Y_-:=\bigcup_{k=0}^\infty {\bf\Phi}_k\,.
\end{equation}
Then, we endow $Y_-$ with the {\it finest} topology that make all
the following identity maps continuous:

\begin{equation}\label{f38}
    i_n\;: {\bf\Phi}_n\longmapsto Y_-\,.
\end{equation}
Roughly speaking, finest means that any other topology in $Y_-$
that makes all the $i_n$ continuous must have less open sets. The
topology in $Y_-$ is usually called  strict inductive limit
topology \cite{J,SC,VKK}. The space $Y_-$ is a nuclear space
\cite{J}.

We denote by $Y_-^\times$ the dual space of $Y_-$. It has the
property that for all natural $n$, ${\mathbb
C}\oplus{\bf\Phi}^\times
\oplus\dots\oplus({\bf\Phi}^\times\otimes\dots\otimes{\bf\Phi}^\times)
\subset Y_-^\times$, where the last tensor product has $n$ factors
\cite{AGPP}.

To complete a rigged Hilbert space, we need a Hilbert space. Let
us start with ${\cal H}_-^2\otimes L^2({\mathbb R}^3)$. This space
is the completion of $({\cal H}_-^2\cap{\cal S})\otimes{\cal
S}({\mathbb R}^3)$ with respect to the Hilbert space topology. It
is naturally isometric to the completion $\cal H$ of $\bf\Phi$
with respect also to the Hilbert space topology. The space $\cal
H$ can be looked as the result of using functions in ${\cal
H}_-^2\otimes L^2({\mathbb R}^3)$ instead of functions in $({\cal
H}_-^2\cap{\cal S})\otimes{\cal S}({\mathbb R}^3)$ in (\ref{f35}).
Then,

\begin{equation}\label{f39}
    {\cal H}_n={\mathbb C}\oplus{\cal H} \oplus \dots\oplus ({\cal
    H}\otimes \dots\otimes{\cal H})\,, \qquad {\cal H}_0={\mathbb
    C}
\end{equation}
and then construct the Fock space

\begin{equation}\label{f40}
    {\cal F}:=\bigcup_{k=0}^\infty {\cal H}_n\,.
\end{equation}
Then, the triplet

\begin{equation}\label{f41}
    Y_-\subset{\cal F}\subset Y_-^\times
\end{equation}
is a rigged Hilbert space. The Gamow vector $\varphi_{\rm
in}^G({\bf k})$ is a continuous functional on $Y_-$ and therefore,
it belongs to $Y_-^\times$. When applied to a vector in $\mathbb
C$ or ${\bf\Phi}\otimes\dots\otimes{\bf\Phi}$, it gives zero,
provided that the tensor product involves two or more copies of
$\bf\Phi$. On $\bf\Phi$ itself, $\varphi_{\rm in}^G({\bf k})$ is
nonzero.

The equation $e^{itP_0} Y_-\subset Y_-$ is valid if and only if
$t\ge 0$ \cite{AGPP}. Therefore, by duality (see Appendix), we
obtain that the time evolution operator $e^{-itP_0}$ acts on
$Y_-^\times$ with $e^{-itP_0}Y_-^\times\subset Y_-^\times$ if and
only if $t\ge 0$. Thus, (\ref{f31}) is correct if and only if
$t\ge 0$.

Also the operator $P_0$ has the property that $P_0Y_-\subset Y_-$
so that it can be extended into the dual $Y_-^\times$. It has the
property that \cite{AGPP}.

\begin{equation}\label{f42}
    P_0 \,\varphi_{\rm in}^G({\bf k})=z_r\,\varphi_{\rm in}^G({\bf
    k})\,.
\end{equation}
Here we conclude this discussion on the construction of the rigged
Hilbert space (\ref{f41}) often called the rigged Fock space.

\subsection{Friedrichs model with virtual transitions.}

The Friedrichs model with virtual transitions is probably the
simplest Friedrichs model for unstable interaction that uses
second quantization techniques. This is an approach which is
completely different from the models discussed so far as we shall
see in the sequel.

\subsubsection{The second quantization of the Friedrichs model.}

As we have see in section 2, the total Hamiltonian in the
Friedrichs model has three terms $H=H_D+H_C+H_I$:

i.) A continuous part, $H_C$, represented by the integral in
(\ref{10}). This integral is like a ``linear combination'' of the
eigenkets, $|\omega\rangle$, of the free Hamiltonian $H_0$ whose
eigenvalues are in the continuous spectrum of $H_0$, which is the
positive semiaxis ${\mathbb R}^+=[0,\infty)$. Thus, the eigenkets
$|\omega\rangle$ fulfill the relation
$H_0|\omega\rangle=\omega|\omega\rangle$. For $|\omega\rangle$, we
can define creation, $b^\dagger_\omega$, and annihilation operators,
$b_\omega$, so that from a hypothetical vacuum $|0\rangle$,
$b^\dagger_\omega|0\rangle=|\omega\rangle$ and
$b_\omega|\omega\rangle=|0\rangle$. Clearly, $b^\dagger_\omega
b_\omega$ are the number operators for states $|\omega\rangle$.
Obviously, $b^\dagger_\omega b_\omega|0\rangle\langle
0|=|\omega\rangle\langle\omega|$, so that we can write the integral
in (\ref{10}) in terms of these operators as

\begin{equation}\label{v1}
    H_C=\int_0^\infty
    d\omega\,\omega\,|\omega\rangle\langle\omega| = \int_0^\infty
d\omega\,\omega\,b^\dagger_\omega b_\omega|0\rangle\langle 0|\,.
\end{equation}
Then, we can omit the dyad $|0\rangle\langle 0|$ in (\ref{v1}) and
simply write

\begin{equation}\label{vv1}
  H_C= \int_0^\infty
d\omega\,\omega\,b^\dagger_\omega b_\omega\,.
\end{equation}

ii.) A discrete part, $H_D$, represented by the first term in
(\ref{10}). This discrete part consists in a single eigenvalue
imbedded in the continuous spectrum of $H_0$. This eigenvalue has
an eigenvector that we call $|1\rangle$. If $a^\dagger$ and $a$
are the respective creation and annihilation operators for
$|1\rangle$, the ``discrete'' part

\begin{equation}\label{v2}
    H_D=\omega_0|1\rangle\langle 1|=\omega_0\; a^\dagger a|0\rangle\langle
    0|\,,
\end{equation}
or simply,
\begin{equation}\label{vv2}
H_D=\omega_0\; a^\dagger a\,.
\end{equation}
Note that $\omega_0>0$.

iii.) The interaction between the discrete and the continuous parts,
$H_I$, depends on both the coupling constant $\lambda$ and the form
factor $f(\omega)$. the explicit form is given in (\ref{11}) (except
for the multiplication by $\lambda$). In terms of the creation and
annihilation operators following the notation as used in (\ref{vv1})
and (\ref{vv2}), $H_I$ is given by

\begin{equation}\label{v3}
    H_I=\lambda\int_0^\infty
    d\omega\,f(\omega)\,(a^\dagger\,b_\omega+a\,b^\dagger_\omega)\,.
\end{equation}

These creation and annihilation operators satisfy commutation
relations corresponding to bosons, so that

\begin{equation}\label{v4}
    [a,a^\dagger]=I\qquad;\qquad
    [b_\omega,b^\dagger_{\omega'}]=\delta(\omega-\omega')\,,
\end{equation}
so that the Friedrichs model can be seen as a boson interacting
with a boson field.

In the sequel, we want to propose a slight modification of the
above model with total hamiltonian given by $H:=H_0+\lambda V$
with

\begin{equation}\label{v5}
    H_0:= \varepsilon_0+\omega_0\, a^\dagger a+\int_0^\infty
    d\omega\,\omega \,b_\omega^\dagger\,b_\omega\,,
\end{equation}
where $\varepsilon_0>0$ is the energy of the vacuum. The potential
$V$ is given by

\begin{equation}\label{v6}
V:= \int_0^\infty d\omega\,f(\omega)\,(b_\omega^\dagger
+b_\omega)(a^\dagger+a)\,.
\end{equation}
We have chosen a real form factor which should be, in addition,
square integrable:

\begin{equation}\label{v7}
    \int_0^\infty d\omega\,f^2(\omega)<\infty\,.
\end{equation}
Note that $H$ has a continuous spectrum given by
$[\varepsilon_0,\infty)$ and a discrete spectrum, which has an
isolated point at $\varepsilon_0+\omega_0$, which is embedded in
the continuum.

In order to obtain a compact form for $H=H_0+\lambda V$, we need
to diagonalize this operator. When we use this second quantized
version of the Friedrichs model it is useful to do it by solving
the eigenvalue problem of the Liouville-von Neumann operator
$L=[H,\cdot]$ acting in the algebra of observables generated by
the operators $a_1$, $a^\dagger_1$, $b_\omega$ and
$b^\dagger_\omega$. This eigenvalue problem can be posed as

\begin{equation}\label{v8}
    LB^\dagger_\omega=\omega\,B^\dagger_\omega \qquad;\qquad
    LB_\omega=-\omega B_\omega\,,
\end{equation}
where $B^\dagger_\omega$ and $B_\omega$ are functions of the
previously given creation and annihilation operators respectively
\cite{KPPP,AGPP}. The final result will be the diagonalized total
Hamiltonian which has the following form

\begin{equation}\label{v9}
    H=E_0+\int_0^\infty d\omega\,\omega\,B^\dagger_\omega
    B_\omega\,,
\end{equation}
where $E_0$ is a renormalized vacuum energy that will be obtained
later.

This problem can be solved as follows: We write the operators
$B_\omega^\dagger$ and $B_\omega$ as linear combinations of
$a^\dagger$, $a$, $b^\dagger_\omega$ and $b_\omega$ with unknown
coefficients that are unknown functions and constants. The
procedure has been explained in \cite{AGPP}. These linear
combinations are then inserted in (\ref{v8}). It results a set of
integral equations for the unknown functions and constants. The
solution of these integral equations give the mentioned
coefficients. The final solution is

\begin{eqnarray}
  (B^\dagger_\omega)_{{\rm in}\atop{\rm out}} =
   b^\dagger_\omega+2\omega_0\lambda\,f(\omega)\,G^\pm(\omega) \nonumber
   \\[2ex]
   \times \left\{\int_0^\infty d\omega'\,
   \lambda\,f(\omega')\;\left(\frac{b^\dagger_{\omega'}}{\omega'-(\omega\pm
   i0)}-\frac{b_{\omega'}}{\omega'+\omega}
   \right)-\frac{(\omega-\omega_0)\,a^\dagger+(\omega-\omega_0)\,a}{2\omega_0}
   \right\}\nonumber\\[2ex]\label{v10} \\[2ex]
  (B_\omega)_{{\rm in}\atop{\rm out}}= b_\omega+2\omega_0\lambda\,f(\omega)\,G^\mp(\omega)
  \nonumber\\[2ex]
   \times \left\{\int_0^\infty d\omega'\,
   \lambda\,f(\omega')\;\left(\frac{b_{\omega'}}{\omega'-(\omega\mp
   i0)}-\frac{b^\dagger_{\omega'}}{\omega'+\omega}
   \right)-\frac{(\omega-\omega_0)\,a^\dagger+(\omega-\omega_0)\,a}{2\omega_0}
   \right\}\,.\nonumber\\[2ex]\label{v11}
\end{eqnarray}

Two sets of solutions called incoming and outgoing are determined
by the boundary values of the Green function $G(z)$ on the real
axis. The boundary values from above to below are denoted as
$G^+(\omega)=G(\omega+i0)$ and those form below to above as
$G^-(\omega)=G(\omega-i0)$. The Green function $G(z)$ for complex
values of $z$ is given by

\begin{equation}\label{v12}
    G(z)=\left[\omega_0^2-z^2-\int_0^\infty
    d{\omega'}^2\,\frac{2\omega_0\,\lambda^2\,f^2(\omega')}{{\omega'}^2-z^2}
    \right]^{-1}\,.
\end{equation}

If, in addition, the form factor $f(\omega)$ satisfies the
following condition

\begin{equation}\label{v13}
    \omega_0^2-\int_0^\infty d\omega'\,
    \frac{2\omega_0\,\lambda^2\,f^2(\omega)}{{\omega'}^2}>0\,,
\end{equation}
then $G(z)$ is analytic in the whole plane without any
singularities except for a branch cut coinciding with the real
semiaxis $[\varepsilon_0,\infty)$. Nevertheless, $G(z)$ admits
analytic continuations through the cut as mentioned before.

The new operators satisfy simple commutation relations:

\begin{equation}\label{v14}
    [(B_\omega)_{\rm in}, (B^\dagger_{\omega'})_{\rm in}] =
    [(B_\omega)_{\rm out}, (B^\dagger_{\omega'})_{\rm out}]
    =\delta(\omega-\omega')\,.
\end{equation}

The next step is to write the original field operators in terms of
the new ones. As we are mainly interested in the decay process we
are in the time sector $t>0$. As the ``in'' operators are
ultimately responsible for creation and annihilation of the
decaying sector, we are from now on assuming that we are using
only incoming operators. Consequently, we shall drop the subscript
``in'' unless necessary. With the help of the properties of the
function $G(z)$ \cite{KPPP}, we obtain that

\begin{equation}\label{v15}
b^\dagger_\omega=B^\dagger_\omega-2\omega_0\,\lambda\,f(\omega)\int_0^\infty
d\omega'\,\lambda\,f(\omega')\,\left[
\frac{G^-(\omega)(\omega')}{\omega'-\omega-i0}\,B^\dagger_{\omega'}
-\frac{G^+(\omega')}{\omega'+\omega}\, B_{\omega'} \right]\,,
\end{equation}

\begin{equation}\label{v16}
b_\omega=B_\omega-2\omega_0\,\lambda\,f(\omega)\int_0^\infty
d\omega'\,\lambda\,f(\omega')\,\left[
\frac{G^+(\omega)(\omega')}{\omega'-\omega+i0}\,B_{\omega'}
-\frac{G^-(\omega')}{\omega'+\omega}\, B^\dagger_{\omega'}
\right]\,,
\end{equation}

\begin{equation}\label{v17}
    a^\dagger=-\int_0^\infty
    d\omega\,\lambda\,f(\omega)\,[(\omega+\omega_0)\,G^-(\omega)\,B_\omega^\dagger
    -(\omega-\omega_0)\,G^+(\omega)\,B_\omega]\,,
\end{equation}

\begin{equation}\label{v18}
    a=-\int_0^\infty
    d\omega\,\lambda\,f(\omega)\,[(\omega+\omega_0)\,G^+(\omega)\,B_\omega
    -(\omega-\omega_0)\,G^-(\omega)\,B^\dagger_\omega]\,.
\end{equation}

We note that there is a substantial difference between the
representation in terms of $a$ and $b_\omega$ and the
representation in terms of $B_\omega$ (and their hermitic
conjugates). The reason is clear: while the free Hamiltonian $H_0$
has a bound state, and we need the operators $a$ and $a^\dagger$
to annihilate and create this state respectively, this situation
does not arises for the total Hamiltonian $H$, which solely has
continuous spectrum. Thus, the Bogolubov transformation that
transfer equations in terms of operators of the type $a$,
$b_\omega$, etc (often called {\it bare} operators) into
$B^\dagger_\omega$ and $B_\omega$ ({\it dressed operators}) are of
the so called {\it improper} type \cite{BER}. In any case, it is
possible to construct a new vacuum state $|\Omega\rangle$ from the
vacuum state $|\Xi\rangle$ annihilated by $a$ and $b_\omega$. This
new vacuum state will satisfy for all
$\omega\in[\varepsilon_0,\infty)$,

\begin{equation}\label{v19}
    B_\omega|\Omega\rangle=0\,.
\end{equation}
This new vacuum state along the commutation relations (\ref{v14})
for the operators $B_\omega^\dagger$ and $B_\omega$ permit us to
construct the Fock space suitable for the space of states of the
new situation. Following the general approach for the Bogolubov
transformation \cite{BER}, the new vacuum state is related with
the old one through the following expression

\begin{equation}\label{v20}
    |\Omega\rangle:= e^V\,|\Xi\rangle\,,
\end{equation}
where $V$ is the so called {\it dressing operator} and is given by
a quadratic form on the bare operators. The unknown constants and
functions that serve as coefficients of this quadratic form are
obtained by inserting it into (\ref{v20}) and then using
(\ref{v19}). The technique was already used in the previous model,
see \cite{AGPP}. The final result is

\begin{eqnarray}
  V=\int_0^\infty d\omega'\int_0^\infty
  d\omega''\,\omega_0\,\lambda^2\,f(\omega')\,f(\omega'')\,
  \eta(\omega')\,\eta(\omega'')\nonumber\\[2ex]\times
\left(
\frac{1}{2(\omega_0+\delta\varepsilon_0)}+\frac{1}{\omega'+\omega''}\,
\right)\,b^\dagger_{\omega'}\,b^\dagger_{\omega''}\nonumber
  \\[2ex]
 +\int_0^\infty d\omega'\,\omega_0\,\frac{\lambda\,f(\omega')\,\eta(\omega')}{2(\omega_0+\delta\varepsilon_0)}
\,b^\dagger_{\omega'}\,a^\dagger-
\frac{\delta\varepsilon_0}{2(\omega_0+\delta\varepsilon_0)}\,(a^\dagger)^2\,,
 \label{v21}
\end{eqnarray}
where $\delta\varepsilon_0$ is the vacuum energy shift. This
result allows us to determine the vacuum energy $E_0$ of the total
Hamiltonian $H$ as

\begin{equation}\label{v22}
    E_0=\varepsilon_0+\int_0^\infty d\omega\,\frac{\omega_0\,
    \lambda^2\,f^2(\omega)\,\eta(\omega)}{\omega_0+\delta\varepsilon_0}=
    \varepsilon_0+\delta\varepsilon_0\,.
\end{equation}

The function $\eta(\omega)$, or for complex values of the argument
$\eta(z)$, is the solution of the factorization of the Green
function $G(z)$, a problem solved in \cite{AGPP}:

\begin{equation}\label{v23}
    \eta(z)\,\eta(-z)=G(z)\,.
\end{equation}
The function $\eta(-z)$ is analytic on the complex plane with a
branch cut in $(-\infty,-\varepsilon_0]$. It obeys the following
expressions:

\begin{equation}\label{v24}
\eta^{-1}(-z)=z-(\omega_0+2\delta\varepsilon_0)-\int_0^\infty
d\omega\,\frac{2\omega_0\,\lambda^2\,f^2(\omega)\,\eta(\omega)}{\omega-z}
\end{equation}
and

\begin{equation}\label{v25}
\eta(-z)=\int_0^\infty
d\omega\,\frac{2\omega_0\,\lambda^2\,f^2(\omega)\,|G(\omega)|^2}{\eta(\omega)\,(\omega-z)}\,.
\end{equation}

\subsubsection{Time evolution for the bare operators.}

From equations (\ref{v15}-\ref{v18}), it is possible to obtain the
time evolution with respect to the total Hamiltonian of the bare
operators. The derivation of next formulas appear in \cite{KPPP} and
the final result is given by
($b^\dagger_\omega(0)=b^\dagger_\omega$)

\begin{eqnarray}
 b^\dagger_\omega(t)= e^{i\omega t}\,b_\omega^\dagger(0) -
 2\omega_0\,\lambda\,f(\omega) \int_0^\infty
 d\omega'\,\lambda\,f(\omega')
\nonumber\\[2ex]
  \times \left([g(\omega,t)-g(\omega',t)]\,\frac{b^\dagger_{\omega'}}{\omega'-\omega-i0}
 -[g(\omega,t)-g(-\omega',t)]\,\frac{b_{\omega'}}{\omega'+\omega} \right) \nonumber\\[2ex]
 -\lambda \,f(\omega)\,\left[\left(i\frac{\partial}{\partial
 t}-\omega_0
 \right)\,a^\dagger + \left( i\frac{\partial}{\partial
 t}+\omega_0\right)\,a
 \right]\, g(\omega,t)
 \label{v26}
\end{eqnarray}
and

\begin{eqnarray}
  a^\dagger(t) =- \left(i\frac{\partial}{\partial t}-\omega_0\right) \int_0^\infty d\omega \,
\lambda \,f(\omega)\,[g(\omega,t)\,b^\dagger_\omega+g(-\omega,t)\,b_\omega]   \nonumber \\[2ex]
+\int_0^\infty d\omega\,\lambda^2\,f^2(\omega)\,|G(\omega)|^2
\,\{[(\omega+\omega_0)^2\,e^{i\omega
t}-(\omega-\omega_0)^2\,e^{-i\omega t}]\,a^\dagger \nonumber
\\[2ex]
 +(\omega^2-\omega_0)^2\, [e^{i\omega t}-e^{-i\omega
 t}]\,a\}\,,\label{v27}
\end{eqnarray}
where

\begin{equation}\label{v28}
g(z,t):=\int_0^\infty d\omega\,
2\omega_0\,\lambda^2\,f^2(\omega)\,|G(\omega)|^2\,\left(
\frac{e^{i\omega t}}{\omega-z}+\frac{e^{-i\omega
t}}{\omega+z}-G(z)\right)\,e^{izt}\,.
\end{equation}

One can prove that $g(z,t)$ is an analytic function of $z$ except
for the infinity, see \cite{KPPP}. In terms of $t$, it has the
following asymptotic behavior:

\begin{equation}\label{v29}
g(\omega,t)\longmapsto -e^{-i\omega t}\,G^\mp(\omega) \hskip2cm {\rm
as}\qquad\qquad t\mapsto\pm\infty\,.
\end{equation}
Also, we can obtain \cite{KPPP} the asymptotic behavior, in terms
of $t$, of the bare operators:

\begin{equation}\label{v30}
b^\dagger_\omega \longmapsto (B^\dagger_\omega)_{{\rm
out}\atop{\rm in}}\hskip2cm {\rm as}\qquad\qquad
t\mapsto\pm\infty\,,
\end{equation}
and

\begin{equation}\label{v31}
a^\dagger(t)\longmapsto {\bf 0}\hskip2cm {\rm as}\qquad\qquad
t\mapsto\pm\infty\,.
\end{equation}

Equations (\ref{v30}) and (\ref{v31}) are valid in a weak sense so
that no conclusion about the products $a^\dagger(t)a(t)$ and
$a(t)a^\dagger(t)$ can be made. Therefore, there is no
contradiction between (\ref{v31}) and the commutation relation

\begin{equation}\label{v32}
[a(t),a^\dagger(t)]=1\,,
\end{equation}
valid for any finite value of $t$.

\subsubsection{Resonance poles.}

In order to find resonance poles, we have to go back to
(\ref{v12}). In fact, $G(z)$ (\ref{v12}) has the same structure as
$\eta^{-1}(z)$ in (\ref{15}) corresponding to the standard
Friedrichs model. Therefore, the analytic continuation of  $G(z)$
has two complex conjugate resonance poles located at the points
$z_R$ and $z_R^*$.  In the first order of approximation in
$\lambda$, we find that

\begin{equation}\label{v33}
z_R\approx \widetilde\omega_0+i\frac{\gamma}{2}\,,
\end{equation}
where

\begin{equation}\label{v34}
\gamma=2\pi \lambda^2\,f^2(\omega_0)
\end{equation}
and

\begin{equation}\label{v35}
\widetilde\omega_0 -PV\int_0^\infty
d\omega\,\frac{\lambda^2\,f^2(\omega)}{\omega^2-\omega_0^2}\,,
\end{equation}
where $PV$ stands for principal value.

\subsubsection{Why Friedrichs model with virtual transitions?}

As an example of the application of the solution of this second
quantized Friedrichs model, we shall consider the time evolution
of the number of {\it bare} photons. Starting with the bare vacuum
$|\Xi\rangle$ as the initial condition, the goal is to calculate
the quantity given by

\begin{equation}\label{v36}
n_\omega(t):=\langle\Xi|b^\dagger_\omega\,b_\omega|\Xi\rangle\,,
\end{equation}
where obviously $b^\dagger_\omega\,b_\omega$ is the number
operator for photons with energy $\omega$. Then, using (\ref{v26})
and its Hermitian conjugate, we obtain the following result:

\begin{eqnarray}
 n_\omega(t)=2\omega_0\,\lambda^2\,f^2(\omega)\int_0^\infty
 d\omega''
 \,\frac{2\omega_0\,\lambda^2\,f^2(\omega'')}{(\omega+\omega'')^2}\,
 |g(-\omega,t)-g(\omega'',t)|^2
  \nonumber\\[2ex]
 +\lambda^2\,f^2(\omega)\left|\left(i\frac{\partial}{\partial
 t}-\omega_0
 \right)\, g(-\omega,t)\right|^2\,.\label{v36}
\end{eqnarray}
This expression is exact and it is valid for any value of
$\lambda$ verifying (\ref{v13}). In the weak coupling case (small
$\lambda$ and constant $t\lambda^2$), we can obtain the following
formula \cite{KPPP,PPP}:

\begin{equation}\label{v37}
n_\omega(t)=2\,\frac{\lambda^2\,f^2(\omega)}{(\omega+\omega_0)^2}
\,[1-\cos((\omega+\omega_0)t)]+O(\lambda^4)\,.
\end{equation}

This formula reflects the initial grow of the number of bare
photons due to virtual process. According a point of view
\cite{PPP}, this corresponds to the formation of a photon cloud
around an atom. Also, the first term in (\ref{v37}) gives the
probability of emission of a virtual photon.

\section{Other types of Friedrichs models.}

New kinds of Friedrichs models have been studied in the
literature. Along the present section, we discuss two more
versions. In the former, we have a discrete Friedrichs model,
which has been used to approximate systems with a continuous
spectrum by systems with a discrete spectrum \cite{GGA}.  In the
other, the internal channel has already continuous spectrum and
this allows to the appearance of a resonant branch cut in the
analytic continuation of the reduced resolvent \cite{RG}, as we
shall see.

\subsection{A discrete Friedrichs model.}

We begin with a harmonic oscillator, the system (in the language
of thermodynamics) or the internal channel (in the language of the
Friedrichs model), interacting with a bath of harmonic oscillators
(the external channel), for which the free Hamiltonian is given by

\begin{equation}\label{b1}
    H_0:= \frac{P^2}{2M}+\frac 12 M\Omega X^2+\sum_{n=1}^N \left(
    \frac{p_n^2}{2m_n}+\frac 12 m_n\omega_n^2 x_n^2\right)\,.
\end{equation}

Following \cite{GGA}, we use capital letters for the variables in
the system and small letters for the elements of the bath. This
makes a proper notational distinction. We need to define an
interaction between both system and bath (or internal and external
channels), and the proposed interaction reads \cite{GGA}:

\begin{equation}\label{b2}
    H_I:= \sum_{n=1}^N c_n\left(Xx_n+\frac{Pp_n}{M\Omega
    m_n\omega_n}\right)\,,
\end{equation}
where the $c_n$ are real (due to the need for hermiticity of $H_I$
these variables must be real) coupling constants. Note that $H_I$ is
non local.  The total Hamiltonian is $H:=H_0+H_I$. If we define the
usual creation and annihilation operators as (we omit the constant
$\hbar$ or make it equal to one):

\begin{eqnarray}
  B &=& \sqrt{\frac{M\Omega}{2}}\, X+i\sqrt{\frac{1}{2M\Omega}}\,
  P\label{b3}
  \\[2ex]
  b_n &=&
  \sqrt{\frac{m_n\omega_n}{2}}\,x_n+i\sqrt{\frac{1}{2m_n\omega_n}}\,
  p_n\,,\label{b4}
\end{eqnarray}
which satisfies the canonical commutation relations given by

\begin{equation}\label{b5}
    [B,B^\dagger]=I\hskip01cm;\hskip1cm
    [b_n,b_m^\dagger]=\delta_{nm}\,,
\end{equation}
where $\delta_{nm}$ is the Kronecker delta and this equation is
valid for any $n,m=1,2\dots,N$. All other commutators vanish. In
terms of these operators, the total Hamiltonian $H$ has the
following form:

\begin{equation}\label{b6}
    H=\Omega\left(B^\dagger B+\frac 12\right) +
    \sum_{n=1}^N\omega_n\left(b_n^\dagger b+\frac 12\right)
    +\sum_{n=1}^N g_n (Bb_n^\dagger+B^\dagger b_n)\,,
\end{equation}
where $g_n:=c_n/\sqrt{M\Omega m_n\omega_n}$. This kind of coupling
has been previously considered, for instance in \cite{LW}.

Let us assume that the vacuum state is defined as

$$
|0\rangle:=|0\rangle\otimes|0,\dots,0\rangle\,,
$$
where the first term $|0\rangle$ in the above tensor product is
the ground state for the system and $|0,\dots,0\rangle$ is the
ground state of the bath, in which we distinguish each harmonic
oscillator. For instance $|a_1, a_2, \dots, a_N\rangle$ is the
state of the bath with the first oscillator in the state $a_1$,
the second in the state $a_2$, etc, up to the latter in the state
$a_N$. In the present case, $a_1=a_2=\dots=a_N=0$. Define now the
following vectors:

\begin{eqnarray}
  |\Omega\rangle &=& B^\dagger|0\rangle=|1\rangle\otimes|0,\dots,0\rangle
 \label{b7} \\[2ex]
  |\omega_k\rangle &=& b_k^\dagger|0\rangle=|0\rangle\otimes
  |0,\dots,1,\dots,0\rangle\,, \label{b8}
\end{eqnarray}
where the $1$ in (\ref{b8}) is placed at the $k-$th site. With
this definition, the total Hamiltonian becomes:

\begin{equation}\label{b9}
    H=\Omega|\Omega\rangle\langle
    \Omega|+\sum_{n=1}^N\omega_n|\omega_n\rangle\langle\omega_n| +
    \sum_{n=1}^Ng_n(|\Omega\rangle\langle\omega_n|+|\omega_n\rangle\langle
    \Omega|)+C\,,
\end{equation}
where $C=\Omega/2+\sum_{n=1}^N(\omega_n/2)$ is a
constant\footnote{Note that, save for the constant $C$, the
Hamiltonian in (\ref{b9}) is given by $\tilde H|0\rangle\langle 0|$,
where $\tilde H$ is the Hamiltonian $H$ given in (\ref{b5}). The
situation is quite similar than those discussed in Section 5.3.}.
The Hamiltonian in (\ref{b9}) has already the structure of the
Hamiltonian of the Friedrichs model (except for the constant that we
may eventually drop). In fact, (\ref{b9}) has a free Hamiltonian
given by $\Omega|\Omega\rangle\langle
    \Omega|+\sum_{n=1}^N\omega_n|\omega_n\rangle\langle\omega_n|$
    plus an interaction that connects the internal and external
    channels given by $\sum_{n=1}^Ng_n(|\Omega\rangle\langle\omega_n|+|\omega_n\rangle\langle
    \Omega|)$. The contants $g_n$ play the role of form factors. In
addition, we impose the obvious condition that $\Omega$ should not
be identical to none of the $\omega_k$.

In order to solve an eigenvalue equation, similar to (\ref{v8}), we
first make the following Bogolubov transformation:

\begin{equation}\label{b10}
    c_k:= A_k B+\sum_{n=1}^N a_{kn} b_n\,,
\end{equation}
where the coefficients $A_k$ and $a_{kn}$, $n=1,2,\dots,N$,
$k=0,1,2,\dots, N$ are chosen in such a way that the
transformation be unitary and the following commutation relations
hold

\begin{equation}\label{b11}
    [c_k,c_l^\dagger]=\delta_{kl}\,,
\end{equation}
and all other vanish. In addition, we require transformation
(\ref{b10}) to diagonalice the total Hamiltonian $H$ as:

\begin{equation}\label{b12}
    H=\sum_{k=0}^N\alpha_k c_k^\dagger c_k+C\,.
\end{equation}
This requirements completely determine the transformation
(\ref{b10}). In fact, if $|\alpha_k\rangle:=c_k^\dagger
|0\rangle$, the coefficients in (\ref{b10}) are given by

\begin{equation}\label{b13}
    A_k=\langle \alpha_k|\Omega\rangle\hskip1cm;\hskip1cm a_{kn}=
    \langle \alpha_k|\omega_n\rangle\,.
\end{equation}
The vectors $|\alpha_k\rangle$ are orthonormal. We want to obtain
the eigenvalues of the renormalized Hamiltonian $H_R:=H-C$, $C$
being the constant that appears in the expression of $H$ in
(\ref{b9},\ref{b12}).

The operators $c_k$, $B$ and $b_n$ do not commute with the total
Hamiltonian $H$ (or $H_R$) and their commutators give:

\begin{eqnarray}
  [c_k,H] &=& \alpha_k c_k= \alpha_k A_k B+\sum_{n=1}^N \alpha_k a_{kn}b_n\,,\label{b14}
  \\[2ex]
  [B,H] &=& \Omega B+\sum_{n=1}^N g_nb_n\,, \label{b15} \\[2ex]
  [b_n,H] &=& \omega_n b_n+g_n B\,. \label{b16}
\end{eqnarray}
These equations give:

\begin{eqnarray}
  \alpha_k A_k &=& \Omega A_k+\sum_{n=1}^N a_{kn} g_n \label{b17}
  \\[2ex]
  \alpha_ka_{kn} &=& A_k g_n+\omega_n a_{kn} \,. \label{b18}
\end{eqnarray}
From (\ref{b18}) one obtains

\begin{equation}\label{b19}
    a_{kn}=\frac{g_n A_k}{\alpha_n-\omega_n}\,.
\end{equation}
If we carry (\ref{b19}) into (\ref{b18}), we obtain

\begin{equation}\label{b20}
    A_k\left(\alpha_k-\Omega-\sum_{n=1}^N\frac{g_n^2}{\alpha_k-\omega_n}\right)
    =0\,,
\end{equation}
which yields

\begin{equation}\label{b21}
\alpha_k-\Omega-\sum_{n=1}^N\frac{g_n^2}{\alpha_k-\omega_n}=0\,,
\hskip1cm k=0,1,\dots,N\,.
\end{equation}

 Equation (\ref{b21}) is transcendental and can be solved by numerical
 methods.

 As in any Friedrichs type model, we are interested in the reduced
 resolvent. Let $P$ be the orthogonal projection into the external
 channel, the one dimensional vector space spanned by the vector
 $|\Omega\rangle$, and let $Q:=I-P$, which is the projection into
 the $N$ dimensional vector space spanned by the $\{|\omega_n\rangle\}$,
 $n=1,2,\dots,N$. In particular, $P=|\Omega\rangle\langle\Omega|$.
The reduced resolvent is given by the operator

 \begin{equation}\label{b22}
    P\frac{1}{H-zI}P=\eta^{-1}(z)P\,.
\end{equation}
As in the ordinary Friedrichs model, $P$ projects into a one
dimensional subspace so that the identity in (\ref{b22}) holds. In
order to find the function $\eta^{-1}(z)$, we shall make use of a
formula that holds in this case and that we show in Appendix C
(see also \cite{E,CG}). This is (\ref{c13}) which reads (the
original form includes the coupling constant $\lambda$ that here
is taken equal to one):

\begin{equation}\label{b23}
P\frac{1}{H-zI}P=[G(z)]^{-1}P
\end{equation}
with

\begin{equation}\label{b24}
    G(z)= (z-H_0)\,P+  P\,V\,P
-\,P\,V\,Q\,\frac{1}{z-H_0}\,Q\,V\,P\,.
\end{equation}
Then, we get $G(z)|\Omega\rangle$ and this immediately gives
(\ref{b23}) as we shall see. Note that the potential $V$ is here
given by
$V=\sum_{n=1}^Ng_n(|\Omega\rangle\langle\omega_n|+|\omega_n\rangle\langle
    \Omega|)$ or in matrix form:

\begin{equation}\label{b25}
    V=\left(
\begin{array}{cc}
  0 & \begin{array}{ccc}
    g_1 & \cdots & g_N \\
  \end{array} \\
  \begin{array}{c}
    g_1 \\
    \vdots \\
    g_N \\
  \end{array} & {\bf 0} \\
\end{array}
\right)\,.
\end{equation}
This matrix is written in the basis $\{|\Omega\rangle,
|\omega_1\rangle, \dots,|\omega_N\rangle\}$. Note that

\begin{equation}\label{b26}
\left(
\begin{array}{cc}
  0 & \begin{array}{ccc}
    g_1 & \cdots & g_N \\
  \end{array} \\
  \begin{array}{c}
    g_1 \\
    \vdots \\
    g_N \\
  \end{array} & {\bf 0} \\
\end{array}
\right)\,\left(
\begin{array}{c}
  1 \\
  0\\
  \vdots \\
  0 \\
\end{array}
\right)=  \left(
\begin{array}{c}
  0 \\
  g_1\\
  \vdots \\
  g_N \\
\end{array}
\right)\,.
\end{equation}

Let us apply the last term in (\ref{b24}) to $|\Omega\rangle$.
This gives:

\begin{eqnarray}
  P\,V\,Q\,\frac{1}{z-H_0}\,Q\,V\,P \left(
\begin{array}{c}
  1 \\
  0\\
  \vdots \\
  0 \\
\end{array}
\right)= P\,V\,Q\,\frac{1}{z-H_0}\,Q\,V\, \left(
\begin{array}{c}
  1 \\
  0\\
  \vdots \\
  0 \\
\end{array}
\right)\nonumber\\[2ex]
=P\,V\,Q\,\frac{1}{z-H_0}\,Q\,\left(
\begin{array}{c}
  0 \\
  g_1\\
  \vdots \\
  g_N \\
\end{array}
\right)=P\,V\,Q\,\frac{1}{z-H_0}\,\left(
\begin{array}{c}
  0 \\
  g_1\\
  \vdots \\
  g_N \\
\end{array}
\right)\,.\label{b27}
\end{eqnarray}

We know that

$$
\left(
\begin{array}{c}
  0 \\
  g_1\\
  \vdots \\
  g_N \\
\end{array}
\right)=\sum_{n=1}^N g_n|\omega_n\rangle
$$
and that $H_0|\omega_n\rangle=\omega_n\,|\omega_n\rangle$. Then,
(\ref{b27}) equals to

\begin{eqnarray}
  P\,V\,Q\,\left(
\begin{array}{c}
  0 \\
  \frac{g_1}{z-\omega_1}\\
  \vdots \\
  \frac{g_N}{z-\omega_N} \\
\end{array}
\right)=P\,V\,\left(
\begin{array}{c}
  0 \\
  \frac{g_1}{z-\omega_1}\\
  \vdots \\
  \frac{g_N}{z-\omega_N} \\
\end{array}
\right) =P\,\left(
\begin{array}{c}
 \sum_{n=1}^N \frac{g_n^2}{z-\omega_n} \\
  0\\
  \vdots \\
  0 \\
\end{array}
\right)\nonumber\\[2ex]= \left(
\begin{array}{c}
 \sum_{n=1}^N \frac{g_n^2}{z-\omega_n} \\
  0\\
  \vdots \\
  0 \\
\end{array}
\right)\,.\label{b28}
\end{eqnarray}
The other terms are:

\begin{equation}\label{b29}
    PVP \,\left(
\begin{array}{c}
  1 \\
  0\\
  \vdots \\
  0 \\
\end{array}
\right)=\left(
\begin{array}{c}
  0 \\
  0\\
  \vdots \\
  0 \\
\end{array}
\right)\;; \hskip0.6cm (z-H_0)P= \left(
\begin{array}{c}
  1 \\
  0\\
  \vdots \\
  0 \\
\end{array}
\right)=\left(
\begin{array}{c}
  z-\Omega \\
  0\\
  \vdots \\
  0 \\
\end{array}
\right)\,.
\end{equation}
Therefore,

\begin{equation}\label{b30}
    G(z)|\Omega\rangle =z-\Omega-\sum_{n=1}^N \frac{g_n^2}{z-\omega_n}
\end{equation}
and then,

\begin{equation}\label{b31}
    [G(z)]^{-1}|\Omega\rangle=\left(z-\Omega-\sum_{n=1}^N
    \frac{g_n^2}{z-\omega_n}\right)^{-1}|\Omega\rangle\,.
\end{equation}
Thus, obviously,

\begin{equation}\label{b32}
    \eta(z)= z-\Omega-\sum_{n=1}^N
    \frac{g_n^2}{z-\omega_n}\,.
\end{equation}
Compare (\ref{b32}) with (\ref{15}). We know that the equation
$\eta(z)=0$ has real solutions given by $\alpha_0,\dots,\alpha_N$.

This model has been used to compare the results given by a system
with discrete spectrum and another similar with continuous spectrum.
The usual argument consists in taking the limit of the sum in
(\ref{b32}) in a way that the sum is transformed into an integral of
the type that it appears in (\ref{15}). Some results are discussed
in \cite{GGA,U,TOM1}.

\subsection{A generalized Friedrichs model with branch cuts.}

In this section, we want to present a generalized form of the
Friedrichs model in which, instead a single or multiple resonance
pole, a continuous set of singular points, a branch cut, appears.
This construction may appear rather artificial, although we intend
to write an explicit example of a continuous resonance and a
method to produce it. This example has already appeared some time
ago \cite{RG}.

In the simplest form of the Friedrichs model, we have a free
Hamiltonian with one eigenstate plus a continuum. Therefore, the
model has two channels, one infinite dimensional corresponding to
the continuous spectrum of $H_0$, the external channel, and the
other with dimension one corresponding to the stationary state of
$H_0$, the internal channel. The mutual interaction of these two
channels produces the resonance. This pattern is shown in any kind
of generalization of the Friedrichs model.

In our case,  both channels are infinite dimensional. The external
channel has also a continuous spectrum coinciding with the
positive semiaxis, although now this continuous spectrum has an
homogeneous infinite multiplicity (or degeneracy). The internal
channel is here a $l_2$ type infinite dimensional Hilbert space.
Both are believed to be minimal requirements to produce a
continuous resonance.

Thus, the Hilbert space is divided into two mutually orthogonal
parts:

\begin{equation}\label{g1}
    {\cal H}={\cal H}^{\rm in}\oplus{\cal H}^{\rm ex}\,.
\end{equation}
Needless to say that the terms ``in'' and ``ex'' stand for
internal and external channels respectively. The Hilbert space for
the internal channel is defined as

\begin{equation}\label{g2}
{\cal H}^{\rm ex}:=l_2({\mathbb Z})\,.
\end{equation}
Here, $l_2({\mathbb Z})$ is a space of sequences of complex
numbers labelled by the set of the integers with the following
condition: if $\{a_n\}$ is one such sequence, it must satisfy:

\begin{equation}\label{g3}
    \sum_{n=-\infty}^\infty|a_n|^2<\infty
\end{equation}
If  $A\equiv\{a_n\}$ and $B\equiv\{b_n\}$ are two sequences with
the property (\ref{g3}), we can define the following inner product
of them:

\begin{equation}\label{g4}
    \langle B|A\rangle = \sum_{n=-\infty}^\infty b_n^*a_n\,.
\end{equation}
With the scalar product (\ref{g4}), $l_2({\mathbb Z})$ is a
Hilbert space \cite{BN}.

In order to construct the Hilbert space corresponding to the
external channel, let $X_n$ be, for each integer value of $n$, a
copy of the Hilbert space $L^2({\mathbb R})$, so that $X_n\equiv
L^2({\mathbb R})$ for all $n=0,\pm1,\pm2, \dots$. Then, let us
define

\begin{equation}\label{g5}
    {\cal H}^{\rm ex}:=\bigoplus _{n-\infty}^\infty X_n\,.
\end{equation}
After this definition, the elements of ${\cal H}^{\rm ex}$ are
sequences $\{f_n(x)\}$ of square integrable functions on the real
line with the property:

\begin{equation}\label{g6}
    \sum_{n=-\infty}^\infty ||f_n(x)||^2<\infty\,.
\end{equation}
The norm in (\ref{g6}) is the well known $L^2({\mathbb R})$ norm.
The scalar product in ${\cal H}^{\rm ex}$ of two such sequences,
${\bf f}\equiv\{f_n(x)\}$ and ${\bf g}\equiv\{g_n(x)\}$ is given
by

\begin{equation}\label{g7}
    \langle {\bold f}|{\bf g}\rangle:= \sum_{n=-\infty}^\infty
    \int_{-\infty}^\infty f_n^*(x)g_n(x)\,dx\,.
\end{equation}

We see that the external channel is multiple valued.
Alternatively, it can be looked as a superposition of infinite
simply valued external channels \cite{RG}. It is just a matter of
language.

As for the Hilbert space, the free Hamiltonian can be split into
two contributions on the internal, $H^{\rm in}$, and on the
external channel, $H^{\rm ex}$, respectively, so that

\begin{equation}\label{g8}
    H_0=H^{\rm in}\oplus H^{\rm ex}\,.
\end{equation}

Let us define $H^{\rm in}$ first. If

\begin{equation}\label{g9}\psi:=\{\psi_n\}\in l_2({\mathbb Z})\equiv {\cal H}^{\rm in}\,,
\end{equation}
$H\psi$ will be the sequence in $l_2({\mathbb Z})$ such that its
$n-$th term is

\begin{equation}\label{g10}
    \{H^{\rm in}\psi\}:= a\psi_n-\psi_{n+1}-\psi_{n-1}\,,
\end{equation}
where $a$ is a fixed real number. We can prove that $H^{\rm in}$
is a bounded (continuous) operator on $l_2({\mathbb Z})$ and that
is has a doubly degenerate continuous spectrum given by the
interval $[a-2,a+2]$ \cite{RG}.

On the other hand,  $H^{\rm ex}$ must be an unbounded operator on
${\cal H}^{\rm ex}$, as we requiere that its spectrum is ${\mathbb
R}^+\equiv[0,\infty)$. For unbounded operators, we need to define
the domain of the operator (the space in which the operator acts)
plus its action on the domain. For $H^{\rm ex}$, the domain is
defined as

\begin{equation}\label{g11}
    {\cal D}(H^{ex})=\bigoplus_{n=-\infty}^\infty K_n\,,
\end{equation}
where, for all $n\in\{\dots ,k, \dots, -2, -1, 0,1, 2,\dots, k,
\dots\}\equiv{\mathbb Z}$, we have

\begin{equation}\label{g12}
    K_n\equiv W_2^2({\mathbb R})\,, \hskip1cm \forall\,n\in{\bf
    Z}\,.
\end{equation}

The Sobolev space $W_2^2({\mathbb R})$ is the space of twice
differentiable functions $f(x)$ from the real line $\mathbb R$
into the set of complex numbers $\mathbb C$ such that

\begin{equation}\label{g13}
    \int_{-\infty}^\infty \{|f(x)|^2+|f''(x)|^2\}\,dx<\infty\,.
\end{equation}
This Sobolev space $W_2^2({\mathbb R})$ with the square norm
 given by (\ref{g13}) is a Hilbert space. In addition $W_2^2({\mathbb
 R})$is dense in $L^2({\mathbb R})$ (being given a function
 $g(x)\in L^2({\mathbb R})$ and an $\epsilon>0$, there is always
 an $g(x)\in W_2^2$ such that $||f-g||<\epsilon$ in the $L^2$
 norm), which implies that the domain of $H^{\rm ex}$, ${\cal D}(H^{\rm ex})$
 is dense in ${\cal H}^{\rm ex}$.

 Once the domain of $H^{\rm ex}$ has been defined, we define the
 operator  $H^{\rm ex}$ by giving its action on an arbitrary  ${\bf f}\in{\cal D}(H^{\rm
 ex})$. If ${\bf f}=\{f_n(x)\}$, we have

 \begin{equation}\label{g14}
    \{(H^{\rm ex}{\bf f})_n\}:= \left\{-\frac{d^2}{dx^2}f_n(x)\right\}\,.
\end{equation}

It is a technical exercise to show that $H^{\rm ex}$ is
essentially self adjoint\footnote{This means that $H^{\rm in}$ has
one and only one self-adjoint extension and that therefore is a
well defined quantum observable.} and that it has an absolutely
continuous spectrum given by $[0,\infty)$ with an homogeneous
infinite degeneracy\footnote{These are technical concepts that we
do not want to discuss in the paper. See \cite{BAU,WE}.}.

As a consequence, the Hamiltonian $H_0=H^{\rm in}+H^{\rm ex}$ is
self adjoint with domain ${\cal H}^{\rm in}\oplus{\cal D}(H^{\rm
ex})$ and a spectrum given by

\begin{equation}\label{g15}
    \sigma(H_0)=\sigma(H^{\rm in})\cup \sigma(H^{\rm
    ex})=[a-2,a+2]\cup [0,\infty)\,,
\end{equation}
with infinite degeneracy.

This $H_0$ is the unperturbed Hamiltonian. We need to find a
perturbation such that the part of $\sigma(H^{\rm in})$ that
intersects $\sigma(H^{\rm ex})\equiv [0,\infty)$, which obviously
is $[0,a+2]$, is ``disolved'' and becomes a resonance in form of a
branch cut in a suitable analytical continuation of the resolvent.

To find this perturbation, we shall use a technique developed in
\cite{KP,KEP}. It has three steps: i.) Finding a dense domain
${\cal D}\subset {\cal D}(H_0)$ such that the restriction, $H_0'$,
of $H_0$ into $\cal D$ admits infinite self-adjoint
extensions\footnote{In the mathematical language, $H_0'$ must have
equal deficiency indices different from zero. See Appendix B.};
ii.) We find the adjoint ${H_0'}^\dagger$ of $H_0'$. This adjoint
does exist as $H_0'$ is densely defined (its domain is dense in
$\cal H$) and this adjoint is an extension of $H_0'$ (for any $f$
in the domain of ${H_0'}^\dagger$, $H_0'f={H_0'}^\dagger f$; iii.)
Providing suitably boundary conditions on the domain of
${H_0'}^\dagger$, we find an extension of $H_0'$ of the form
$H_0+\gamma V$ (which is also a restriction of ${H_0'}^\dagger$,
i.e., it acts as ${H_0'}^\dagger$ on a domain smaller than the
domain of ${H_0'}^\dagger$). This operator $H_\gamma:= H_0+\gamma
V$ is self adjoint and $V$ is the desired potential. Note that
$H_\gamma$ has the typical form of the total Hamiltonian in a
Friedrichs like model.

In order to fix $H_0'$ it is enough fixing its domain ${\cal
D}(H_0')$, since $H_0'$ is a restriction of $H_0$ (and therefore
$H_0{\bf f}=H_0'{\bf f}$, for all ${\bf f}\in {\cal D}(H_0')$).
This domain is giving by

\begin{equation}\label{g16}
{\cal D}(H_0'):= {\cal H}^{\rm ex}\oplus D\,; \hskip1cm
D:=\bigoplus_{n=-\infty}^\infty D_n\,,
\end{equation}
where $D_n:=C_0^\infty ({\mathbb R})$, for all $n\in\mathbb Z$, so
that all the spaces $D_n$ are copies of the same $C_0^\infty
({\mathbb R})$. This $C_0^\infty ({\mathbb R})$ is the vector
space of all infinitely differentiable functions that vanish on a
certain neighborhood of the origin, which does not need to be the
same for all functions in $C_0^\infty ({\mathbb R})$. One can
prove that ${\cal D}(H_0')$ is dense in the total Hilbert space
$\cal H$ and therefore, the adjoint ${H_0'}^{\dagger}$ of $H_0'$
is well defined, although $H_0'$ is not essentially self adjoint
 because it has both deficiency indices equal to $\infty$. In order to
prove this latter statement, it seems convenient to finding
${H'_0}^\dagger$ first. Its domain is given by

\begin{equation}\label{g17}
    {\cal D}({H'_0}^\dagger)={\cal H}^{\rm in}\oplus{\cal C}\,;
    \hskip1cm {\cal C}:= \bigoplus_{n=-\infty}^\infty {\cal C}_n
\end{equation}
with

\begin{equation}\label{g18}
    {\cal C}_n:= W_2^2({\mathbb R}/\{0\})\,; \hskip1cm \forall\,
    n\in\mathbb Z\,.
\end{equation}

The functions in the space $W_2^2({\mathbb R}/\{0\})$ are like the
functions in $W_2^2$, now allowing our functions and their first
derivatives to have a finite jump in the origin. The norm square
in $W_2^2({\mathbb R}/\{0\})$ is exactly as in (\ref{g13}) and
obviously $W_2^2\subset W_2^2({\mathbb R}/\{0\})$. Thus, ${\bf f}$
is in $\cal C$ if and only if ${\bf f}$ is a sequence of the form
$\{f_n(x)\}$ with $f_n(x)\in\cal C$ for all $n\in\mathbb Z$.

Once we know the domain of ${H_0'}^\dagger$, we can define $H_0'$
itself. For any $\psi\in{\cal H}^{\rm in}$, ${H_0'}^\dagger\psi=
H^{\rm in}\psi=H_0\psi$. For any ${\bf f}\equiv\{f_n(x)\}\in\cal
C$,

\begin{equation}\label{g19}
    \{({H_0'}^\dagger{\bf f})_n\}:=\left\{-\frac{d^2 f_n(x)}{dx^2}\right\}\,,
\end{equation}
exactly as in (\ref{g14}). Observe that ${H_0'}^\dagger$ is an
extension of $H_0$. In addition, one can show that there exists a
relation between the domain of $H_0'$ and the domain of its
adjoint, which is given by

\begin{equation}\label{g20}
    {\cal D}({H_0'}^\dagger)={\cal D}(H_0')\dot +{\cal N}_+\dot
    +{\cal N}_-\,,
\end{equation}
where: i.) $\dot +$ means direct algebraic sum, in general
nonorthogonal and ii.) ${\cal N}_\pm$ are the deficiency indices
of $H_0'$. Both are infinite dimensional and thus equal and
therefore, $H_0'$ admits an infinite number of self adjoint
extensions. The form of ${\cal N}_\pm$ is obtained and discussed
in \cite{RG}.

Note that an arbitrary vector $\bf\Phi$ in ${\cal H}={\cal H}^{\rm
in}\oplus{\cal H}^{\rm ex}$ can be written in a column vector form
as:

\begin{equation}\label{g21}
    {\bf\Phi}=\left(
\begin{array}{c}
  \varphi \\[2ex]
  {\bf f} \\
\end{array}
\right)=\left(
\begin{array}{c}
  \{\varphi_n\} \\[2ex]
  \{f_n(x)\} \\
\end{array}
\right)\,.
\end{equation}

Now let us assume that ${\bf\Phi}\in {\cal D}({H_0'}^\dagger)$. In
this case, $\varphi$ is arbitrary in ${\cal H}^{\rm in}\equiv
l_2({\mathbb Z})$ and ${\bf f}\in {\cal C}$. As in the case of the
ordinary Friedrichs model, the perturbation will put in interaction
the external with the internal channels. As in the formalism given
in \cite{KP,KEP}, we need to impose certain type of boundary
conditions\footnote{As a matter of fact, in order to obtain the self
adjoint extensions of a symmetric non self-adjoint operator with
equal deficiency indices, we usually do it by imposing boundary
conditions \cite{RSII}. What is characteristic in this theory is
that the extension gives a potential (in general singular) to be
added to the original unperturbed Hamiltonian $H_0$} to obtain
$H_\gamma$. These boundary conditions, for each $n\in\mathbb Z$, are
the following:

\begin{eqnarray}
  f_n(0+)-f_n(0-) &=& -\gamma\varphi_n \label{g22}\\[2ex]
  f'_n(0+) &=& f'_n(0-)\,.\label{g23}
\end{eqnarray}
In (\ref{g22}) and (\ref{g23}), $f_n(0\pm)$ and $f'_n(0\pm)$ are
the limits to the left (with $+$) and to the right (with $-$) of
the functions $f_n(x)$ and its first derivative $f'_n(x)$ for all
$n\in\mathbb Z$. These limits must exist, but they could be
different as each $f_n(x)\in W_2^2({\mathbb R}/\{0\})$.

For each real $\gamma$, these boundary conditions define one
self-adjoint extension of $H_0'$ that we call $H_\gamma$ with
domain:

\begin{equation}\label{g24}
    {\cal D}({H_\gamma):= \{\bf\Phi}\subset {\cal
    D}({H_0'}^\dagger)\; {\rm fulfilling \; (\ref{g22})\; and\;
    (\ref{g23})}\}\,.
\end{equation}

On ${\bf\Phi}\in {\cal D}(H_\gamma)$, the action of $H_\gamma$ is
given by

\begin{equation}\label{g25}
    H_\gamma\,\left(
\begin{array}{c}
  \varphi \\[2ex]
  {\bf f} \\
\end{array}
\right)= \left(
\begin{array}{c}
  \widetilde\varphi \\[2ex]
  \widetilde{\bf f} \\
\end{array}
\right)\,,
\end{equation}
with

\begin{eqnarray}
  \widetilde\varphi_n &=& (H^{\rm in}\varphi)_n+\gamma f'_n(0)
  \label{g26}
  \\[2ex]
  \widetilde f_n(x) &=& -f_n''(x)\,,\label{g27}
\end{eqnarray}
for each $n\in\mathbb Z$. Here, $\widetilde\varphi_n$ and
$\widetilde f_n(x)$ are the $n-$th component of $\widetilde
\varphi$ and $\widetilde{\bf f}$ respectively.

Each of the $H_\gamma$ is certainly self-adjoint \cite{RG}. To
give meaning to $H_\gamma$ as $H_0+\gamma V$, let us consider
${\bf\Phi}\in{\cal D}(H_\gamma)$ and ${\bf\Xi}\in{\cal D}(H_0)$
and the following difference:

\begin{equation}\label{g28}
    \langle H_\gamma{\bf\Phi},{\bf\Xi}\rangle-
    \langle{\bf\Phi},H_0\bf\Xi\rangle\,.
\end{equation}
Obviously, as $H_\gamma$ is self adjoint, $H_\gamma$ can be
written in the form $H_0+\gamma V$ if and only if (\ref{g28}) is
equal to

\begin{equation}\label{g29}
    \gamma\,\langle V{\bf\Phi},{\bf\Xi}\rangle\,.
\end{equation}
If

$$
{\bf\Xi}=\left(
\begin{array}{c}
  \xi \\[2ex]
  {\bf g} \\
\end{array}
\right)=\left(
\begin{array}{c}
  \{\xi_n\} \\[2ex]
  \{g_n(x)\} \\
\end{array}
\right)\,,
$$
where $\xi\in l_2({\mathbb Z})$ and ${\bf g}\in D$, one gets
\cite{RG}

\begin{equation}
\label{g30}
  \langle H_\gamma{\bf\Phi},{\bf\Xi}\rangle-
    \langle{\bf\Phi},H_0{\bf\Xi}\rangle = \gamma\sum_{n=-\infty}^\infty
    {f'_n}^*(0)\,\xi_n +\gamma \sum_{n=-\infty}^\infty \varphi_n^*\,
    \xi'_n(0)\,.
\end{equation}
Note that the boundary condition (\ref{g27}) gives a unique value
for each $f_n'(0)$ and that $\xi_n(0)$ for each $n\in{\mathbb Z}$
is well defined because $\xi_n(x)\in W_2^2$. We can obtain an
explicit for of $V$ using the following matrix kernel form: the
difference $\langle H_\gamma{\bf\Phi},{\bf\Xi}\rangle-
    \langle{\bf\Phi},H_0\bf\Xi\rangle$ is a sum in $M$ and $n$ of
    terms of the form:

\begin{equation}\label{g31}
\left(\varphi_m^*,f^*_m(y) \right) \left(
\begin{array}{cc}
  V_{mn} & V_{mn}(y) \\[2ex]
  V_{mn}(x) & V_{mn}(x,y) \\
\end{array}
\right)\left(
\begin{array}{c}
  \xi_n \\[2ex]
  g_n(x) \\
\end{array}
\right)={f_n'}^*(0)\,\xi_n+\varphi_n^*\,g'_n(0)\,.
\end{equation}
This yields:

\begin{equation}\label{g32}
\left(
\begin{array}{cc}
  V_{mn} & V_{mn}(y) \\[2ex]
  V_{mn}(x) & V_{mn}(x,y) \\
\end{array}
\right)= \left(
\begin{array}{cc}
  0 & \delta_{mn}\,\delta'_{mn}(y) \\[2ex]
  \delta_{mn}\,\delta'_{mn}(x) & 0 \\
\end{array}
\right)\,.
\end{equation}

Thus, the construction of the generalized Friedrichs model is
finished and it remains to find its resonance behavior. This will
be done in the next subsection.

\subsubsection{The resonance as a branch cut.}

The procedure to find the resonance branch cut follows exactly the
same procedure as in the case of the simplest Friedrichs model. In
the latter, we studied the singularities of the analytic
continuation of the complex function given by (\ref{13},\ref{14}),
which expectation values of the operator reduced resolvent
(\ref{12}). In this form of the Friedrichs model, the internal
channel is the one dimensional Hilbert space spanned by the vector
$|1\rangle$. Thus, the generalization of (\ref{12}) to our case is
clear: here $P$ must be the projection into the internal channel
${\cal H}^{\rm in}$. Then, the reduced resolvent is now given by
$P(H_\gamma-zI)^{-1}P$.

To carry out our analysis, we shall use the kernel form of the
resolvent as follows:

\begin{equation}\label{g33}
(H_\gamma-zI)^{-1}\left(
\begin{array}{c}
  \varphi \\[2ex]
  {\bf f} \\
\end{array}
\right)=\left(
\begin{array}{c}
  \xi \\[2ex]
  {\bf g}\\
\end{array}
\right)=\left(
\begin{array}{c}
  \{\xi_n\}_{n=-\infty}^\infty \\[2ex]
  \{g_n(x)\}_{n=-\infty}^\infty \\
\end{array}
\right)\,.
\end{equation}

\medskip
If $\varphi=\{\varphi_n\}_{n=-\infty}^\infty\in l_2({\mathbb
Z})\equiv {\cal H}^{\rm in}$ and ${\bf
f}=\{f_n(x)\}_{n=-\infty}^\infty\in{\cal H}^{\rm out}$, we write:

\begin{eqnarray}
  \xi_n &=& \sum_{m=-\infty}^\infty R_{mn}^{(1)} \,\varphi_n+\sum_{m=-\infty}^\infty \int_{-\infty}^\infty
   R_{mn}^{(2)}(y)\,f_m(y)\,dy\label{g34}\\[2ex]
  g_n(x) &=& \sum_{m=-\infty}^\infty R_{mn}^{(3)} (x)\,\varphi_n+\sum_{m=-\infty}^\infty \int_{-\infty}^\infty
   R_{mn}^{(4)}(x,y)\,f_m(y)\,dy\,.\label{g35}
\end{eqnarray}

The kernels $ R_{mn}^{(1)}$, $ R_{mn}^{(2)}(y)$, $
R_{mn}^{(3)}(x)$ and $ R_{mn}^{(4)}(x,y)$ all depend on the
complex variable $z$, although we must stress that the variables
of these kernels, here denoted as $x$ and $y$, do no have anything
to do with the real and imaginary part of $z$.

For every $k\in{\mathbb Z}$, we define the following sequences:

\begin{eqnarray}
  R_k :&=& \{R_{nk}^{(1)}\}_{n=-\infty}^\infty \label{g36}\\[2ex]
 {\bf R}_k(x) : &=& \{R_{nk}^{(3)}(x)\}_{n=-\infty}^\infty \label{g37}
\end{eqnarray}
and the following vector:

\begin{equation}\label{g38}
    {\bf \Xi}_k:= \left(
\begin{array}{c}
  R_k \\[2ex]
  {\bf R}_k(x) \\
\end{array}
\right)\,.
\end{equation}

It is not difficult to show that, for each $k\in\mathbb Z$, ${\bf
\Xi}_k$ is in the domain of $H_\gamma$ \cite{RG}. For our
purposes, it is sufficient to consider $k=0$ and this gives the
vector ${\bf\Xi}_0$. For simplicity, we can write

\begin{equation}\label{g39}
{\bf\Xi}:={\bf\Xi}_0= \left(
\begin{array}{c}
  R_0 \\[2ex]
  {\bf R}_0(x) \\
\end{array}
\right)\equiv \left(
\begin{array}{c}
  R \\[2ex]
  {\bf R}(x) \\
\end{array}
\right)
\end{equation}
 Let $\delta$ be the sequence
$\{\delta_{n0}\}_{n=-\infty}^\infty$, where $\delta_{n0}$ is the
Kronecker delta. This sequence has all its entries equal to zero
except for $\delta_{00}=1$. Then, we can prove that \cite{RG}

\begin{equation}\label{g40}
    (H_\gamma-zI){\bf\Xi}= \left(
\begin{array}{c}
  \delta \\[2ex]
  {\bf 0} \\
\end{array}
\right)\,,
\end{equation}
which yields:

\begin{eqnarray}
  (H^{\rm in}R)_n+\gamma R'_n(0)-zR_n &=& \delta_{n0}\,, \hskip0.6cm \forall\,n\in\mathbb Z\,,\label{g41}
  \\[2ex]
  -R_n''(x)-zR_n(x) &=& 0 \,, \hskip0.9cm \forall\,n\in\mathbb
  Z\,.\label{g42}
\end{eqnarray}

We first solve the differential equation given by (\ref{g42}) and
find a solution in $W_2^2({\mathbb R}/\{0\})$ with the boundary
conditions (\ref{g22},\ref{g23}). This solution, with $k=\sqrt z$,
is

\begin{equation}\label{g43}
    R_n(x)=\left\{\begin{array}{c}
      C_n\, e^{ikx}\;; \hskip1cm x>0 \\[2ex]
      C_n'\,e^{-ikx} \;; \hskip1cm x<0\\
    \end{array}
    \right.\,.
\end{equation}
The boundary condition (\ref{g23}) reads $R_n'(0+)=R'_n(0-)$ in
this case. It implies that $C_n=C_n'$. Condition (\ref{g24}) is
here $2C_n=-\gamma R_n$. This gives:

\begin{equation}\label{g44}
    R'_n(0)=ikC_n=-\frac{i\gamma}{2}R_n\sqrt z\,.
\end{equation}
If we carry (\ref{g44}) into (\ref{g41}), we have that

\begin{equation}\label{g45}
    [(H^{\rm in}-z){\bf R}]_n-\frac i2 \gamma^2R_n\sqrt
    z=\delta_{n0}\,,
\end{equation}
which after the definition of $H^{\rm in}$ given in (\ref{g10}),
one gets

\begin{equation}\label{g46}
    \left\{ a-z-\frac i2\gamma^2\sqrt z\right\} R_n
    -R_{n-1}-R_{n+1}=\delta_{n0}\,.
\end{equation}

The identity (\ref{g46}) is valid for any integer $n$ and has the
following solution:

\begin{equation}\label{g47}
    R_n=\frac{e^{i\nu |n|}}{2i\sin\nu}\,,
\end{equation}
where

\begin{equation}\label{g48}
    2\cos\nu= a-z-\frac i2\gamma^2\sqrt z\,.
\end{equation}
Observe that the equation (\ref{g48}) has complex solutions in the
indeterminate $\nu$. The dependence on the variable $z$ is
contained in $\nu=\nu(z)$.

Once we have obtained the solution for $k=0$ in (\ref{g38}), it is
immediate to obtain the solution for arbitrary $k$ and one gets

\begin{equation}\label{g49}
    R_{nk}=\frac{e^{i\nu |n-k|}}{2i\sin\nu}\,.
\end{equation}

Let us go back to the reduced resolvent written in the operator
form $P(H_\gamma-z)^{-1}P$. Since $P$ is the projection onto the
space ${\cal H}^{\rm in}$, the operator $P(H_\gamma-z)^{-1}P$ acts
only on ${\cal H}^{\rm in}$. Then, if
$\varphi:=\{\varphi_n\}_{n=-\infty}^\infty\in{\cal H}^{\rm in}$,
we have the following expression for the $n-$th component of
$P(H_\gamma-z)^{-1}P\varphi$, which is in ${\cal H}^{\rm in}$ if
$z\notin\sigma(H_\gamma)$:

\begin{equation}\label{g50}
    [P(H_\gamma-z)^{-1}P\varphi]_n =\sum_{m=-\infty}^\infty
    R_{mn}(z)
    \varphi_m\,.
\end{equation}
Let us call $\delta_k$ to the sequence
$\{\delta_{mk}\}_{m=-\infty}^\infty$, where $\delta_{mk}$ is the
Kronecker delta. This sequence has all its terms equal to zero
except the $k-$th which equals to one. In this case, (\ref{g50})
should be written as:

\begin{equation}\label{g51}
[P(H_\gamma-z)^{-1}P\delta_k]_n =\sum_{m=-\infty}^\infty
    R_{mn}(z)
    \delta_{mk}=R_{nk}(z)\,,
\end{equation}
where $R_{nk}(z)$ is as in (\ref{g49}). Formula (\ref{g51}) can be
written in compact form as:

\begin{equation}\label{g52}
P(H_\gamma-z)^{-1}P\delta_k={\bf R}_k\;; \hskip1.5cm \forall
\,k\in\mathbb Z\,.
\end{equation}

The set $\{\delta_k\}$, $k\in\mathbb Z$ of sequences in ${\cal
H}^{\rm in}\equiv l_2({\mathbb Z})$ form a orthonormal basis in
$l_2({\mathbb Z})$ \cite{BN}. Then, as $P(H_\gamma-z)^{-1}P$ is a
bounded operator on ${\cal H}^{\rm in}$ if
$z\notin\sigma(H_\gamma)$, the reduced resolvent
$P(H_\gamma-z)^{-1}P$ is completely determined by its action on
the sequences $\{\delta_k\}$, for all $k\in\mathbb Z$. For $k=0$,
this formula gave (\ref{g45}). Same manipulations give for any
$k\in\mathbb Z$ the following equation:

\begin{equation}\label{g53}
    \left(H^{\rm in}-z-\frac i2\gamma^2\sqrt z\right) {\bf
    R}_k=\delta_k \;; \hskip1.5cm \forall
\,k\in\mathbb Z\,,
\end{equation}
or equivalently:

\begin{equation}\label{g54}
    {\bf R}_k= \left(H^{\rm in}-z-\frac i2\gamma^2\sqrt
    z\right)^{-1}\delta_k \;; \hskip1.5cm \forall
\,k\in\mathbb Z\,.
\end{equation}
If we combine (\ref{g54}) with (\ref{g52}), we find that

\begin{equation}\label{g55}
P(H_\gamma-z)^{-1}P= \left(H^{\rm in}-z-\frac i2\gamma^2\sqrt
    z\right)^{-1}\,.
\end{equation}
We know that $(H^{\rm in}-zI)^{-1}$ ia analytic at all points
except for those in the spectrum of $H^{\rm in}$ which coincides
with $[a-2,a+2]$. Then,if

\begin{equation}\label{g56}
    g(z):= z+\frac i2\gamma^2\sqrt z\,,
\end{equation}
the operator function given by $P(H_\gamma-z)^{-1}P$ is analytic
except for

\begin{equation}\label{g57}
    g^{-1}[a-2,a+2]\,.
\end{equation}

Therefore, the points for which the partial resolvent
$P(H_\gamma-z)^{-1}P$ is not analytic are given by the equation

\begin{equation}\label{g58}
    t= z+\frac i2\gamma^2\sqrt z\;; \hskip1.5cm
    \forall\,t\in[a-2,a+2]\,.
\end{equation}

To study formula (\ref{g58}), it is better to come back to the
momentum representation through the transformation $k=\sqrt z$.
Then, (\ref{g58}) can be written as

\begin{equation}\label{g59}
    t=k^2+\frac i2\gamma^2 k \;; \hskip1.5cm
    \forall\,t\in[a-2,a+2]\,
\end{equation}
which gives

\begin{equation}\label{g60}
    k_\pm (t)=-\frac{i\gamma^2}{4}\pm \sqrt{t-\frac{\gamma^4}{16}}
    \;; \hskip1.5cm
    \forall\,t\in[a-2,a+2]\,.
\end{equation}

If the spectrum of $H^{\rm in}$, $\sigma(H^{\rm in}=[a-2,a+2]$,
has a negative part given by $[t_1,t_2]$, where $t_1=a-2$ and
$t_2=a+2$ if $a+2<0$ and $0$ otherwise, the possible values of
$z=k^2$ are real, as $k$ is purely imaginary, and belong to the
interval $[k_-^2(t_1),k_+^2(t_2)]$. Also, those $z$ belonging to
the intervals $[k_-^2(t_1),k_+^2(t_1)]$ and
$[k_-^2(t_2),k_+^2(t_2)]$ have multiplicity one and those to
m$[k_-^2(t_1),k_+^2(t_2)]$ have multiplicity equal to two.

The values of $t$ in the interval $[t_0,a+2]$, where
$t_0=\max\{\gamma^4/16, a-2\}$ give a pair of branch cuts, as we
can deduce from (\ref{g60}). These are {\it resonant branch cuts}
and this expression can be understood in the following way: The
internal block of the resolvent $(H^{\rm in}-zI)^{-1}$ is an
analytic bounded operator function on the complex plane in which
the spectral cut at $[a-2,a+2]$ is removed. The total resolvent
has no bounded operator limit on the cut \cite{RG} and, hence, it
cannot be extended through the cut in general. However, the
internal block of the resolvent has a bounded limit on this cut as
Im$\, z>0$ and can be continued from above to below. This
continuation has singularities. In this model, one finds instead
the usual resonance poles, branch cuts. On these cuts, the
internal block of the resolvent becomes unbounded. These cuts are
associated, by analogy to the standard Friedrichs model, to
resonance phenomena.

\section*{Acknowledgements.}

Partial financial support is indebt to the Spanish  Ministry of
Science (Project MTM2009-10751), and to the Russian Science
Foundation (Grants 10-01-00300 and 09-01-12123). Part of this work
was done while on of us (M. Gadella) was on leave with a sabbatical
licence bestowed by the University of Valladolid.

\appendix

\section{Rigged Hilbert Spaces}

Following the spirit of \cite{3}, we are going to present the
Gamow vectors for the Friedrichs model as functionals in a rigged
Hilbert space (RHS). By RHS we mean the following:

Let $\cal H$ be an infinite dimensional Hilbert space. Let
${\bf\Phi}$ be a subspace of $\cal H$ with the following
properties:

1.- The subspace $\bf\Phi$ is dense in $\cal H$. This means that
for any vector $\psi\in\cal H$ and any neighborhood $V$ of $\psi$
there exists $\varphi\in\bf\Phi$ in $V$. In other words, any
vector in $\cal H$ can be approximated, with arbitrary accuracy,
by another vector in $\bf\Phi$ with the topology on $\cal H$.

2.- The subspace $\bf\Phi$ has its own topology, $\tau_{\bf\Phi}$,
(compatible with the structure of vector space), which is finer
than the topology that $\bf\Phi$ inherits as subspace of $\cal H$
(we shall denote by $\tau_{\cal H}$ the Hilbert space topology on
$\cal H$). This means that with $\tau_{\bf\Phi}$, $\bf\Phi$ has
more open sets than with $\tau_{\cal H}$. Equivalently, any
convergent sequence by $\tau_{\bf\Phi}$ is convergent under
$\tau_{\cal H}$, but the converse is not true.

3.- Once the space $\bf\Phi$ has been determined, let us consider
the set of mappings $F$ from $\bf\Phi$ onto the field of complex
numbers $\mathbb C$, with the following properties:

i.) Each of the $F:{\bf\Phi}\longmapsto \mathbb C$ is antilinear:
For any $\psi,\varphi\in{\bf\Phi}$ and any pair of complex numbers
$\alpha,\beta$, we have:

\begin{equation}
F(\alpha\psi+\beta\varphi)=\alpha^*F(\psi)+\beta^*F(\varphi)\,,\label{a1}
\end{equation}
where the star denotes complex conjugation.

ii.) The mapping $F$ is continuous. In particular, this means that
if $\{\varphi_n\}$ is a sequence of vectors in $\bf\Phi$ which
converges with $\tau_{\bf\Phi}$ to the vector
$\varphi\in{\bf\Phi}$, then:

$$
F(\varphi_n)\longmapsto F(\varphi)\,,
$$
in the field of complex numbers $\mathbb C$\footnote{This
conditions is necessary although not sufficient for the continuity
of $F$. It may happen that $\bf\Phi$ be a nonmetrizable space,
i.e., a space in which the topology cannot be given by a distance.
Also this condition suggest that, at least, $\bf\Phi$ should be
sequentially complete in the sense that any Cauchy sequence have a
limit in $\bf\Phi$.}.

iii.) We shall call {\it functionals} to the set of mappings with
the above properties. The set of functionals is a vector space,
provided that we define its sum and product by complex numbers as
follows:

$$
(\alpha F+\beta G)(\psi):=\alpha F(\psi)+\beta G(\psi)\,,
$$
for all $\psi\in\bf\Phi$. Here, $F$ and $G$ are functionals and
$\alpha$ and $\beta$ arbitrary complex numbers. The vector space
of functionals is denoted as ${\bf\Phi}^\times$.

iv.) For convenience, the space ${\bf\Phi}^\times$ is usually
endowed with a topology (weak, strong, Mackey), a fact that we
shall not discuss her. See \cite{6} and references quoted therein.

v.) Let $\varphi$ be an arbitrary vector in the Hilbert space
$\cal H$. Then, define the mapping $F_\varphi$ as follows:

\begin{equation}
F_\varphi(\psi):=\langle\psi|\varphi\rangle\,,\label{a2}
\end{equation}
where $\psi$ is an arbitrary vector in $\bf\Phi$ and
$\langle-|-\rangle$ denotes the scalar product in $\cal H$. One
can show that $F_\varphi$ is a functional in ${\bf\Phi}^\times$
(the antilinearity is in fact straightforward). Furthermore, if
$\varphi\ne\phi$, then $F_\varphi\ne F_\phi$ as a consequence of
the Riesz theorem \cite{BN,BS}. Then, we shall identify the
functional $F_\varphi$ and the vector $\varphi$ for notational
convenience.

Observe that, after this identification, the space of functionals
${\bf\Phi}^\times$ contains the Hilbert space $\cal H$, so that we
have the following relation between these three spaces:

\begin{equation}
{\bf\Phi}\subset{\cal H}\subset {\bf\Phi}^\times\,.\label{a3}
\end{equation}
This triplet of spaces is called a {\it rigged Hilbert space} or a
{\it Gelfand triplet}. The topology on ${\bf\Phi}^\times$,
$\tau_\times$, is chosen so that, on $\cal H$, $\tau_{\cal H}$ is
finer than $\tau_\times$. The space $\bf\Phi$ is often called the
space of {\it test vectors} or {\it test functions} (if $\bf\Phi$
is a space of functions).

So far, the definition of RHS. It is interesting to comment
briefly some of the properties of operators on RHS.

First of all a question concerning notation. Following the usual
habit of physicists, we shall use the Dirac notation and write for
any $F\in{\bf\Phi}^\times$ ad any $\psi\in\bf\Phi$:

\begin{equation}
\langle F|\psi\rangle=F(\psi)\,.\label{a4}
\end{equation}

Note that this notation is consistent with (\ref{a2}).

Let $H$ be a densely defined operator on $\cal H$ (and therefore
with a well defined adjoint operator $H^\dagger$) such that

i.) The space $\bf\Phi$ reduces $H$, i.e., for every
$\psi\in\bf\Phi$, $H\psi\in\bf\Phi$ (we can describe this property
as $H{\bf\Phi}\subset\bf\Phi$).

ii.) The operator $H$ is continuous on $\bf\Phi$ with
$\tau_{\bf\Phi}$.

Then, $H$ can be extended as a (weakly continuous) linear operator
into the dual ${\bf\Phi}^\times$ by means of the {\it duality
formula}:

\begin{equation}
\langle F|H\psi\rangle =\langle HF|\psi\rangle\,, \label{a5}
\end{equation}
where $F$ and $\psi$ are arbitrary in ${\bf\Phi}^\times$ and
$\bf\Phi$ respectively\footnote{Some authors distinguish this
extension with the notation $H^\times$ \cite{B}}.

It is important to remark that the eigenvalue problem can be posed
for the extension of $H$ on the antidual ${\bf\Phi}^\times$. In
particular, Gamow vectors will be, in this context, eigenvectors
of the extended Hamiltonian, to a certain antidual, with
eigenvalues given by the resonance poles.

\medskip

It is also interesting to recall here a Theorem due to Gelfand and
Maurin on the completeness of generalized eigenvectors of a self
adjoint operator \cite{GIV,MAU}. This result gives a rigorous
meaning to the Dirac generalized expansion of a self adjoint
operator \cite{DI}. It states that being given an arbitrary self
adjoint operator on an infinite dimensional Hilbert space, a
rigged Hilbert space can be found in which this generalized
expansion exists. A more precise form of this theorem can be
presented as follows:

Let $\cal H$ be an infinite dimensional Hilbert space and $A$ a
self adjoint operator on $\cal H$. Without loss of generality, we
can assume that the spectrum of $A$ is purely continuous. Then,
there exists a RHS ${\bf\Phi}\subset{\cal H}\subset{\bf
\Phi}^\times$ such that

\smallskip
i.) The space $\bf\Phi$ reduces $A$. This means that for any
vector $\varphi\in\bf\Phi$, $A\varphi\in{\bf\Phi}$. Equivalently,
$A{\bf\Phi}\subset{\bf\Phi}$. The operator $A$ is continuous on
$\bf\Phi$ and it can be extended, using the duality formula
(\ref{a5}) into the antidual ${\bf\Phi}^\times$. The extension to
the antidual of $A$, that we shall also call $A$, is continuous
with respect to the (weak) topology on ${\bf\Phi}^\times$.

\smallskip
ii.) There exists a measure $\mu(\lambda)$ supported on the
spectrum of $A$, $\sigma(A)$, (it is zero outside $\sigma(A)$),
such that for almost all $\lambda\in\sigma(A)$ (with respect to
the measure $\mu(\lambda)$), there exists a functional
$|\lambda\rangle\in{\bf\Phi}^\times$ such that

\begin{equation}\label{a14}
A|\lambda\rangle=\lambda\,|\lambda\rangle\,.
\end{equation}
We can apply this $|\lambda\rangle$ to any $\varphi\in\bf\Phi$.
When $\lambda$ runs out the spectrum of $A$, $\sigma(A)$, we
obtain a function $\langle\varphi|\lambda\rangle \in
L^2[\sigma(A),d\mu(\lambda)]$. The mapping $U:{\bf
\Phi}\longmapsto L^2[\sigma(A),d\mu(\lambda)]$ given by
$U\varphi=\langle\varphi|\lambda\rangle$ can be extended to be
unitary from $\cal H$ onto $L^2[\sigma(A),d\mu(\lambda)]$.

\smallskip

iii.) For any pair of vectors $\varphi,\psi\in{\bf \Phi}$, the
following completeness relation holds:

\begin{equation}\label{a15}
(\varphi,\psi)=\int_{\sigma(A)}
\langle\varphi|\lambda\rangle\langle\lambda|\psi\rangle\,d\mu(\lambda)\,,
\end{equation}
where $\langle\lambda|\psi\rangle=\langle\psi|\lambda\rangle^*$
and the star denotes complex conjugation.

Furthermore, if $f(\lambda)$ is a (measurable) bounded complex
function on $\sigma(A)$, we have in addition that for any
$\varphi,\psi\in{\bf\Phi}$

\begin{equation}\label{a16}
(\varphi,f(A)\,\psi)=\int_{\sigma(A)} f(\lambda)\,
\langle\varphi|\lambda\rangle\langle\lambda|\psi\rangle\,d\mu(\lambda)\,.
\end{equation}

Formula (\ref{a16}) provides the functional calculus for $A$. It
can be also extended to the case that $f(\lambda)$ be a
polynomial. In this case, $f(A)$ is the corresponding polynomial
in the variable $A$.

\subsection{Examples of Rigged Hilbert spaces.}

There are several examples of RHS that are useful in Quantum
Mechanics as well as in chaotic systems \cite{AG}. To begin with,
we introduce the simplest and more fundamental of all, the
Schwartz space.

The Schwartz space $\cal S$ is the vector space of all
continuously indefinitely differentiable (derivable to all orders)
functions from the real line $\mathbb R$ into the set of complex
numbers $\mathbb C$ with the following property: If $f(x)\in\cal
S$, then:

$$
\lim_{x\mapsto\pm\infty}\left|\frac{d^n}{dx^n}\,x^mf(x)\right|=0\,,
$$
for all $m,n=0,1,2,\dots$.

The space $S$ has the following properties:

\smallskip
1.- Each function $f(x)\in \cal S$ is square integrable. In
addition, $\cal S$ is a dense subspace of $L^2({\mathbb R})$.

2.- $\cal S$ has its own metrizable topology inhereted from
$L^2({\mathbb R})$.

3.- The antidual space of $\cal S$, ${\cal S}^\times$ contains
$L^2({\mathbb R})$ and, with the weak topology on ${\cal
S}^\times$, $L^2({\mathbb R})$ is dense in ${\cal S}^\times$.

4.- Consequently, the triplet

\begin{equation}
{\cal S}\subset L^2({\mathbb R}) \subset {\cal S}^\times\label{a6}
\end{equation}
is a RHS.

The notion of Schwartz space can be extended to more dimensions in
the configuration space. Thus, let us consider the vector space of
the indefinitely differentiable functions $f(x_1,\dots,x_q)$ from
${\mathbb R}^q$ into $\mathbb C$, such that

\begin{equation}
\lim_{||{\bf x}||\mapsto\infty}\left|
\frac{\partial^\alpha}{\partial x_1^{\alpha_1}\dots
\partial x_q^{\alpha_q}}x_1^{\beta_1}\dots x_q^{\beta_q}\,f(x_1,\dots,x_q)
\right|=0 \,,\label{a7}
\end{equation}
where $\alpha_i$ and $\beta_j$ are arbitrary nonegative integers
with $\alpha=\alpha_1+\dots+\alpha_q$. This new Schwartz space is
often denoted ${\cal S}({\mathbb R}^q)$ (observe that ${\cal
S}({\mathbb R})\equiv \cal S$). The properties of ${\cal
S}({\mathbb R}^q)$ are discussed in standard textbooks
\cite{RSII,R} and give that the triplet

$$
{\cal S}({\mathbb R}^q)\subset L^2({\mathbb R}^q)\subset {\cal
S}^\times({\mathbb R}^q)\,,
$$
where ${\cal S}^\times({\mathbb R}^q)$ is the antidual of ${\cal
S}({\mathbb R}^q)$, is a new RHS.

The Schwartz space in all its versions has a topological advantage
over other topological vector spaces: it is metrizable, i.e., its
topology can be derived from a distance compatible with the
structure of vector space. There are other interesting topological
features on ${\cal S}({\mathbb R}^q)$ such as nuclearity that may
be interesting like, for instance, the Dirac representation theory
for some operators \cite{AN,GG}, although this properties will not
be discussed in here. One of the interests of the Schwartz space
is that can be used to construct other types of RHS useful for the
precise definition of the Gamow vectors for resonances. We shall
discuss this construction in the next subsection.

Other type of RHS includes nonmetrizable test vector spaces. The
most popular example in this category is

$$
{\cal D}\subset L^2({\mathbb R})\subset{\cal D}^\times\,,
$$
where $\cal D$ is the vector space of functions in $\cal S$ that
vanish outside of a bounded interval. The space $\cal D$ is dense
in $L^2({\mathbb R})$ (and hence in $\cal S$) with the Hilbert
space topology $\tau_{\cal H}$ and in $\cal S$ with its metrizable
topology. Therefore, in order to distinguish $\cal D$ from $\cal
S$, we have to endow $\cal D$ with a topology different from the
inherited from $\cal S$. This topology is nonmetrizable but still
keeps some properties that make it tractable almost as if it were
metrizable\footnote{This is a strict inductive limit
topology\cite{VKK} of metrizable spaces. Problems of continuity of
operators on $\cal D$ can be investigated in each one of the
metrizable spaces that construct the inductive limit, because of
the Dieudonn\'e-Schwartz theorems \cite{VKK}. That is why we said
that the topology is ``almost metrizable''.}.

This type of RHS can be generalized to higher dimensions in the
configuration space. The vector space ${\cal D}({\mathbb R}^p)$ is
the subspace of the functions ${\cal S}({\mathbb R}^p)$  that
vanish outside a compact set in ${\mathbb R}^p$, endowed with a
nonmetrizable topology of the strict inductive limit type
\cite{VKK}. We he the following RHS

$$
{\cal D}({\mathbb R}^p)\subset L^2({\mathbb R}^p)\subset {\cal
D}({\mathbb R}^p)^\times\,,
$$
for each positive integer $p$. Note that for $p=1$, we obtain the
previously discussed case.

\subsection{Rigged Hilbert Spaces of Complex Analytic Functions.}

So far, we have introduced the most basic examples of rigged
Hilbert spaces of functions. We shall not extend on their
properties into here, addressing the interested reader to the
specialized literature \cite{GII,GIV}.  With the help of those, we
can construct the triplets relevant for the definition of Gamow
vectors.

\subsubsection{Rigged Hilbert Spaces of Hardy functions.}

To begin with, let us denote the open upper half plane (the set of
complex numbers with positive imaginary part) as ${\mathbb C}^+$.
Analogously, ${\mathbb C}^-$ will be the open lower half plane,
containing the complex numbers with negative imaginary part.

A Hardy function $f_+(z)$ on the upper half plane ${\mathbb C}^+$
is a complex analytic function on ${\mathbb C}^+$ such that
\cite{H,D,KI,KII}:

\begin{equation}
\sup_{y>o}\int_{-\infty}^\infty
|f(x+iy)|^2\,dx<K<\infty\,,\label{a8}
\end{equation}
where $K$ is certain positive real number. The inequality
(\ref{a8}) means that for each $y>0$ the function on the variable
$x$, $f(x+iy)$, is square integrable and all the integrals for
$y>0$ are uniformly bounded by $K$.

The set of Hardy functions on the upper half plane form a vector
space that we denote as ${\cal H}^2_+$. Functions in ${\cal
H}^2_+$ have well defined (save for a set of zero Lebesgue
measure) boundary values on the real line $\mathbb R$. The
function $f_+(x)$ of these boundary values is also is also square
integrable. The Hardy function $f_+(z)$ determines uniquely (save
for a set of zero measure) the boundary function $f_+(x)$. The
reciprocal is also true and is a consequence of the well known
Titchmarsh theorem \cite{T}, so that $f_+(x)$ determines uniquely
the Hardy function $f_+(z)$ and, therefore, we can identify both.

The definition of Hardy function on the lower half plane is
similar, changing $y>0$ by $y<0$. The vector spaces of Hardy
functions on the upper and lower half planes are denoted by ${\cal
H}_+^2$ and ${\cal H}_-^2$ respectively. Relevant properties of
these spaces can be found in the standard literature
\cite{H,D,KI,KII}. We list a few of them in here:

\smallskip
i.) It is very convenient to identify each function in
$f_\pm(z)\in{\cal H}_\pm^2$ with the function $f_\pm(x)$ of its
boundary values on the real line ${\mathbb R}$ (allowed by the
Titchmarsh theorem that also applies for ${\cal H}_-^2$). The
functions $f_\pm(x)$ are square integrable and therefore ${\cal
H}_\pm^2\subset L^2({\mathbb R})$.

ii.) There is a simple way to construct both ${\cal H}_+^2$ and
${\cal H}_-^2$ which is a straightforward consequence of the {\it
Paley-Wienner} theorems \cite{PW,KII,R}, which state that the
Fourier transformation $\cal F$ is a {\it unitary mapping} between
the Hilbert spaces $L^2({\mathbb R}^\mp)$ and ${\cal H}_\pm^2$,
i.e.,

\begin{eqnarray}
{\cal F}:\, L^2({\mathbb R}^+)\equiv {\cal
H}_-^2\hskip0.5cm;\hskip0.5cm {\cal F}:\, L^2({\mathbb R}^-)\equiv
{\cal H}_+^2\,.\label{a9}
\end{eqnarray}

iii.) Consequently, ${\cal H}_\pm^2$ are closed subspaces of
$L^2({\mathbb R})$. Furthermore:

\begin{equation}
L^2({\mathbb R})={\cal H}_+^2\oplus {\cal H}_-^2\,.\label{a10}
\end{equation}
Thus, the Hilbert space of square integrable functions on the real
line is a direct orthogonal sum of the spaces of Hardy functions
${\cal H}_+^2$ and ${\cal H}_-^2$. Then, any square integrable
function $f(x)$ can be uniquely written as

$$
f(x)=f_+(x)+f_-(x)
$$
with $f_\pm(x)\in{\cal H}_\pm^2$.

iv.) Hardy functions can be recovered from its boundary values on
a semiaxis of the real line. As a consequence of a result by van
Winter \cite{VW} all values of any function in ${\cal H}_\pm^2$
(including boundary values on ${\mathbb R}^-$) can be obtained
from its boundary values on the positive semiaxis ${\mathbb R}^+$,
by means of a formula that uses the Mellin transform \cite{VW}.

v.) All the above properties (and some further results) are
essential in the construction of new rigged Hilbert spaces
\cite{BG,G}:

\begin{equation}
{\cal H}_\pm^2\cap {\cal S}\subset {\cal H}_\pm^2 \subset ({\cal
H}_\pm^2\cap {\cal S})^\times\,,\label{a11}
\end{equation}
where $\cal S$ is the Schwartz space. The space ${\cal
H}_\pm^2\cap {\cal S}$ is endowed with the topology on $S$ and is
a closed subspace of $\cal S$.

The van Winter formula establishes a one to one correspondence
between Hardy functions and the functions of their boundary values
on ${\mathbb R}^+$. Thus, if we denote by ${\cal H}_\pm^2\cap
{\cal S}\Big|_{{\mathbb R}^+}$ the space of these boundary value
functions, we have that

\begin{equation}
{\cal H}_\pm^2\cap {\cal S}\Big|_{{\mathbb R}^+} \subset
L^2({\mathbb R}^+)\subset \left({\cal H}_\pm^2\cap {\cal
S}\Big|_{{\mathbb R}^+}\right)^\times \label{a12}
\end{equation}
is a pair of new RHS. The spaces ${\cal H}_\pm^2\cap {\cal
S}\Big|_{{\mathbb R}^+}$ are endowed with an special topology
which is derived from the topology on $\cal S$ \cite{G,BG}.

\subsubsection{RHS of entire analytic functions.}

Let us consider the vector space of the Schwartz functions that
vanish outside a compact set included in the negative semiaxis
(these functions are zero on the positive semiaxis ${\mathbb
R}^+$), here denoted as ${\cal D}({\mathbb R}^-)$. The space of
the Fourier transformations ${\cal Z}_-\equiv {\cal F}({\cal
D}({\mathbb R}^-))$ is a vector space of entire analytic functions
which are Hardy on the upper half plane. Analogously ${\cal
Z}_+\equiv {\cal F}({\cal D}({\mathbb R}^+))$ is a space of entire
analytic functions which are Hardy on the lower half plane. The
spaces ${\cal Z}_\pm$ have a nonmetrizable topology obtained from
the topology on ${\cal D}({\mathbb R})$ \cite{GII}\footnote{The
spaces ${\cal Z}_\pm$ are not discussed in this reference,
although their properties are obvious from those of $\cal
Z\equiv{\cal F}({\cal D}({\mathbb R}))$ discussed in the given
reference and the properties of Hardy functions.}. We have new RHS
given by

\begin{equation}
{\cal Z}_\pm\subset {\cal H}_\pm^2\subset {\cal
Z}_\pm^\times\label{a12}
\end{equation}
and

\begin{equation}
{\cal Z}_\pm\Big|_{{\mathbb R}^+}\subset L^2({\mathbb R}^+)
\subset \left({\cal Z}_\pm\Big|_{{\mathbb
R}^+}\right)^\times\,.\label{a13}
\end{equation}

These RHS have not been of great use in the description of
resonances, although they are of some interest in the discussions
of irreversibility or time reversal invariance of the current
formalisms on resonance phenomena \cite{CGL}.

\subsection{A rigged Hilbert space for the Friedrichs model.}

In the case of the standard Fridrichs model as presented in
Section 2, we can easily construct the rigged Hilbert space in
which all the generalized objects that appear in this situation
make sense.

The point of departure is the free Hamiltonian $H_0$ in
(\ref{10}). According to the Gelfand-Maurin theorem, introduced in
Appendix A, there is a rigged Hilbert space ${\bf\Phi}\subset
{\cal H}\subset {\bf\Phi}^\times$, such that i.)
$H_0{\bf\Phi}\subset {\bf\Phi}$, ii.) $H_0$ is continuous on
$\bf\Phi$ with its own topology and iii.) for any
$\omega\in{\mathbb R}^+$, the absolutely continuous spectrum of
$H_0$, there exists a $|\omega\rangle\in{\bf\Phi}^\times$ with
$H_0|\omega\rangle=\omega\,|\omega\rangle$.

Let us recall (\ref{a16}), let us omit the arbitrary
$\varphi,\psi\in\bf\Phi$ and let us choose $f(\omega)=\omega$ and
$A\equiv H_0$ (in order to be consistent with the usual notation
for the eigenvalues of $H_0$, we shall use $\omega$ instead of
$\lambda$ in (\ref{a15}) and (\ref{a16})). Then, according to the
Gelfand-Maurin theorem,  one gets for the continuous part of $H_0$
the following generalized expansion

\begin{equation}\label{r1}
\int_0^\infty |\omega\rangle\langle\omega|\,d\mu(\omega)\,.
\end{equation}
As the continuous spectrum of $H_0$ is purely absolutely
continuous and non-degenerated, (\ref{r1}) can be written as
\cite{BGW}

\begin{equation}\label{r2}
\int_0^\infty |\omega\rangle\langle\omega|\,d\omega\,.
\end{equation}
In addition to the continuous spectrum, $H_0$ has an eigenvector
that we have called $|1\rangle$
($H_0|1\rangle=\omega_0\,|1\rangle$). Then, the complete expansion
of $H_0$ in terms of its generalized eigenvectors with eigenvalues
on its Hilbert space spectrum is given by (\ref{10}), i.e.,

\begin{equation}\label{r3}
H_0=\omega_0\,|1\rangle\langle 1|+\int_0^\infty
|\omega\rangle\langle\omega|\,d\omega\,.
\end{equation}

The explicit form of a RHS that implements in our case the
conditions of the Gelfand-Maurin theorem is easy to find. In fact,
we shall have a pair of RHS that satify these conditions. In
addition, both members of the pair of Gamow vectors that describes
the resonance will be functionals on each of these RHS. These RHS
are very easy to construct. In fact, as $H_0$ has a bound state
and a non-degenerate continuous spectrum covering the positive
semiaxis, the minimal Hilbert space in which $H_0$ acts is given
by

\begin{equation}\label{r4}
{\cal H}={\mathbb C}\oplus L^2({\mathbb R}^+)\,,
\end{equation}
where $\mathbb C$ denotes the set of complex numbers and $\oplus$
direct sum. All vectors in (\ref{r4}) have the form (\ref{3}).

In our case, the vector space of test functions should have certain
analyticity conditions so that the antiduals contain the Gamow
vectors. Several motivations (see
\cite{BG,AG,B,BKW,BGM,BGW,BGMI,MBG}) suggest the use of functions
which are at the same time Hardy and Schwartz, so that the spaces of
test vectors are given by

\begin{equation}\label{r5}
{\bf\Phi}_\pm:={\mathbb C}\oplus \left({\cal S}\cap {\cal
H}_\pm^2\Big|_{{\mathbb R}^+}\right)\,,
\end{equation}
where $\cal S$ denotes the Schwartz space, ${\cal H}_\pm^2$ the
space of Hardy functions on the upper (with $+$) and lower (with
$-$) half planes. These functions are restricted to the positive
semiaxis because they represent wave functions in the energy
representation and we assume that the energy is always positive.
Note that the values of a Hardy function on the positive semaixis
determine all its values as a consequence of a Theorem due to van
Winter \cite{VW}. Finally, the RHS for the basic Friedrichs model
is the following:

\begin{equation}\label{r6}
    {\mathbb C}\oplus \left({\cal S}\cap {\cal
H}_\pm^2\Big|_{{\mathbb R}^+}\right) \subset {\mathbb C}\oplus
L^2({\mathbb R}^+)\subset {\mathbb C}\oplus \left({\cal S}\cap
{\cal H}_\pm^2\Big|_{{\mathbb R}^+}\right)^\times\,.
\end{equation}
Note that the duals ${\bf\Phi}_\pm^\times\equiv \left({\mathbb
C}\oplus \left({\cal S}\cap {\cal H}_\pm^2\Big|_{{\mathbb
R}^+}\right)\right)^\times\equiv {\mathbb C}\oplus \left({\cal
S}\cap {\cal H}_\pm^2\Big|_{{\mathbb R}^+}\right)^\times$. The
Gamow vectors $|f_n\rangle$, $n=0,1$, belong to
${\bf\Phi}_-^\times$ and the Gamow vectors $|\widetilde
f_n\rangle$, $n=0,1$, belong to ${\bf\Phi}_+^\times$.

We recall that the potential is given by (\ref{11}), i.e.,

\begin{equation}\label{r7}
V=\int_0^\infty (f^*(\omega)\,|\omega\rangle\langle
1|+f(\omega)\,|1\rangle\langle \omega|)\,d\omega\,.
\end{equation}

Note that if $f(\omega)$ is a square integrable function, then $V$
is a bounded operator from $\cal H$ into $\cal H$. The proof goes
as follows: Take $\psi$ as in (\ref{3}), then

\begin{eqnarray}
  ||V\psi||^2 &=& \left|\int_0^\infty f(\omega)\,\varphi(\omega)\,d\omega\right|^2+
   |\alpha|^2\,\int_0^\infty |f(\omega)|^2\,d\omega \nonumber
   \\[2ex]
   &\le & ||f||^2\,||\varphi||^2+|\alpha|^2\,||f||^2=||f||^2(|\alpha|^2+||\varphi||^2)
   \nonumber\\[2ex]
  &=& ||f||^2\,||\psi||^2\,. \label{r8}
\end{eqnarray}
Since $\psi$ is arbitrary, we conclude that $V$ is bounded. In
addition, $V$ is Hermitian. Then, a consequence of the
Kato-Rellich theorem  \cite{RSII} establishes that $H:=H_0+V$ is
self adjoint on $\cal H$. This is true for any square integrable
form factor $f(\omega)$. If, in addition, $f(\omega)\in{\cal
S}\cap{\cal H}_\pm^2\Big|_{{\mathbb R}^+}$, $H$ is continuous on
${\bf\Phi}^2_\pm$ respectively and can be extended into a weakly
continuous operator on the antiduals. Then, the action of $H$ on
the Gamow vector makes sense.

It could happen that $f(\omega)$ is not in this class, still being
square integrable, or even that $f(\omega)$ were not square
integrable, being for instance a polynomial in the variable
$\omega$. In this cases and in some others (like for instance if
the form factor $f(\omega)$ is a Dirac delta), $H$ is a continuous
mapping from the space of test functions into the antiduals:

\begin{equation}\label{r9}
H:{\bf\Phi}_\pm\longmapsto {\bf\Phi}_\pm^\times\,.
\end{equation}
In some cases, the domain of $H$ can be extended to some subspaces
of ${\bf\Phi}^\times_\pm$ like for instance to the spaces spanned
by the Gamow vectors. This happens for instance in the two
situations mentioned before of square integrable or polynomial
form factor, also the form factor $\sqrt{\omega}/P(\omega)$ for
the double pole resonances in the Friedrichs model is of this
form. Then, the extension of $H$ into the antiduals cannot be
produced by the duality formula (\ref{a5}), but instead by
completion \cite{AS}. Note that this type of extensions may not be
extended into the whole antiduals.

Other types of more sophisticated Friedrichs model need different
RHS construction. Along the present work, we make a case by case
study of these different types of constructions.

\section{Calculation of $\eta(z)$ in the ordinary Friedrichs model.}

In this appendix, we want to obtain the expression of the reduced
resolvent for the simplest form of the Friedrichs model leading to
equation (\ref{15}). We shall make use of operator theory and
follow the guidelines of \cite{E}. In fact, we want to show that

\begin{equation}\label{c1}
    \langle 1|\frac{1}{z-H}|1\rangle=\left(-z+\omega_0 +\lambda^2\int_0^\infty
\frac{|f(\omega)|^2}{z-\omega}\,d\omega \right)^{-1}\,.
\end{equation}

We start with the demonstration of the so called Friedrichs
condition that reads

\begin{equation}\label{c2}
    QVQ=O\,,
\end{equation}
where $O$ is the zero operator. The action of $O$ on any vector
$\psi\in\cal H$ is the zero vector $0\in\cal H$. Taking
$\psi\in\cal H$ as in (\ref{3}), we have that

\begin{eqnarray}
QVQ\psi=
QVQ \,\left(\begin{array}{c}\alpha\\[2ex] \varphi(\omega)
\end{array}
\right)\nonumber\\[2ex] =QV\, \left(\begin{array}{c}0\\[2ex] \varphi(\omega)
\end{array} \right) =Q\, \left(\begin{array}{c} \int_0^\infty
f(\omega)\,\varphi(\omega)\,d\omega \\[2ex] 0  \end{array}\right)=
\left(\begin{array}{c} 0\\[2ex] 0 \end{array}\right)\,. \label{c3}
\end{eqnarray}
Thus, (\ref{c2}) is proven.

The second step in the proof of (\ref{c1}) is noting that the
projectors $P$ and $Q$ commute with the free Hamiltonian $H_0$.
Since $Q=I-P$, this claim would be shown if we can see that
$PH_0=H_0P$. In fact, if the arbitrary vector $\psi\in\cal H$ is
as in (\ref{3}), we have

\begin{eqnarray}
ph_0\Psi=PH_0\,\left(\begin{array}{c}\alpha\\[2ex] \varphi(\omega)
\end{array}
\right)=P\,\left(\begin{array}{c}\omega_0\alpha\\[2ex] \omega\varphi(\omega)
\end{array}
\right)= \left(\begin{array}{c}\omega_0\alpha\\[2ex] 0
\end{array}
\right) \nonumber\\[3ex]
H_0P\psi=H_0P\,\left(\begin{array}{c}\alpha\\[2ex] \varphi(\omega)
\end{array}
\right)=H_0\,\left(\begin{array}{c}\alpha\\[2ex] 0
\end{array}
\right) = \left(\begin{array}{c}\omega_0\alpha\\[2ex] 0
\end{array}
\right) \,, \label{c4}
\end{eqnarray} as we claimed.

Now, we use the second resolvent identity as given in \cite{WE}:

\begin{equation}\label{c5}
    R(z,H)=R(z,H_0)-\lambda \,R(z,H_0)\,V\, R(z,H)\,,
\end{equation}
where

$$
R(z,H)=\frac{1}{z-H}\equiv (zI-H)^{-1}\;,\hskip1cm
R(z,H_0)=\frac{1}{z-H_0}\equiv (zI-H_0)^{-1}
$$
and $H=H_0+V$. This gives:

\begin{equation} P\,\frac{1}{z-H}\, P = P\,\frac{1}{z-H_0}\, P -\lambda
P\,\frac{1}{z-H_0}\, V \frac{1}{z-H}\,P\,. \label{c6}
\end{equation}
Now, if $I$ is, the identity operator, as usual, we can insert
$P+Q=I$ in the last term of (\ref{c6}) as follows:

\begin{eqnarray}
 P\,\frac{1}{z-H_0}\,(P+Q)\, V \,(P+Q)\,\frac{1}{z-H}\,P\nonumber\\[2ex]
= P\,\frac{1}{z-H_0}\,P\, V \,P\,\frac{1}{z-H}\,P +
P\,\frac{1}{z-H_0}\,Q\,
V \,P\,\frac{1}{z-H}\,P\nonumber\\[2ex]
+P\,\frac{1}{z-H_0}\,P\, V \,Q\,\frac{1}{z-H}\,P +
P\,\frac{1}{z-H_0}\,Q\, V \,Q\,\frac{1}{z-H}\,P\,. \label{c7}
\end{eqnarray}
As $P$ and $Q$ commute with $H_0$ and $PQ=O$, we have that

$$
P\,\frac{1}{z-H_0}\,Q=O
$$
and therefore the second and forth terms in the right hand side of
(\ref{c7}) vanish. We also have that

\begin{equation}
 Q\,\frac{1}{z-H}\,P = Q\,\frac{1}{z-H_0}\,P-\lambda
Q\,\frac{1}{z-H_0}\,V\,\frac{1}{z-H}\,P\,. \label{c8}
\end{equation} The first term of the right hand side of
(\ref{c7}) also vanishes. Then, we have:

\begin{eqnarray}
 Q\,\frac{1}{z-H}\,P= -\lambda\,
Q\,\frac{1}{z-H_0}\,(P+Q)\,V\,(P+Q)\,\frac{1}{z-H}\,P \nonumber\\[2ex]
=-\lambda\, Q\,\frac{1}{z-H_0}\,Q\,V\,Q\,\frac{1}{z-H}\,P
-\lambda\,
Q\,\frac{1}{z-H_0}\,P\,V\,P\,\frac{1}{z-H}\,P \nonumber\\[2ex]
-\lambda\, Q\,\frac{1}{z-H_0}\,P\,V\,Q\,\frac{1}{z-H}\,P
-\lambda\, Q\,\frac{1}{z-H_0}\,Q\,V\,P\,\frac{1}{z-H}\,P
\label{c9}
\end{eqnarray}
 Again, the second and the third term in the right hand side of
(\ref{c9}) vanish. Due to the Friedrichs condition (\ref{c2}),
also the first term vanishes. Thus,

\begin{equation} Q\,\frac{1}{z-H}\,P= -\lambda\,
Q\,\frac{1}{z-H_0}\,Q\,V\,P\,\frac{1}{z-H}\,P \label{c10}
\end{equation}
Inserting (\ref{10}) into (\ref{7}) and then, (\ref{7}) into
(\ref{6}), we have

\begin{eqnarray} P\,\frac{1}{z-H}\, P = P\,\frac{1}{z-H_0}\, P -\lambda
P\,\frac{1}{z-H_0}\,
P\,V\,P\,\frac{1}{z-H}\,P \nonumber\\[2ex]+\lambda^2
P\,\frac{1}{z-H_0}\,V\,Q\,\frac{1}{z-H_0}\,
Q\,V\,P\,\frac{1}{z-H}\,P\,. \label{c11} \end{eqnarray}
 If we
multiply (\ref{c11}) to the left by $(z-H_0)\,P$, we can write

$$
(z-H_0)\,P\,\frac{1}{z-H_0}\,P=P
$$
and therefore,

\begin{equation} P=[(z-H_0)\,P+\lambda\,  P\,V\,P
-\lambda^2\,P\,V\,Q\,\frac{1}{z-H_0}\,Q\,V\,P]\,P\,
\frac{1}{z-H}\,P\,, \label{c12} \end{equation} or

\begin{equation} P\, \frac{1}{z-H}\,P = [G(z)]^{-1}\,P\,, \label{c13} \end{equation}
 where

\begin{equation} G(z)= [(z-H_0)\,P+\lambda\,  P\,V\,P
-\lambda^2\,P\,V\,Q\,\frac{1}{z-H_0}\,Q\,V\,P]\,. \label{c14}
\end{equation}
 Since,

 \begin{eqnarray}
P\,V\,Q\,\frac{1}{z-H_0}\,Q\,V\,P \left(\begin{array}{c} 1\\[2ex]
0\end{array}\right) = P\,V\,Q\,\frac{1}{z-H_0}\,Q\,V \left(\begin{array}{c} 1\\[2ex]
0\end{array}\right) \nonumber\\[2ex]
= P\,V\,Q\,\frac{1}{z-H_0}\,Q \left(\begin{array}{c} 0\\[2ex]
f^*(\omega)\end{array}\right) =  P\,V\,Q\,\frac{1}{z-H_0} \left(\begin{array}{c} 0\\[2ex]
f^*(\omega)\end{array}\right) \nonumber\\[2ex]
= P\,V\,Q \left(\begin{array}{c} 0\\[2ex]
\frac{f^*(\omega)}{z-\omega}\end{array}\right)=P
\left(\begin{array}{c} \int_0^\infty
\frac{|f(\omega)|^2}{z-\omega}\,d\omega
\\[2ex]0\end{array}\right) = \left(\begin{array}{c}
\int_0^\infty \frac{|f(\omega)|^2}{z-\omega}\,d\omega
\\[2ex]0\end{array}\right)\,, \label{c15}
\end{eqnarray} and $P\,V\,P\,|1\rangle= 0$, we have that

\begin{equation} G(z)|1\rangle= \left[z-\omega_0-\lambda^2 \int_0^\infty
\frac{|f(\omega)|^2}{z-\omega}\,d\omega \right]\,|1\rangle\,.
\label{c16} \end{equation} Consequently,

\begin{eqnarray}
\langle 1|\frac{1}{z-H}|1\rangle= \langle
1|P\,\frac{1}{z-H}\,P|1\rangle= \langle 1|[G(z)]^{-1}|1\rangle
\nonumber\\[2ex] =\left[z-\omega_0-\lambda^2 \int_0^\infty
\frac{|f(\omega)|^2}{z-\omega}\,d\omega \right]^{-1}\,,
\label{c17}
\end{eqnarray}
which proves (\ref{c1}).

\end{document}